\documentclass[onecolumn, 12pt, floatfix, altaffilletter, superscriptaddress, preprintnumbers,showpacs,nofootinbib, showkeys]{revtex4-1}

\pdfoutput=1
\usepackage[utf8]{inputenc}
\usepackage[colorlinks=true, citecolor=purple, linkcolor=blue]{hyperref}
\usepackage{amsmath,amssymb}
\usepackage{pifont}
\usepackage{epsfig}
\usepackage{graphicx}               
\usepackage{url}
\usepackage{color}
\usepackage{slashed}
\usepackage{multirow}
\usepackage{placeins}
\usepackage[dvipsnames]{xcolor}
\usepackage{epstopdf}
\usepackage{soul}
\usepackage{tikz}
\usepackage{mathtools}
\usepackage{xcolor}
\usepackage{adjustbox}
\usepackage{float}

\usepackage[medium]{titlesec}

\newcommand{\beq}{\begin{equation}}
\newcommand{\eeq}{\end{equation}} 
\newcommand{\bea}{\begin{eqnarray}}
\newcommand{\eea}{\end{eqnarray}}
\newcommand{\ba}{\begin{array}}
\newcommand{\ea}{\end{array}}

\definecolor{pink}{RGB}{255,105,180}

\begin{document}

\title{\Large\textbf{
Enhancing Sensitivities to Long-lived Particles  with \\ High Granularity Calorimeters at the LHC
}}

\author{Jia Liu}
\email{liuj1@uchicago.edu}
\affiliation{Enrico Fermi Institute, University of Chicago, Chicago, IL 60637, USA}

\author{Zhen Liu}
\email{zliuphys@umd.edu}
\affiliation{Maryland Center for Fundamental Physics, Department of Physics, University of Maryland, College
	Park, MD 20742, USA}

\author{Lian-Tao Wang}
\email{liantaow@uchicago.edu}
\affiliation{Enrico Fermi Institute, University of Chicago, Chicago, IL 60637, USA}
\affiliation{Department of Physics, University of Chicago, Chicago, IL 60637, USA}

\author{Xiao-Ping Wang}
\email{xia.wang@anl.gov}
\affiliation{High Energy Physics Division, Argonne National Laboratory, Argonne, IL 60439, USA}

\date{\today}

\preprint{    EFI-20-8 }

\begin{abstract}
The search for long-lived particles (LLP) is an exciting physics opportunity in the upcoming runs of the Large Hadron Collider. 
In this paper, we focus on a new search strategy of using the High Granularity Calorimeter (HGCAL), part of the upgrade of the CMS detector, in such searches. In particular, we demonstrate that the high granularity of the calorimeter allows us to see ``shower tracks" in the calorimeter, and can play a crucial role in identifying the signal and suppressing the background. We study the potential reach of  the HGCAL using a signal model in which the Standard Model Higgs boson decays into a pair of LLPs, $h \to XX$. After carefully estimating the Standard Model QCD and the misreconstructed fake-track backgrounds, we give the projected reach for both an existing vector boson fusion trigger and a novel displaced-track-based trigger. 
Our results show that the best reach for the Higgs decay branching ratio, BR$(h \to XX)$, in the vector boson fusion channel is
about $\mathcal{O}(10^{-4})$ with lifetime  $c\tau_X \sim 0.1$--$1$ meters, while for the gluon gluon fusion
channel it is about $\mathcal{O}(10^{-5}\text{--}10^{-6})$ for similar lifetimes. 
For longer lifetime $c\tau_X \sim 10^3$ meters, our search could probe  BR$(h\to XX)$ down to a few $\times 10^{-4}$($10^{-2}$) in the gluon gluon fusion (vector boson fusion) channels, respectively. 
In comparison with these previous searches, our new search shows enhanced sensitivity in complementary regions of the LLP parameter space. We also comment on many improvements can be implemented to further improve our proposed search.
\end{abstract}

\maketitle
\tableofcontents

\setcounter{secnumdepth}{2} 
\setcounter{tocdepth}{2}

\section{Introduction}

Models of new physics beyond the Standard Model (BSM) often predict the existence of long-lived particles (LLPs), 
giving rise to distinct signatures at colliders (see ~\cite{Alimena:2019zri} for a recent review). 
There have been many searches for LLPs at the ATLAS, CMS, and LHCb experiments at the Large Hadron Collider (LHC). 
The signatures of the LLP depend on its charge,  lifetime, and decay products. 
Accordingly, various search strategies and detection techniques can be used, including 
the non-prompt photon detection using the electromagnetic (EM) calorimeter \cite{Aad:2014gfa, CMS-PAS-EXO-19-005}, 
the disappearing track searches
based on the tracking system \cite{Aad:2013yna, Aaboud:2017mpt, Sirunyan:2018ldc}, and the displaced leptons or lepton jets searches based on the tracking system \cite{Aad:2015rba, Aaboud:2017iio, CMS:2014hka, Sirunyan:2018pwn, Sirunyan:2018vlw, CMS-PAS-EXO-16-022, CMS-PAS-EXO-19-021},
the calorimeter \cite{Aaboud:2019opc, Sirunyan:2017sbs},  as well as the muon system \cite{Aad:2015uaa, Aaboud:2018jbr, Aaboud:2018aqj}. Many new search targets and strategies for the LLPs based on the LHC experiment have also been proposed \cite{Heisig:2012zq, Barry:2013nva, Helo:2013esa, Cui:2014twa, Anelli:2015pba, Alekhin:2015byh,
Co:2015pka, Liu:2015bma, Evans:2016zau, Evans:2016zau, Accomando:2016rpc, Dev:2016vle, Aaij:2016isa, Coccaro:2016lnz, Antusch:2016vyf, 
Buchmueller:2017uqu, Mahbubani:2017gjh, Khoze:2017ixx,	Ghosh:2017vhe, Dev:2017dui, Gligorov:2017nwh, Feng:2017uoz,
Abada:2018sfh, Evans:2018jmd, Kribs:2018ilo, Berlin:2018jbm, Lara:2018rwv, Nemevsek:2018bbt, Helo:2018qej, Cottin:2018kmq,  Kilic:2018sew, Dev:2018kpa,
Calibbi:2018fqf, Ariga:2018uku, Curtin:2018ees,Curtin:2018mvb,  Gligorov:2018vkc,Aaboud2018, Das:2018tbd,
Drewes:2019fou, Du:2019mlc, Liu:2019ayx, Lubatti:2019vkf, Drewes:2019vjy, Aielli:2019ivi, Bauer:2019vqk, Bondarenko:2019tss, Filimonova:2019tuy,Serra:2019omd, Arguelles:2019ziu, Cheung:2019qdr, Bhattacherjee:2019fpt, Chiang:2019ajm, Das:2019fee,
Li:2020aoq, Yuan:2020eeu, Acharya:2020uwc, Bhattacherjee:2020nno, Shuve:2020evk, Alimena:2020web,Felea:2020cvf}. 
 
In this work, we focus on a new sub-detector HGCAL, a highly granular and
silicon-based calorimeter, which is the Phase-2 upgrade of the CMS endcap calorimeter~\cite{Collaboration:2293646}. 
It consists of a sampling calorimeter with silicon and scintillators as active material, including both the electromagnetic and the hadronic sections with unprecedented fine segmentation.
In particular, each section consists of silicon cells of size (0.5 -- 1 $ {\rm cm^2}$) and the remainder of the hadron calorimeter will use highly-segmented plastic scintillators of size  (4 -- 30 $ {\rm cm^2}$)~\cite{Collaboration:2293646}. 
It has an intrinsic high-precision timing capability from silicon sensors with a resolution of $\sim 25$ ps. Due to its fine transverse granularity,
the HGCAL has an angular resolution of about $5 \times 10^{-3}$ radians for electromagnetic shower with $p_T > 20$ GeV, after taking into account the broadening effect from the shower. 
The HGCAL can handle different LLPs signatures. It also serves as a semi-forward detector different from most LLP studies at LHC main detectors that are mainly based on central detectors.~\footnote{For the consideration of non-pointing photon at HGCAL for the triggering, see \cite{Alimena:2020web}.}

\begin{figure}[h!]
    \includegraphics[width=0.49 \columnwidth]{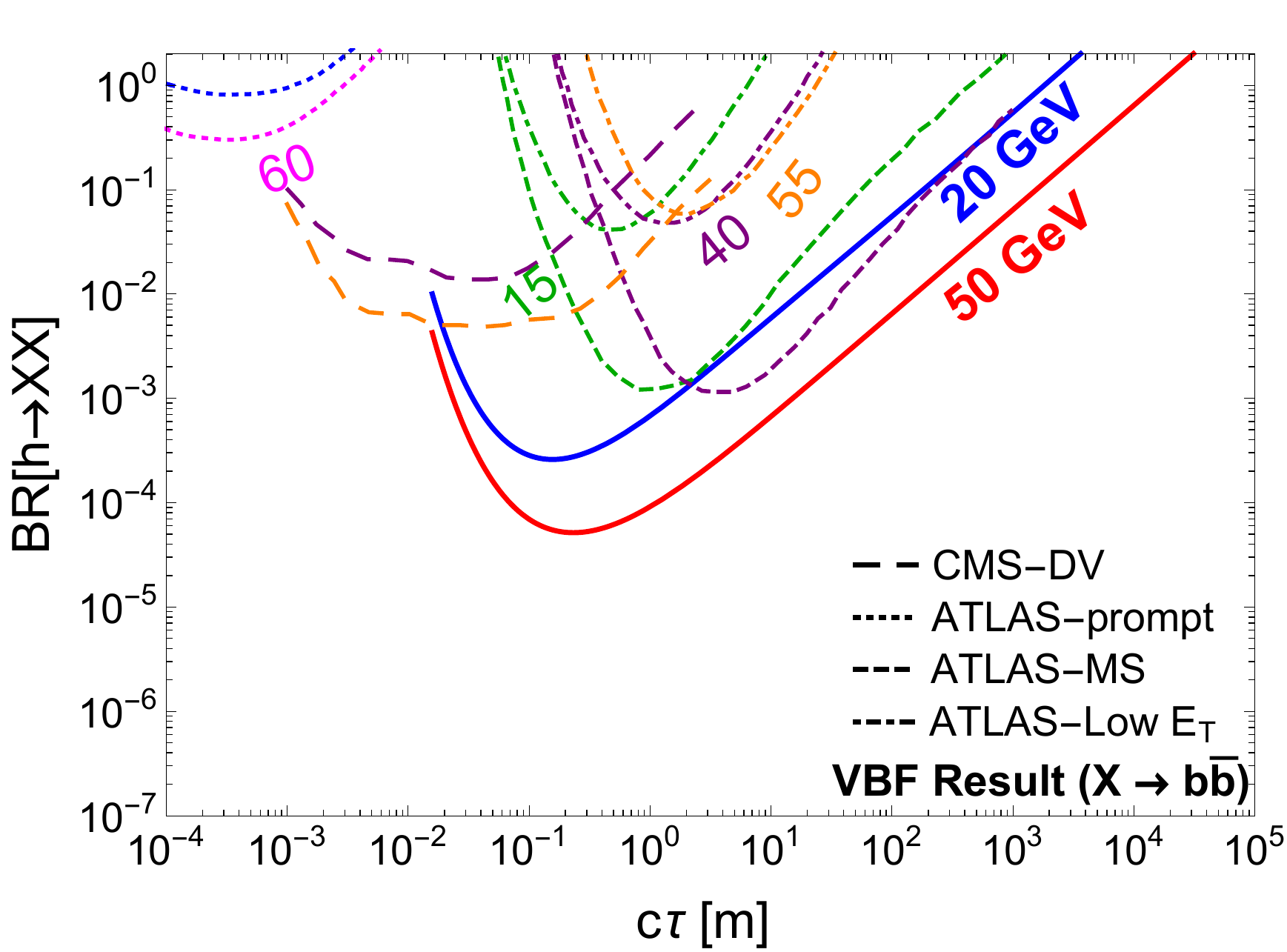}
    \includegraphics[width=0.49 \columnwidth]{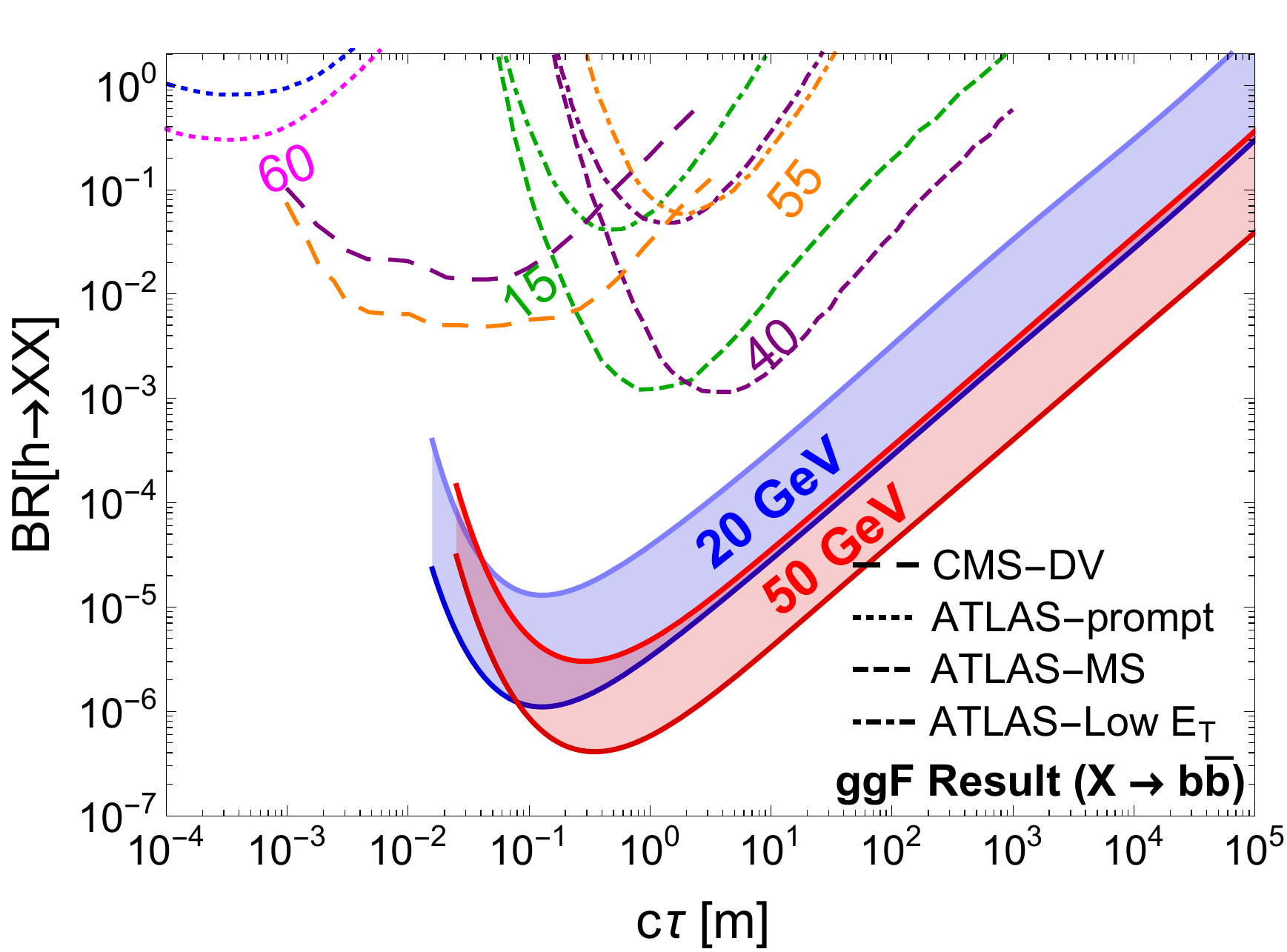}
    \caption{The projected sensitivity for Higgs decays to long-lived particles with VBF trigger (left panel) and a displaced track trigger for the ggF channel (right panel) at the HL-LHC ($3~ \text{ab}^{-1}$) as a function of proper lifetime of $X$ using our proposed HGCAL LLP search. 
    We consider two scenarios of the displaced track trigger. The solid line on the top of the shaded region corresponds to the reach with a trigger requirement of  $H_T>100$~GeV, while the 
     solid line on the bottom of the shaded region is obtained without such additional requirement. 
    The existing limits for BR($h\to XX$) from ATLAS Run 2 searches based on prompt VH \cite{Aaboud:2018iil} (dotted), the muon spectrometer \cite{Aaboud:2018aqj} (dashed), the calorimeter \cite{Aaboud:2019opc} (dot-dashed),  with integrated luminosity of $36~\text{fb}^{-1}$,  and the CMS search based on displaced vertex in the tracker system \cite{CMS-PAS-EXO-19-021} (long dashed) with integrated luminosity of $132~\text{fb}^{-1}$,  are also shown for comparison.  The numbers on different colored lines indicate the mass of the LLP in units of GeV for the corresponding searches. 
    }
    \label{fig:ggF}
\end{figure}

We carefully simulated and estimated the SM background for generic LLP signals, which contains prompt and displaced QCD background
and non-prompt misconnected fake-track background. Based on these, we design a set of cuts that take advantage of the unique features of the signal and the capabilities of the HGCAL detector.  
We use a signal model in which  scalar LLPs ($X$)  are produced from SM Higgs decay ($h\to X X$).  This simple model is quite representative \cite{Curtin:2013fra}, covering a broad range of new physics scenarios, such as the hidden valley models~\cite{Strassler:2006im,Strassler:2006ri,Han:2007ae}, and more recent proposals motivated by neutral naturalness~\cite{Chacko:2005pe,Chacko:2005un,Burdman:2006tz,Craig:2015pha, Csaki:2015fba, Curtin:2015fna}.
Two production channels of SM Higgs are considered.  One is the vector boson fusion (VBF) channel, together with
the existing VBF trigger. 
The other is the gluon-gluon fusion (ggF) channel with a potential displaced track trigger, enabled by new trigger considerations from the tracker and HGCAL.
The sensitivity of HL-LHC is given as a function of the proper lifetime of $X$, shown in  Fig.~\ref{fig:ggF}. The best reach for VBF channel is
about BR$(h\to XX) \sim \mathcal{O}(10^{-4})$ with a lifetime of  $c\tau_X \sim 0.1$--$1$ meters, while for the ggF
channel it is about BR$(h \to XX ) \sim \mathcal{O}(10^{-5}\text{--}10^{-6})$ for similar lifetime. Alternatively, for an LLP with $c\tau_X \sim 10^3$ meters, the HGCAL based search should be able to probe  BR$(h\to XX)$ down to a few $\times 10^{-4}$($10^{-2}$) in the ggF (VBF) channels, respectively.

The paper is organized as follows. In Sec.~\ref{sec:signalmodel}, we discuss the signal model and the
trigger considerations for the signal. In Sec.~\ref{sec:sigbkgGen}, we describe signal and  background generation. 
In Sec.~\ref{sec:kinematics}, the distributions of kinematic variables are discussed, and the corresponding cuts are applied.
Finally, we show our results in Sec.~\ref{sec:results} and conclude in Sec.~\ref{sec:conclusion}.

\section{Analysis framework}

\subsection{Signal model: long-lived particles from Higgs decay}
\label{sec:signalmodel}

To demonstrate the potential of our proposed search, we use a signal model in which the LLP couples to the SM through the Higgs portal. 
For $m_X < m_h / 2$, the LLP will be produced through the Higgs boson decay
\begin{align}
h \to X X.
\end{align} 
We assume $X$ is a neutral and meta-stable scalar which will further decay via $X \to \bar{b}b$. The free parameters in this simplified model
are mass $m_X$, lifetime $c\tau_X$, and the decay branching ratio ${\rm BR}(h \to XX)$. 

\begin{figure}[h!]
	\includegraphics[width=0.45 \columnwidth]{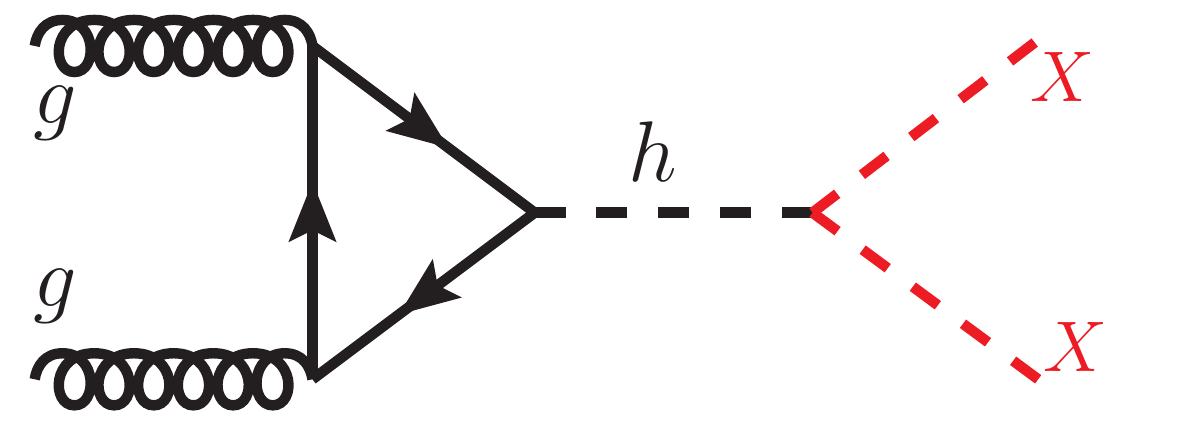}
	\includegraphics[width=0.32 \columnwidth]{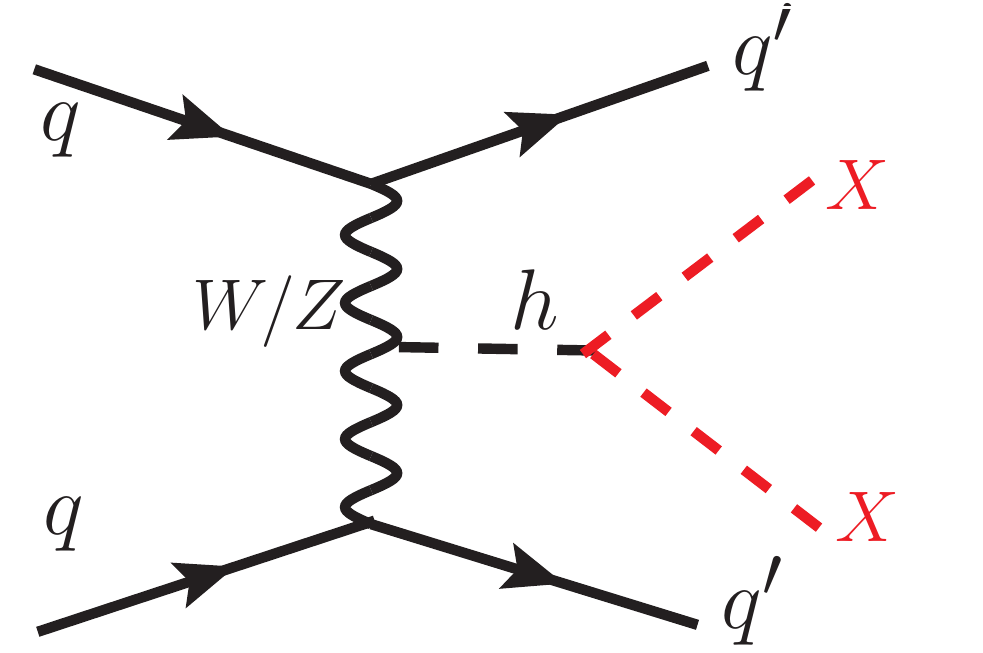}
	\caption{The processes of producing the long-lived particle $X$ from SM Higgs decay  considered in this study.
		Left panel: gluon-gluon fusion Higgs production. Right panel: vector boson
		fusion Higgs production. 
	}
	\label{fig:feyn}
\end{figure}

We consider two Higgs production channels, namely, the VBF production and ggF production, shown in Fig.~\ref{fig:feyn}. The VBF channel is motivated by the possibility of using an existing VBF trigger that does not rely on the properties of the LLP.  In the ggF channel, we will explore the physics potential of using displaced track triggers after LHC Phase-2 upgrades, e.g., Ref.~\cite{Gershtein:2017tsv}. Since the metastable particle $X$ is neutral, it does not leave a track as it travels through the detector.  Subsequently, $X$ decays to $ \bar{b}b$. For our work, we do not use the tagging information of  whether the jets are initiated by heavy or light flavor quarks.~\footnote{In principle, the secondary displacement from the heavy-light mesons, such as $B$ mesons and Kaons can help to identify the specific property of the LLPs. } 

\subsection{Modeling the HGCAL detector}

Our study focuses on the potential of the LLP search of the HGCAL detector~\cite{Collaboration:2293646}. Due to the novelty of the detector and the signature, we cannot perform a full-fledged detector simulation. Instead, we make assumptions based upon the HGCAL performance document. We describe here the relevant detector parameters used in our study.

\begin{figure}[htb]
	\includegraphics[width=0.9\columnwidth]{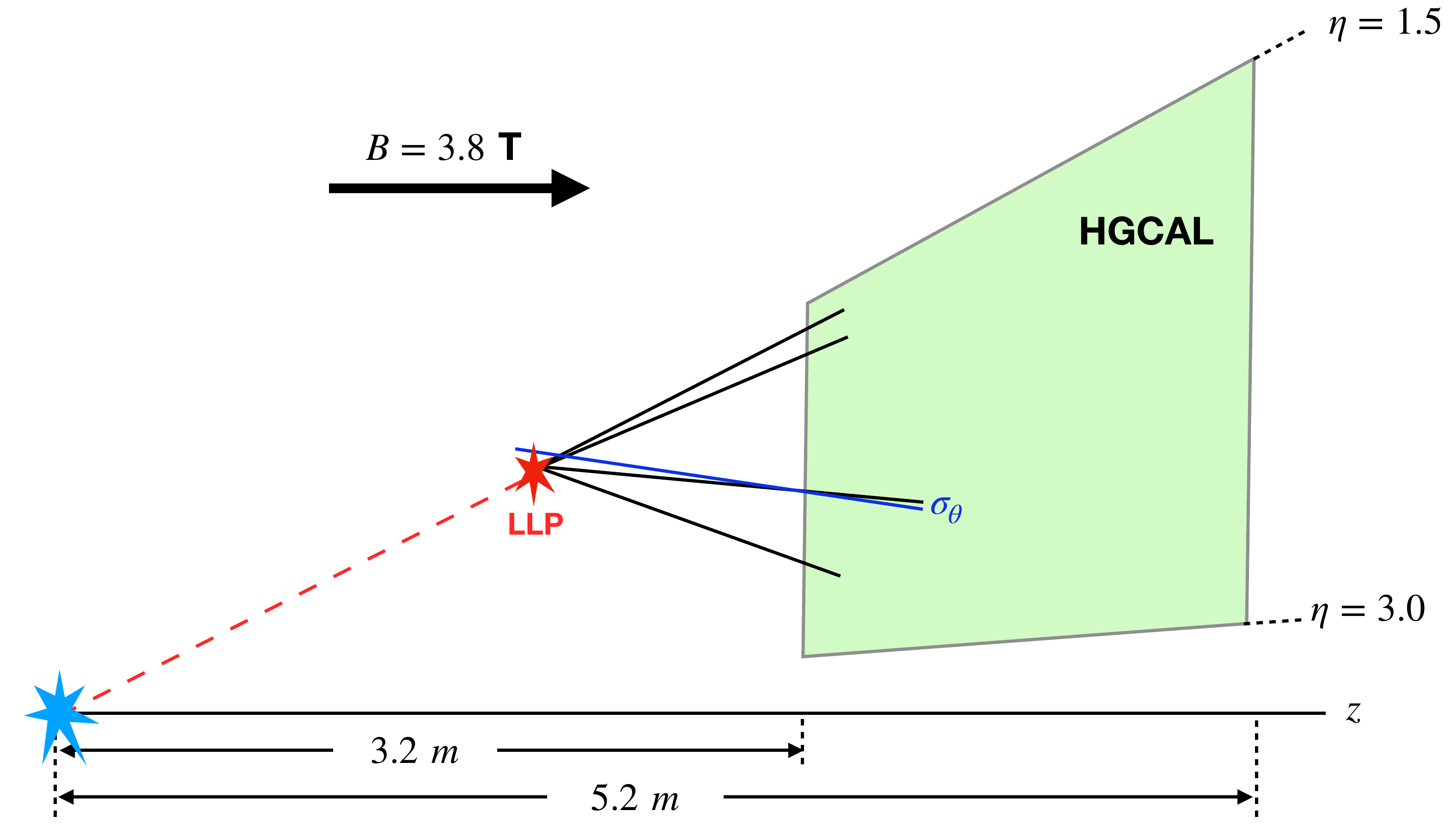}
    \caption{A schematic drawing for the decay products of the  long-lived particle  arriving the HGCAL. The direction of the momentum of the decay products can be measured by the HGCAL with an angular resolution of $\sigma_\theta$,  resulting in an error in reconstructing the displaced vertex. 	
    }
	\label{fig:art}
\end{figure}

The  HGCAL detector locates at $|z| = 3.2$ m and extends to $|z| = 5.2$ m. The angular coverage of the detector
is $1.5 < |\eta| < 3.0$. Its stand-alone angular resolution on the shower direction is taken to be $\sigma_\theta \sim5\times 10^{-3} $ radians,
with possible improvement when combining with the information from the inner detectors. We note here in the text we will not distinguish tracks and shower when discussing HGCAL, since the shower can be viewed as a ``fat track''. In general, the energy deposit pattern in EM calorimeter for photons, electrons and positrons are indistinguishable. Since the EM calorimeter of HGCAL has integrated 28 tracking layers, the resolution for electron and positron should be better. 
For charged hadrons like pions, the track extends from EM part to hadron part of HGCAL, which passes even more tracking layers. 
However, the hadronic tracks come in clusters and may degrade the performance.
We assume that the angular resolution of the hadronic shower is the same as the EM shower.
A subtle difference is HGCAL will be able to see neutral particles shower as well, which traditionally do not 
correspond to tracks.

A schematic plot for the long-lived particle signal arriving the HGCAL is shown in Fig.~\ref{fig:art}.
The particle will travel in a magnetic field of $B=3.8$~T along the $z$ direction, therefore, it would follow a helical trajectory. We require the tracks to go through the first layer of HGCAL at $|z| = 3.2$~m.   The tracks with $p_T$ above 1 GeV can be reconstructed at L1 level~\cite{Collaboration:2293646}. %
Each point on the track trajectory has a 4D coordinate, $(t,x,y,z)$. Once the momentum of a particle at a point on the track  is known, the 4D trajectory of the full track can be calculated.

The directions of particles reaching HGCAL can be measured with an angular resolution of $\sigma_\theta $.
The inaccuracy in measuring its direction is a main source of the error in the measurement of the track direction, which can fake our signal. 
We smear the direction of the momentum using a Gaussian function with a spread equal to the angular resolution $\sigma_\theta $. With this new momentum for the particle at the first layer of HGCAL, we then recalculate its 4D spiral trajectory.

\subsection{Signal and Background generation}
\label{sec:sigbkgGen}

\subsubsection{The long-lived particle signal}

The signal events at parton level are generated using {\tt MadGraph5\textunderscore aMC$@$NLO}~\cite{Alwall:2014hca}, and the parton shower is performed by {\tt Pythia8}~\cite{Sjostrand:2006za, Sjostrand:2007gs}. The charged particles with $p_T >$ 1 GeV are kept as track candidates.

For the signal, the displaced tracks dominantly come from the displaced decay of the LLP $X$, which will give a displaced vertex (DV). The location and time of this DV results from a convolution of $X$ momentum distribution and the lifetime of $X$.
We also require $X$ to decay within $|z| < 1.5$ m to ensure the tracks have five stubs in the tracker.
Given the  4D vertex information and the 4-momentum of each charged particle at that vertex, one can reconstruct its 4D helical trajectory in the magnetic field. From this, we obtain the 3-momentum of the particle when it arrives at the HGCAL. We then smear the direction of its momentum and recalculate the 4D trajectory.

A further improvement of the HGCAL coverage can be achieved by considering LLPs decaying inside HGCAL. The LLP signal would appear as showers with an anomalous  shape in the HGCAL. However, given the difficulty of modeling the showering pattern in this material-dense area and the lack of understanding of the background, we take the rather \textit{conservative} class of signals in which $X$ decay before entering HGCAL. In this case, we use HGCAL to \textit{only} pick out the displaced tracks. These tracks are identified via the showering of the hadronic particles from the LLP decay. Hence, they have a degraded angular resolution than the HGCAL physical limitations due to the broadening caused by interaction with materials.~\footnote{We take this into account by using a degraded angular resolution.} We also require these tracks to match hits in the outer part of the tracking system, which picks only the charged components of the signal. This is clearly a very conservative use of the HGCAL capability and leaves a large room for future improvement with a full understanding of the HGCAL performance.

\subsubsection{SM QCD background}

The main SM prompt backgrounds are the QCD dijet events, including bottom quark pair $b\bar{b}$.
A main feature of the signals is the presence of tracks with large transverse impact parameters. 
There are two reasons for such a QCD background to also have displaced tracks.
The first one is the finite lifetime of mesons and baryons. 
The second one is from the finite angular resolution of HGCAL.

We use {\tt MadGraph5\textunderscore aMC$@$NLO}~\cite{Alwall:2014hca} and {\tt Pythia8}~\cite{Sjostrand:2006za, Sjostrand:2007gs} to generate the SM background events, which properly include the finite lifetime effect of SM mesons and baryons. The displaced tracks come primarily from $K_{S}^0$ meson ($c\tau \sim 2.7$ cm), with some addition contribution from heavy baryons like $\Lambda^0$ ($c\tau \sim 7.8$ cm).

After applying generator level cuts such as $p_T > 20$ GeV at the parton level, the cross-sections of $b\bar{b}$ and $jj$ are $3.6\times 10^6$ pb and $1.7 \times 10^8$ pb~\footnote{Here we use the 4-flavor PDF scheme.}, respectively. The jet matching has been applied with one extra jet added and the minimal $k_t$ is set to be 
30 GeV. 
After hadronization, charged tracks with $p_T> 1$ GeV are kept. Among the tracks arriving at HGCAL, we kept the five leading ones to be smeared~\footnote{This procedure tends to overestimate the suppression provided by our vertexing cuts. However, since our results essentially do not rely on the vertexing cuts for suppressing the SM QCD background, we keep only the five leading tracks for simplicity.}.

\subsubsection{Fake-track background}

We denote as fake-track background the events with mis-reconstructed tracks from the accidental connections of the hits in the tracker system. 
They can easily have very large $d_0$, similar to those from the signal. 
There are $\mathcal{O}(30)$ such tracks per bunch crossing. This high combinatorics makes it possible for a selection of a few tracks to approximately form a vertex.

We follow Ref.~\cite{Gershtein:2019dhy,Hook:2019qoh} to generate events with mis-reconstructed tracks. We also add the timing information to the tracks, which can potentially further reduce the background~\cite{trackreconstruction}.  To generate a fake-track, we use a set of kinematical variables following a flat distribution within the ranges indicated below, which was reproduced by CMS with a full simulation \cite{CMS-PAS-FTR-18-018} and from the estimates of the expected occupancy of the trigger system \cite{Contardo:2020886}.   
\begin{itemize}
    \item $\phi_0 \in \left[0, {2 \pi} \right]$: the azimuthal angle of a reference point from the beam spot in the $x$--$y$ plane.
        \item $z_0 \in \left[-0.15, ~ 0.15\right]$ m:  the $z$ coordinate of the reference point.
    \item $t_0 \in \left[-6, 6\right]$ nanosecond: the time coordinate of the reference point.
    \item $d_0 \in \left[10^{-3}, ~ 0.15\right]$ m: the transverse impact parameter of the track. 
    \item $q/R \in \left[0, \left( 1.75\ \rm{m} \right)^{-1} \right] $: the inverse of the track curvature in $x$--$y$ plane.
        \item $\eta \in  \left[-3, 3\right]$: the pseudo-rapidity of the direction of the track at the reference point. 
\end{itemize}
The reference point is defined at the location of the transverse impact parameter of a given track.
The  curvature of the track and the transverse momentum of the presumed particle responsible for it satisfy $R = |p_T/(q\times B)| = (p_T/{\rm GeV})\times 0.88 \ {\rm m}$. $q$ is the charge of the particle, assumed to be $\pm e$ with equal probability.  
From the range of the curvature, the tracks generated must have $p_T \geq 2$ GeV, with a flat probability in $p_T^{-1}$. 
In the $x$-$y$ plane, the trajectory of the track is a circle with a radius equals to $R$. However, the origin (the beam spot) can be either inside or outside the circle. The distance between the center of the circle and the origin can be either $R - d_0$ or $R + d_0$.  We assume the two cases occur with equal probability. With these parameters, the 4D trajectory of the fake tracks can be determined.

\subsection{Triggering strategy}

For the VBF channel,  we require at least one forward jet $p_T > 110$ GeV, and
both of the forward jets $p_T > 35$ GeV with an invariant mass $m_{jj} > 620 $ GeV \cite{Amendola:2288356}.

For the ggF channel, we try two different trigger strategies. First, we use  a proposed L1 displaced track trigger cuts with $H_T > 100$ GeV,
which has been demonstrated with two displaced tracks with $p_T>2$~GeV within an L1 jet~\cite{Gershtein:2017tsv}. This L1 trigger rate is about 10 kHz in the central region and about a factor of 2--3 higher in the endcap region~\cite{Gershtein:2017tsv}. We require our signals to have more than five displaced tracks and $H_T>100$~GeV, which is more stringent than Ref.~\cite{Gershtein:2017tsv}. Nevertheless, we still assume the same level of L1 trigger rate of 10 kHz. Because that displaced track selection and vertex reconstruction do provide suppression of the L1 rate, the average number of multiple track bundles passing all these trigger requirements should be around one per triggered event. Given that the HL-LHC will run for $10^8$ seconds, the total number of such fake-track bundle events is about $1\times 10^{12}$. 

The second trigger strategy for the ggF channel is a displaced track trigger without the $H_T$ cut. 
It makes use of five displaced tracks with a vertex fitting, rather than the two displaced tracks~\cite{Gershtein:2017tsv}. This should reduce the low-level trigger rate and allow for the removal of the $H_T$ requirement.
We also emphasize that these randomly connected tracks may not be corresponding calorimeter energy deposits in the HGCAL. Even if our estimate of the tracking alone suppression is not sufficient, consistency matching between different sub-detectors of the experiment will provide  sufficient  suppression.

\section{The kinematics of signal and backgrounds}
\label{sec:kinematics}

There are two main characteristics of the signal. First, the signal tracks tend to have large impact parameter, $d_0$. Hence, requiring a number of tracks (five in our case) to have large  $d_0$ allows us to effectively separates the signal from the QCD background, which is mostly prompt. On the other hand, the fake-track background have a flat distribution in a large range of  $d_0$.  This is where the second main characteristic comes into play. Namely,  the signal tracks all originate from a single vertex. Since each fake-track is independent of each other, they have a small probability of reconstructing a common vertex.  In the following, we will define a set of variables to quantify this feature. We note that if the tracks are generated via interaction with detector material, there would be a reconstructable displaced vertex. One could veto all the displaced vertices in the materiel-dense region, as has been done by many LLP searches~\cite{Aad:2011zb,Aaboud:2017iio,Alimena:2019zri}.

\subsection{The Displaced Vertex fitting variables}
\label{sec:precuts}

We fit the candidate tracks to a displaced vertex and define associated fitting variables as follows. We begin with five leading (in $p_T$) tracks and calculate their 4D trajectories. We  perform a 2D vertex fit in the transverse plane by minimizing the following quantity, 
    \begin{align}
    \Delta {\rm D} \equiv \sqrt{\sum_{i=1}^5 \left( \sqrt{(x-x_i^{\rm cen})^2 + (y-y_i^{\rm cen})^2} - R_i \right)^2},
    \label{eq:deltad}
    \end{align} 
    where $\{x_i^{\rm cen}, y_i^{\rm cen} \}$ are the $x$-$y$ coordinates of the center of the circle for the $i$th track and $R_i$ is the transverse radius of the track helix. The minimization gives the best-fit  coordinates , $x$ and $y$, for a presumed DV.  Of course, this fit won't be perfect in reality and the tracks will miss the DV by some amount. To quantify this, we begin by identifying a point, with coordinate $(t_i, x_i, y_i, z_i)$, as the one corresponding to the DV on the $i$th track. Since we have the full 4D trajectory of the track, we only need one parameter to identify this point. To this end, we choose to use the azimuth angle $\phi$ of the direction of the DV, with respect to the center of the circle of the $i$th track. Comparing the $\phi$ change between the best fit DV $(x, y)$ and the hitting point at HGCAL layer, one can use the transverse velocity $v_T$ to fully determine $(t_i, x_i, y_i, z_i)$ for the DV on this particular track. Of course, in the ideal case in which all tracks originating from a DV are perfectly reconstructed, all of the $x_i$ and $y_i$ will coincide with $x$ and $y$.     We can define the following variables associated with a fitted DV.
    \begin{itemize}
    \item The displacement of the vertex in the transverse plane ${\rm r_{DV}} \equiv \sqrt{x^2+y^2}$ that minimizes $\Delta \rm D$ in Eq.~\ref{eq:deltad}. 
    \item The imperfectness or the spread of vertex fitting,  $\Delta  {\rm D}_{\min}$, based on the best-fit 2D vertex coordinates $x$ and $y$ that minimizes $\Delta \rm D$ in Eq.~\ref{eq:deltad}.
 \item Based on the set of $\{z_i, t_i\}$ for each track that form a DV, we can define the mean value $\bar{ z }$ and $\bar{t}$, and their standard deviations $\rm{\sigma_z}$ and $\rm{\sigma_t}$.

 \item For the $i$th track, we define the time delay as $\Delta t_i \equiv t_i - \sqrt{x_i^2+y_i^2+z_i^2}/c$. 
    We define the time delay of the displaced vertex ($\overline{\Delta t}$) as the average of the $\Delta t_i$ of the five leading tracks (in $p_T$) and the standard deviation $\rm{\sigma_{\Delta t}}$. For a slow-moving LLP which decays at the DV, $\Delta t$ would be its time delay in comparison to the prompt particles propagating from the interaction point to the DV. 
\end{itemize}
In summary, we can define the following kinematic variables using the above 2D-4D displaced vertex fitting procedure,
\begin{align}
\rm{r_{DV}}, ~ \Delta  {\rm D}_{\min}, ~ \bar{t}, ~\bar{z}, ~ \overline{\Delta t}, ~\rm{\sigma_t}, ~\rm{\sigma_z}, ~ \rm{\sigma_{\Delta t}}.
\label{eq:7pars}
\end{align}

\begin{figure}[h!]
    \centering
    \begin{tabular}{ccc}
        \includegraphics[width=0.33 \columnwidth]{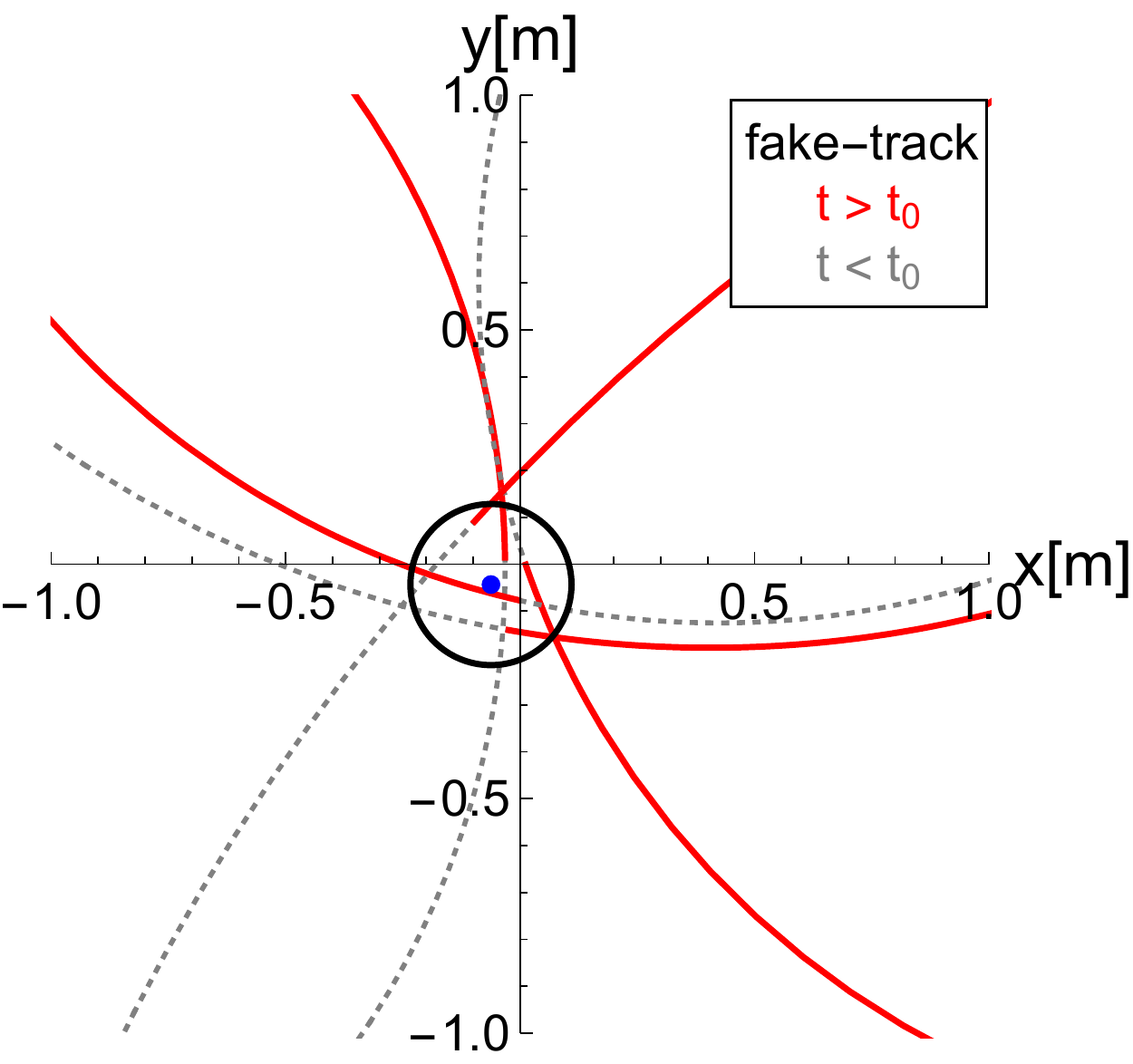} &
        \includegraphics[width=0.33 \columnwidth]{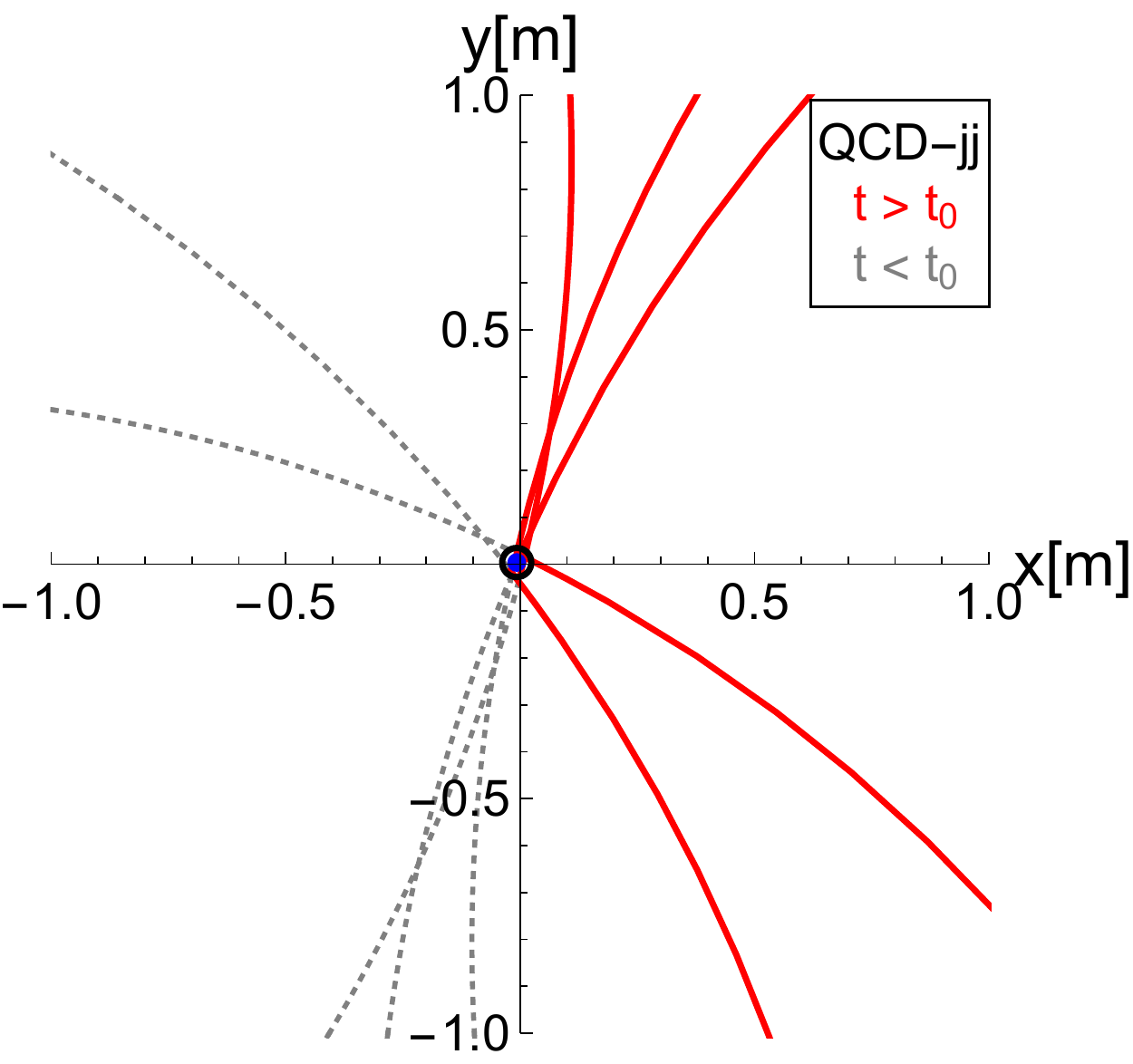} &   
        \includegraphics[width=0.33 \columnwidth]{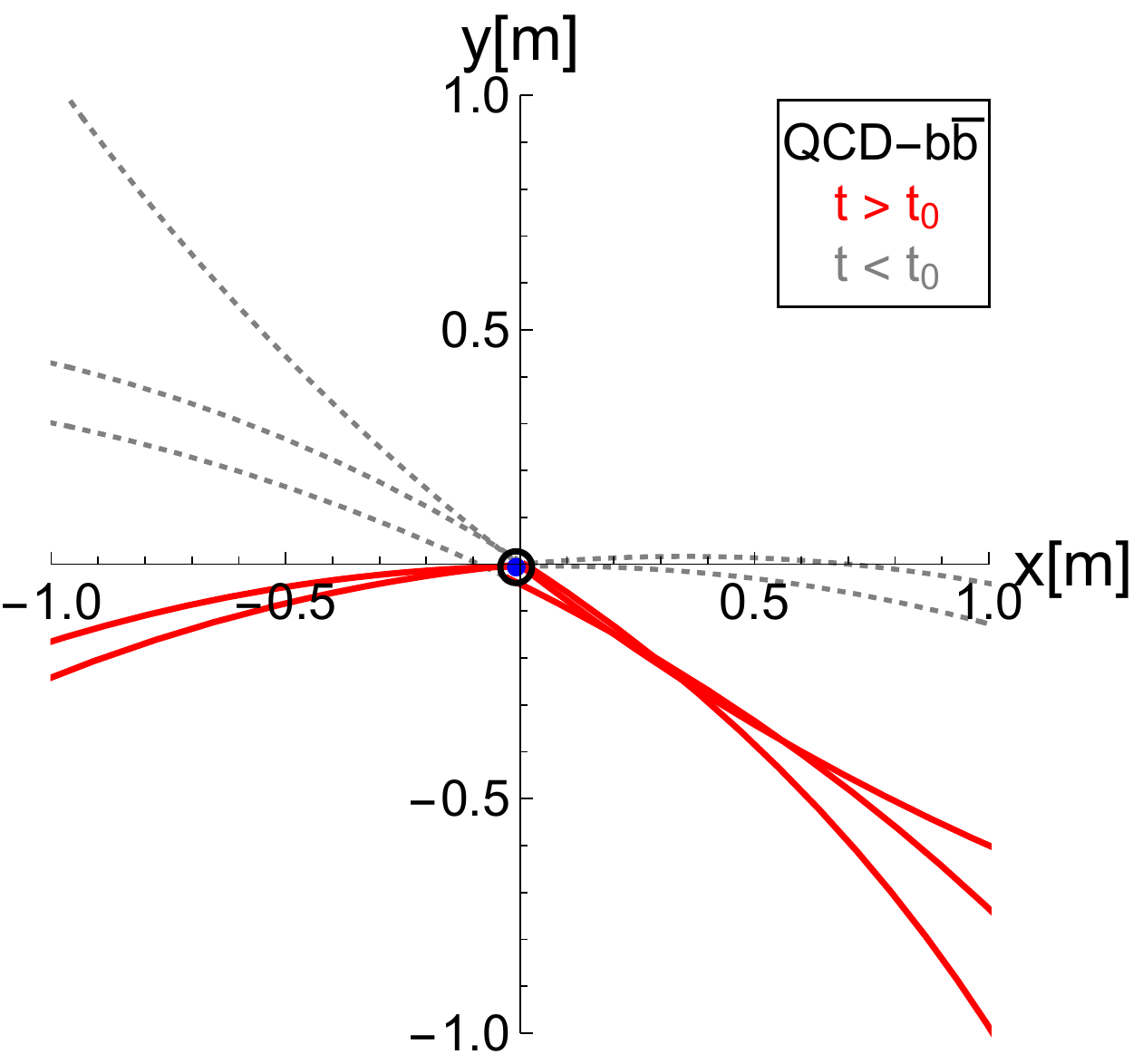} \\
        (a) & (b) & (c)
            \end{tabular}
    \begin{tabular}{cc}
            \includegraphics[width=0.33 \columnwidth]{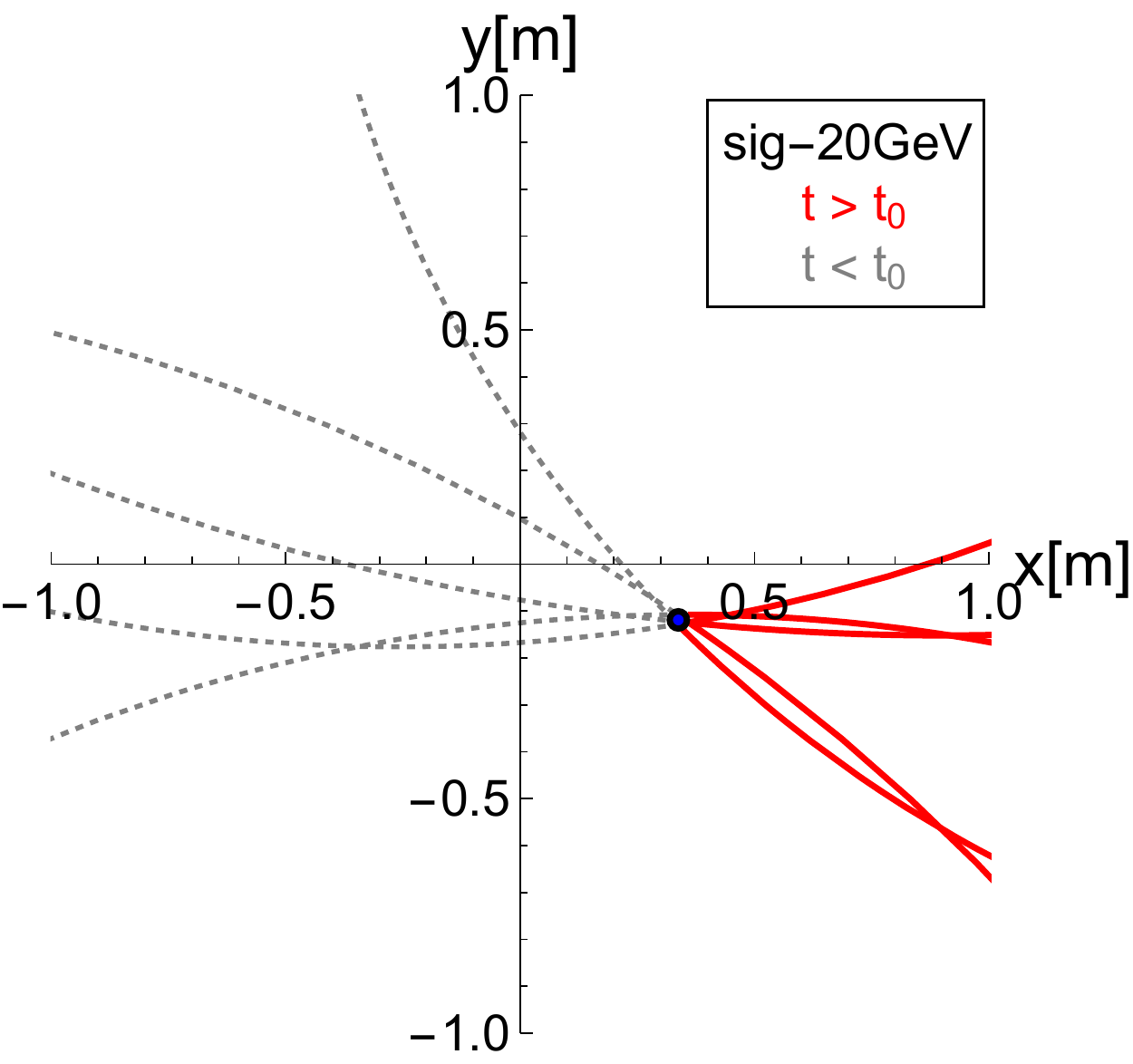}&
        \includegraphics[width=0.33 \columnwidth]{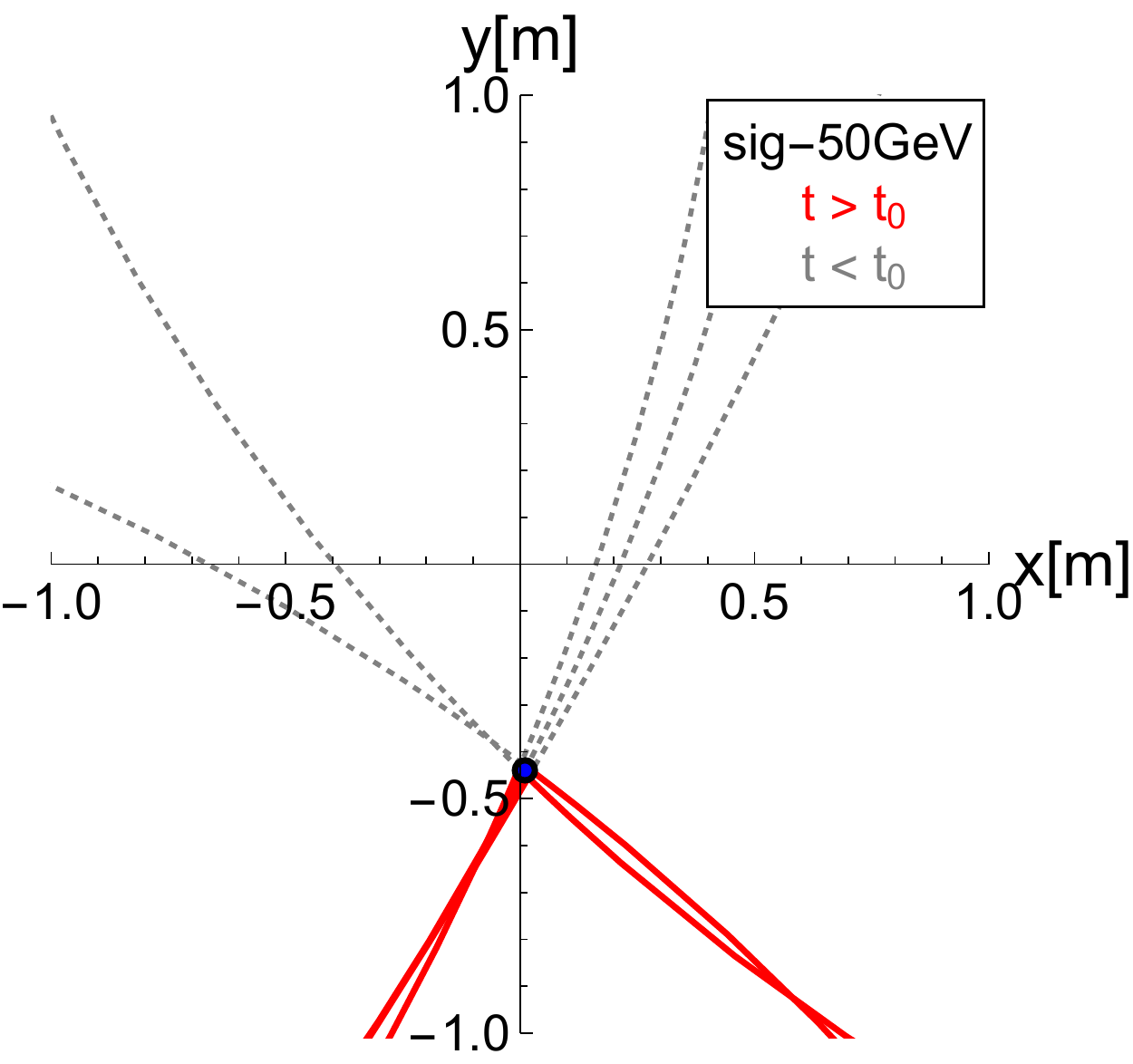} 
         \\
        (d) & (e)  \\
    \end{tabular}
    \caption{Illustrative event displays in the $x$--$y$ plane for the fitting algorithm with leading five displaced tracks. The blue dot is the fitted DV. The solid red (dashed gray) lines are the charged track trajectories in the event, after (before) the fitted DV. From left to right, the plots are for (a) fake-track background, (b) SM QCD light jet background, (c) SM QCD heavy-flavor jet background, (d) LLP signal with $m_X=20$ GeV and (e) LLP signal with $m_X=50$ GeV.
    }
    \label{fig:xyplanefit}
\end{figure}

In Fig.~\ref{fig:xyplanefit}, we illustrate the fitted DV location in the $x$--$y$ plane and the five leading tracks in an event from the fake-track background, the QCD background, and the LLP signal. For the backgrounds, shown in (a), (b), and (c), we use solid red (dashed gray) lines for the trajectories after (before) the fitted DV. 
For the signal, shown in (d) and (e), we use solid red (dashed gray) lines for their trajectories after (before, extrapolated) the LLP decay. The fitted DV location is represented by a blue dot. The black circles have a radius representing the fitted vertex spread, $\Delta \rm D_{min}$. A smaller black circle indicates the vertex fitting algorithm successfully identifies the location of the displaced vertex. This figure shows the different behavior of the various types of background and the signal. For the SM QCD background, the vertices have small displacement, and the fitted vertex has sizable spread. For the fake-track background, the vertices can have large displacement, and the fitted vertices have a much larger spread,  since the tracks are not correlated. For the signals, the fitted vertices would have small spread. For a lighter LLP (hence more boosted), shown in panel (d), the resulting tracks would be more collimated. For a heavier LLP, the resulting tracks spread like two sub-jets, as shown in panel (e).

\begin{figure}
    \centering
    \begin{tabular}{cc}
                (a) & (b)  \\
        \includegraphics[width=0.33 \columnwidth]{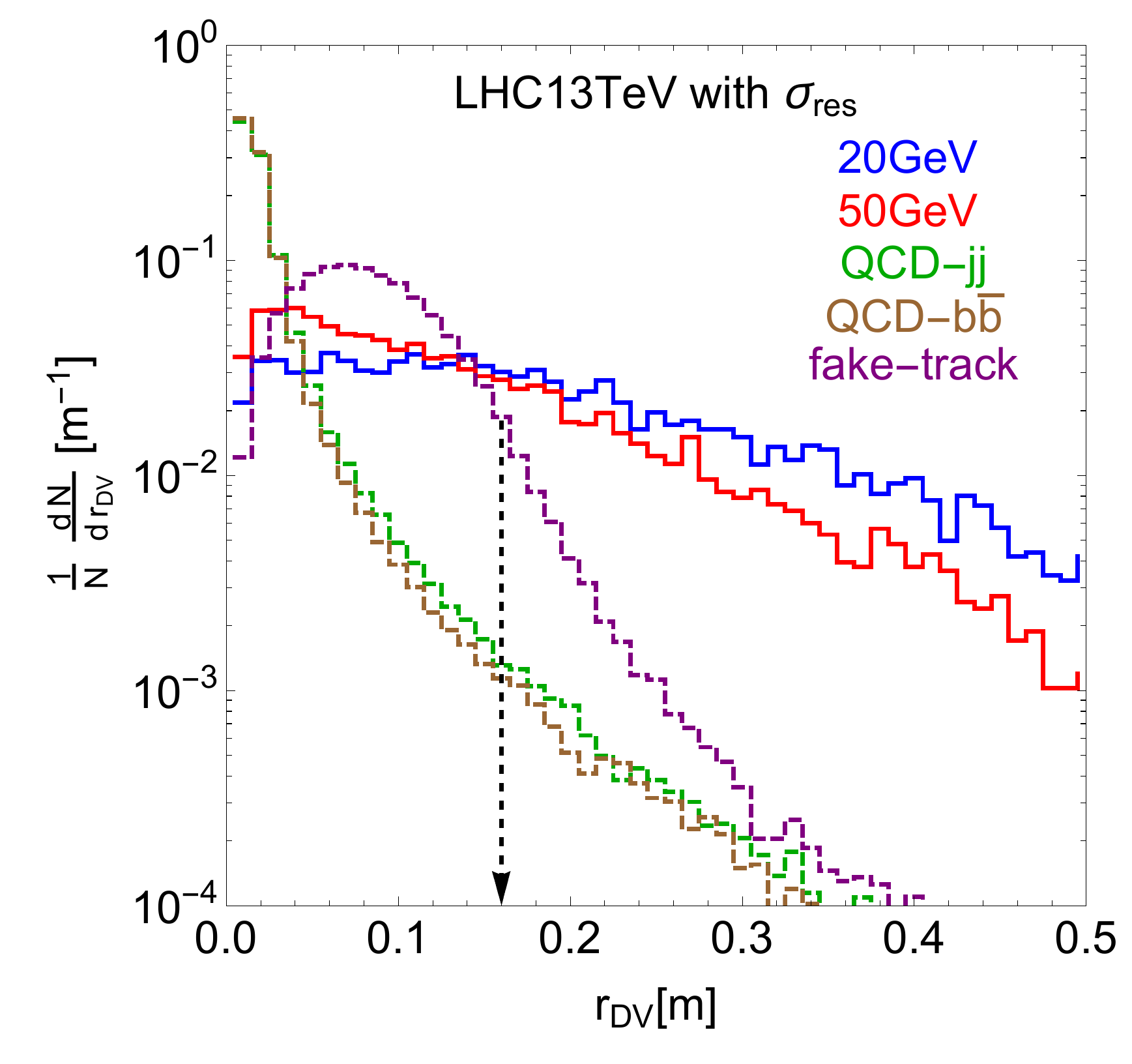} &
        \includegraphics[width=0.33 \columnwidth]{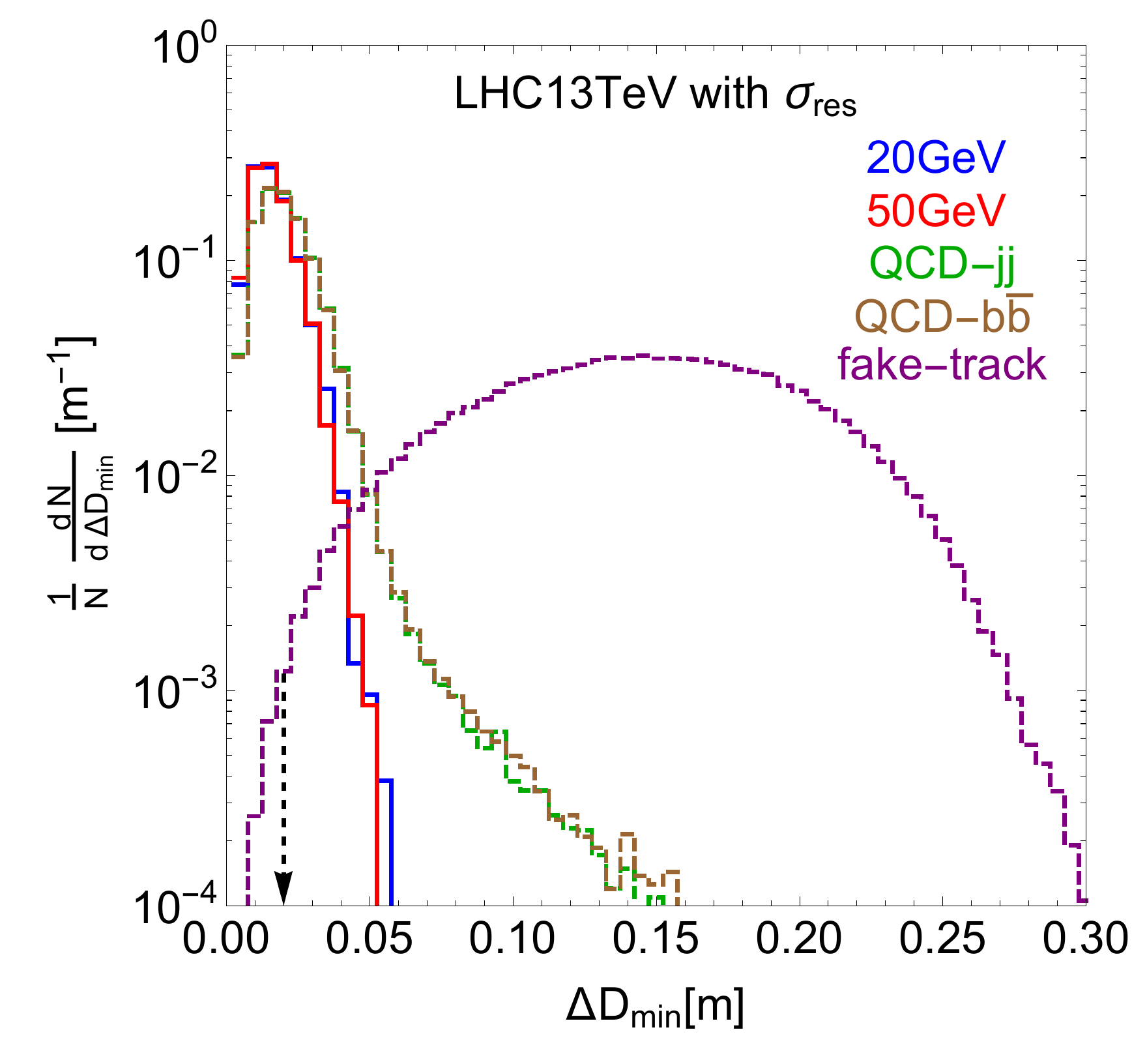}    
 		\end{tabular}   
		\begin{tabular}{ccc}
                (c) & (d) & (e) \\
        \includegraphics[width=0.33 \columnwidth]{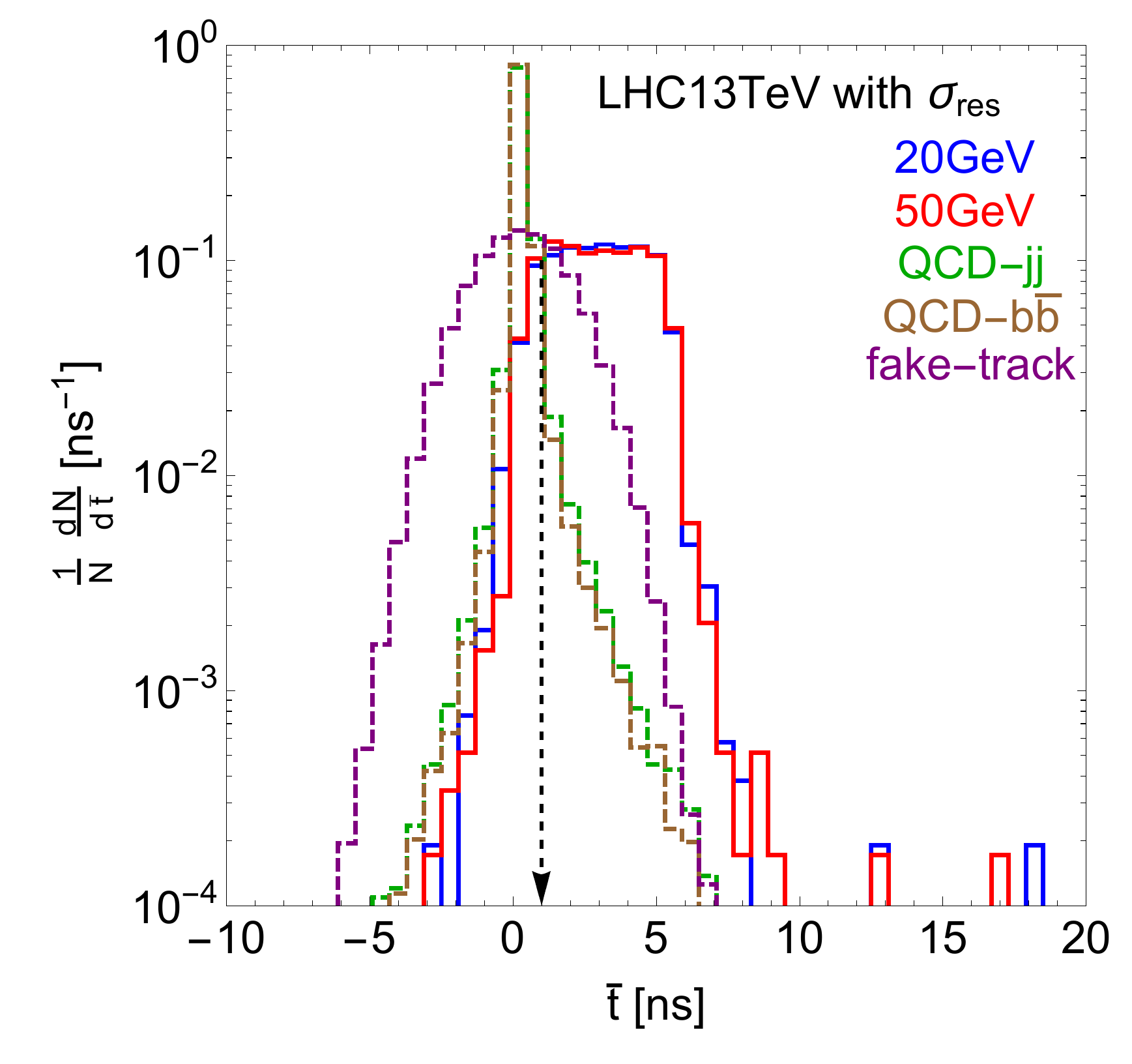} &
        \includegraphics[width=0.33 \columnwidth]{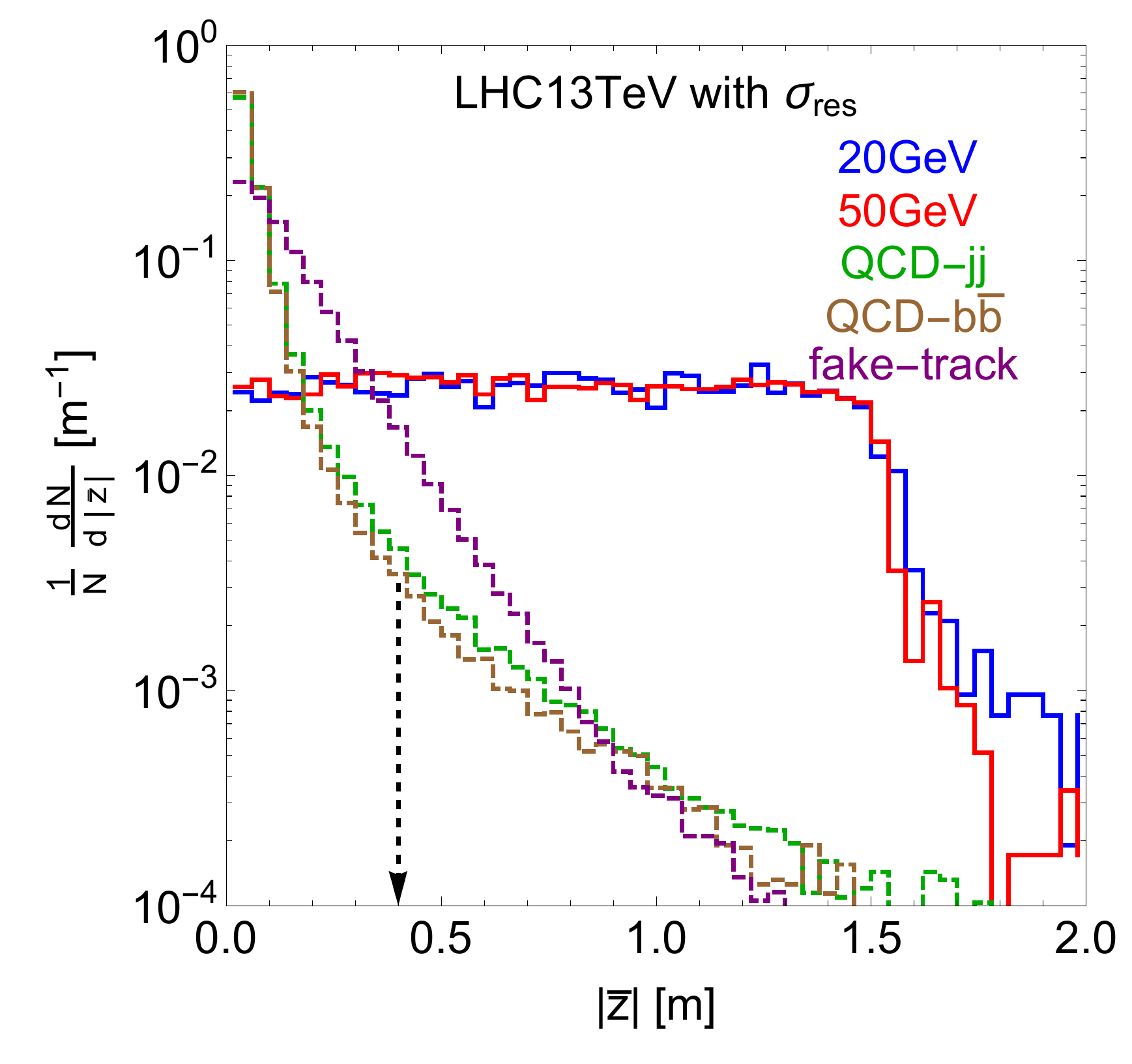}  &
        \includegraphics[width=0.33 \columnwidth]{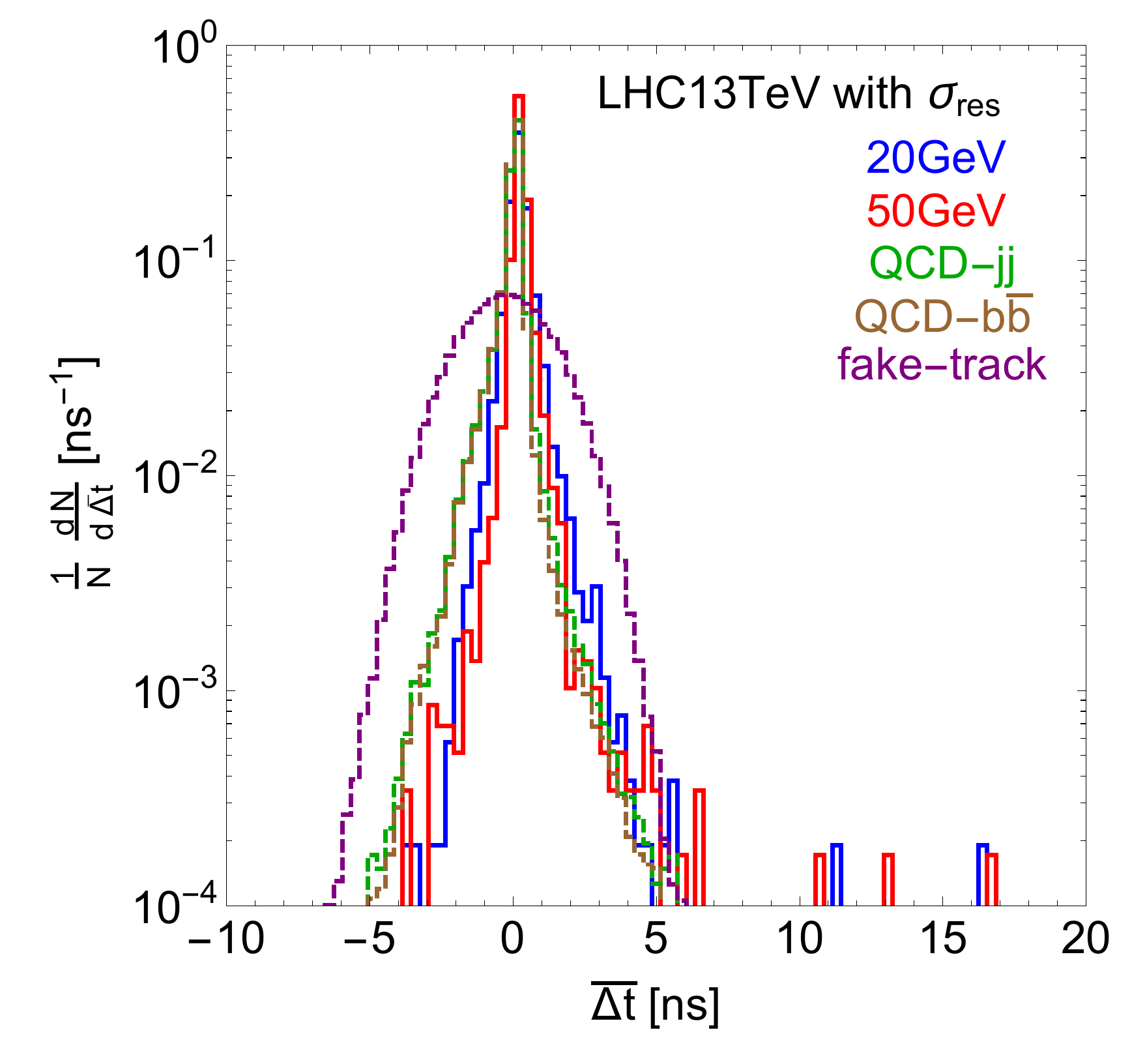}  \\     
                (f) & (g) & (h) \\
        \includegraphics[width=0.33 \columnwidth]{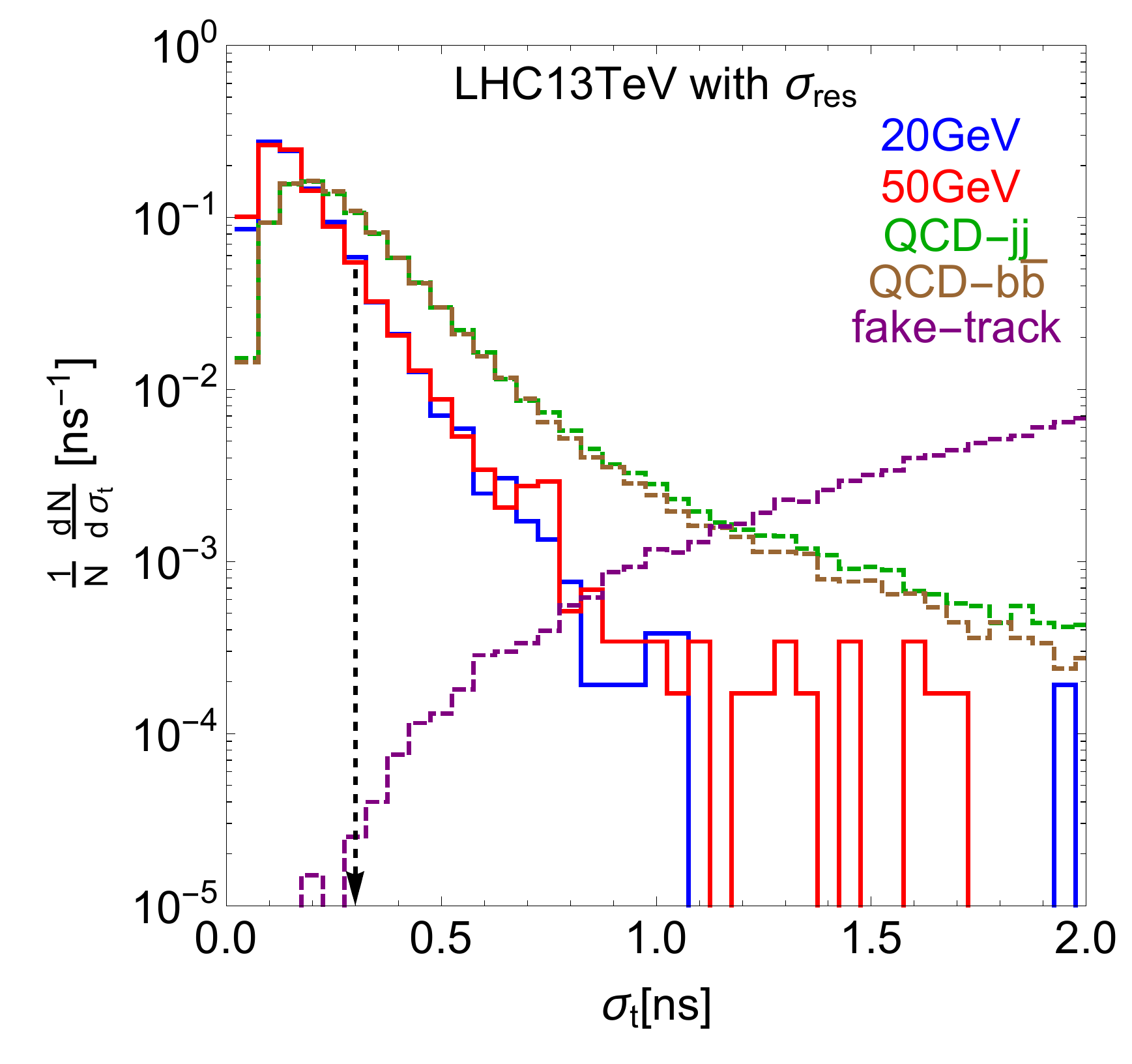} &
        \includegraphics[width=0.33 \columnwidth]{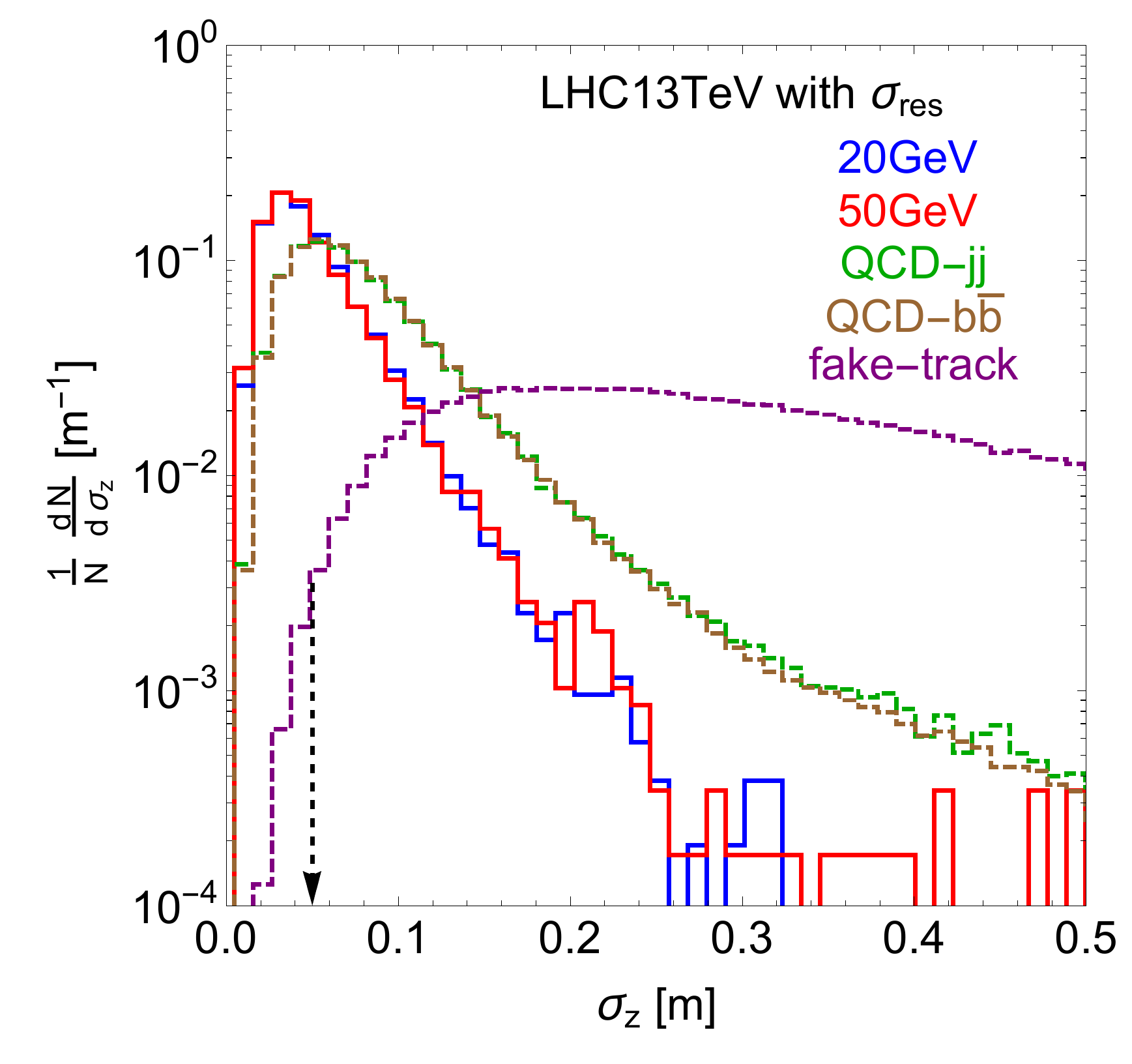} &
        \includegraphics[width=0.33 \columnwidth]{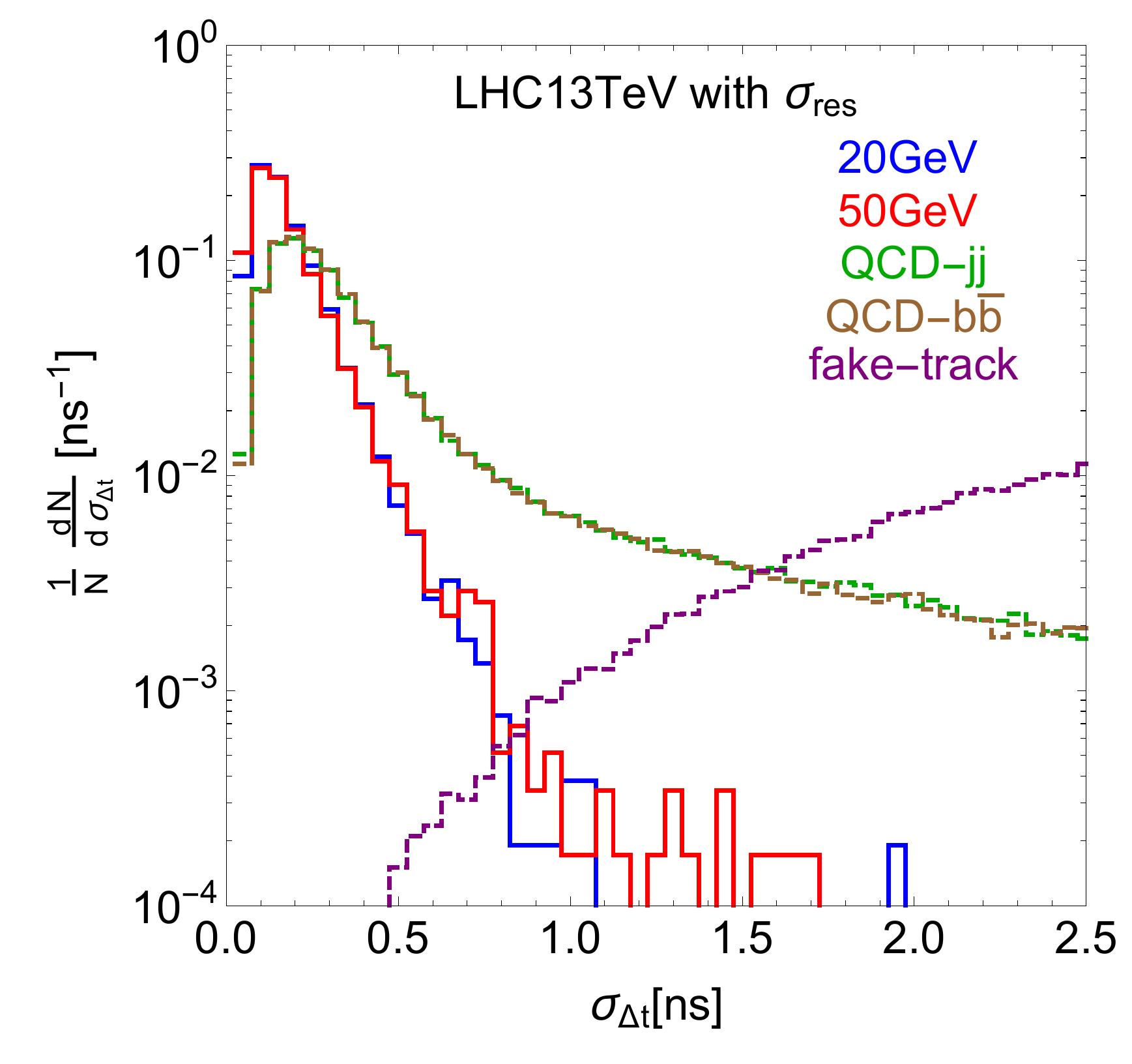}   
    \end{tabular}
    \caption{The kinetic distributions of the leading five tracks for the QCD background, fake-track background and the gluon fusion signal \textit{with} the  angular resolution effect applied. The variables are $\rm{r_{DV}}$, $\rm{\Delta D_{min}}$ in the top row, $\rm{\bar{t}}$, $\left| \rm{\bar{z}} \right|$, $\rm{\overline{\Delta t}}$ in the middle, and $\rm{\sigma_t}$, $\rm{\sigma_z}$, $\rm{\sigma_{\Delta t}}$ in the bottom row. 
    For the signal, we take $m_X$ to be $20$ and 50 GeV respectively, with the same lifetime of $c\tau_X = 1$ m. 
    The vertical dotted black line indicates the cut proposed on the variables.
    }
    \label{fig:9varswithRES}
\end{figure}

The distribution of kinetic variables in Eq.~\ref{eq:7pars} are shown in Fig.~\ref{fig:9varswithRES} for the signal, QCD background and fake-track background. 
For the signal, we show two examples with $m_X=20$ and 50 GeV, with a lifetime $c \tau_X = 1$ m. 
To better understand the effect of angular resolution, we show the distribution of variables without the angular smearing effect in Fig.~\ref{fig:9varswithoutRES} in Appendix~\ref{sec:app}.  
In general, since the fake tracks are randomly generated with a large spread in track parameters, we do not expect the angular resolution effect to change fake-track background significantly, which can be clearly seen in Fig.~\ref{fig:9varswithRES} and Fig.~\ref{fig:9varswithoutRES}. 
Next, we explain the distribution for each variable in detail.

\begin{itemize}
\item{$\rm r_{DV}$}: the distance between DV and origin in the transverse plane.

The QCD backgrounds $jj$ and $b\bar{b}$ peak around zero, which means the fitted DV locates near the origin in the $x$--$y$ plane as most of the tracks from QCD are prompt.  
The distribution of $\rm{r_{DV}}$ extends up to $\sim 0.3$~m. Comparing the results with (Fig.~\ref{fig:9varswithRES}) and without (Fig.~\ref{fig:9varswithoutRES}) $\sigma_{\theta}$, we see that the angular resolution does lead to a broader shape. However, turning off angular resolution does not lead to exact $\rm{r_{DV}} = 0$ m. Some charged tracks start from displaced vertexes from long lifetime mesons decay, e.g., $K_S^0$.
There is no significant difference between QCD backgrounds $jj$ and $b\bar{b}$. The reason is that B-meson has a proper lifetime of $\sim 0.045$ cm, which is too small to generate a difference between $jj$ and $b\bar{b}$.

The fake-track background peaks at around 0.1 m. 
The related variable from fake-track generation is $d_0$, which have a typical value of $0.1$ m.
The fitted DV is not too far from the reference point for each track, because the reference point is the closest point on the track to the beam spot. 
This feature can be seen in Fig.~\ref{fig:xyplanefit} (a) as well. 

For the signal, $\rm{r_{DV}}$ is approximately the position where $X$ particle decays in the $x$--$y$ plane. Its distribution has a very long tail, due to the lifetime of $X$ particle. 

\item{$\rm  \Delta D_{\min}$}: a measure of how well the set of candidate tracks fit in a common vertex.

Both QCD background and the signal should have a distribution of ${\rm \Delta D_{\min}}$ peaks near  zero.  
As shown in Fig.~\ref{fig:9varswithRES} (b), both of them have a similar shape and a spread of about 0.05~m, mostly from the angular resolution of HGCAL. The size of ${\rm \Delta D_{\min}}$ can be estimated as $\sim \sigma_{\theta} R \sqrt{5} = 0.03 \ {\rm m}$, with $R \simeq 3 $ m, and the factor $\sqrt{5}$ comes from the sum of five tracks. This is consistent with Fig. \ref{fig:9varswithRES} (b). 
${\rm \Delta D_{\min}}$ of the fake tracks peaks around 0.2 m, since the tracks have a spread in $d_0$ of $\mathcal{O}(0.1\ \rm m)$ and they do not fit well into a common vertex. 

Turning off the angular resolution in Fig.~\ref{fig:9varswithoutRES} (b), the signal events all have exactly ${\rm \Delta D_{\min}} = 0$ m, which also shows that our algorithm correctly finds the DV where $X$ decays. For the QCD distributions,
there are still a few percent of events with non-zero ${\rm \Delta D_{\min}} $, due to long-lived SM hadrons. 

\item{$\bar{t}$}: the average of the time coordinate of the tracks at the fitted DV. 

The QCD background peaks around zero as shown in Fig.~\ref{fig:9varswithoutRES} (c). 
The spread of $\bar{t}$ dominantly comes from the angular resolution, and it can be estimated to be $\Delta \phi R /v_T/\sqrt{5} $, where $v_T$ is the transverse velocity of the particle responsible for the track and $\Delta \phi$ is the azimuthal angle change when the track evolved from the fitted DV to the HGCAL. 
The geometrical acceptance of the HGCAL selects forward tracks, leading to smaller  $v_T \sim 0.2~c$, as shown in Fig.~\ref{fig:otherparameters} in the Appendix. 
For $\Delta \phi $, its $1\ \sigma$ spread is about 0.02 in Fig.~\ref{fig:deltaPhi} in the Appendix. Therefore, for a typical track radius of $R=3$~m, the spread of $\bar{t}$ for QCD background is about 0.5 ns,  agreeing with Fig.~\ref{fig:9varswithRES}.

For the signal, $\bar t$ peaks around a few ns, due to the delayed decay of $X$. In both Fig. \ref{fig:9varswithRES} and \ref{fig:9varswithoutRES}, we have chosen $c\tau_X = 1$ m 
which corresponds to 3 ns. Moreover, decay products from a lighter LLP (hence with a larger boost) has a larger $\bar{t}$ than that of a heavier LLP. 

For the fake tracks, the $t_i$ for each track should be determined mainly by the random seed time $t_0$, ranging from $\{-6, 6 \}$ ns with a flat distribution. 
The distribution can be approximated by a Gaussian function peaking around zero, with a standard deviation of $3.5/\sqrt{5} = 1.6$ ns, as shown in Fig.~\ref{fig:9varswithRES} (c). Here $3.5$ is an \textit{ad hoc} standard deviation of the flat distribution of each track. In the limit of a large number of tracks, Gaussian function can be used to estimate the spread of the fitted vertex.
From Fig.~\ref{fig:9varswithoutRES}~(c), we see that the distribution from the fake-track background  is not affected by angular resolution,  as expected.

\item{${\rm \sigma_t}$}: the standard deviation of the time-coordinates of the constituent tracks at the fitted DV.

For the signal and QCD background, the distribution is expected to be concentrated at small values, as shown in Fig.~\ref{fig:9varswithoutRES}.
The spread dominantly comes from the angular resolution, as shown in Fig.~\ref{fig:9varswithRES}~(f). The spread can be estimated  by $\Delta \phi R /v_T $. 
As explained for $\bar{t}$, it is $\sim 1$ ns for QCD background, which agrees with the broad distribution up to a few ns. 
In addition, some QCD events have large 
 separation between the displaced tracks of the long-lived mesons and the prompt tracks. For the fake-track background, the spread is largely due to the uncorrelated large spread of the track seed time $t_0$ distributions.

\item{$\bar{z}$}: the averaged z-coordinate of the tracks at the fitted DV.

We first look at the distribution without $\sigma_{\theta}$ in  Fig.~\ref{fig:9varswithoutRES}. The signals have a very flat distribution because of the long lifetime of $X$. There is a hard cut because $X$ is required to decay in the region $|z|  < 1.5$ m to ensure five stubs for the signal track.  

The finite angular resolution $\sigma_{\theta}$ effects on the distributions of the signal and QCD background are shown in Fig.~\ref{fig:9varswithRES}. 
The signal changes very little because the lifetime and the limited decay region are the dominant factors. The distribution of the QCD background  is broadened in a similar fashion  as its $\bar{t}$ distribution, the 1 $\sigma$ spread is roughly $0.5~ {\rm ns} \times v_z \sim 0.15 $ m with $|v_z|  \sim c$ as shown in Fig.~\ref{fig:otherparameters}. 
The tail in the QCD background extends up to $\sim 2$~m with less than  $10^{-4}$ probability, in agreement with the $\bar{t}$ distribution which extends to around $\sim 8$ ns with similar probability. 
Since $8 ~{\rm ns}\times c \sim 2.4 ~ {\rm m}$, this shows a correlation between $\bar{t}$ and $\bar{z}$.
The distribution of the fake-track background follows the exponential shape $e^{-|z|/\sigma}$ with a spread of $\sim 0.15 $ m.

\item{${\rm \sigma_z}$}: the standard deviation of the z-coordinates of tracks from the fitted DV. 

Starting with Fig.~\ref{fig:9varswithoutRES} without $\sigma_{\theta}$,  it is exactly zero for the signal and almost zero for QCD background for the similar reason as $\rm{\sigma_t}$. 
$\sigma_{\theta}$  broadens the distributions up to $0.15$~m for the signal and QCD background, which is in agreement with the previous estimate $(\Delta \phi R /v_T) v_z \sim 0.15$ m. The QCD background has a larger spread than signal, for  the same reason as $\rm{\sigma_t}$. For the fake-track background, the large spread in the seed $z_0$ of the constituent tracks leads to a large spread.

\item{$\overline{\Delta t}$}: the average of the time delay for the tracks.

In Fig.~\ref{fig:9varswithoutRES}, $\Delta t$ of the signal comes from the slow-moving LLP $X$. 
Thus, the values of $\Delta t_i$ are always positive. Moreover, a heavier $X$ moves slower than a lighter $X$, thus the tail of heavier $X$ is longer than that of the lighter $X$ and the QCD background. 
The QCD background has a peak around 0 since the track is prompt. The spread around 0 is due to smearing effects and the fact that some tracks come from long-lived meson.
The fake-track background distribution is Gaussian-like with 1-$\sigma$ spread of about 1.5 ns. It is almost symmetric around zero since its 4D parameters are random and independent from each other. The largest spread comes from random $t_i$, thus $\overline{\Delta t}$ is very similar to $\overline{t}$.

 \item{${\rm \sigma_{\Delta t}}$}: the standard deviation of the time delay of the tracks.
 
Starting with Fig.~\ref{fig:9varswithoutRES} without $\sigma_{\theta}$,  the signal has exactly ${\rm \sigma_{\Delta t}}=0$, while the QCD background peaks at zero with a spread from long-lived meson decay. 
The fake background is similar as in $\rm{\sigma_{ t}}$ because the dominant spread comes from random $t_0$.
After including $\sigma_{\theta}$, the distributions are broadened as expected but without qualitative change. 

\end{itemize}

We see that the distributions of fake tracks are quite different from signal in general. Based on this, 
we propose the six cuts according to the distributions and the cut flow table is given in Table~\ref{tab:DValg}. Explicitly, the cuts for DV fitting variables are,
\begin{align}
 \rm{r_{DV}}>0.16~m , ~\Delta  {\rm D}_{\min} <0.02 ~m ,~ \bar{t} >1  ~ns,~  \rm{\sigma_t} < 0.3 ~ns ,
 ~\left| \rm{\bar{z}} \right| >0.4 ~m , ~\rm{\sigma_z} < 0.05 ~m,
\end{align}
which we denoted them collectively as \textit{vertexing-cuts}.

\subsection{The transverse impact parameter distribution}
\label{sec:d0Tdistribution}

\begin{figure}[h!]
    \centering
    \begin{tabular}{cc}
            (a) & (b) \\
    \includegraphics[width=0.48 \columnwidth]{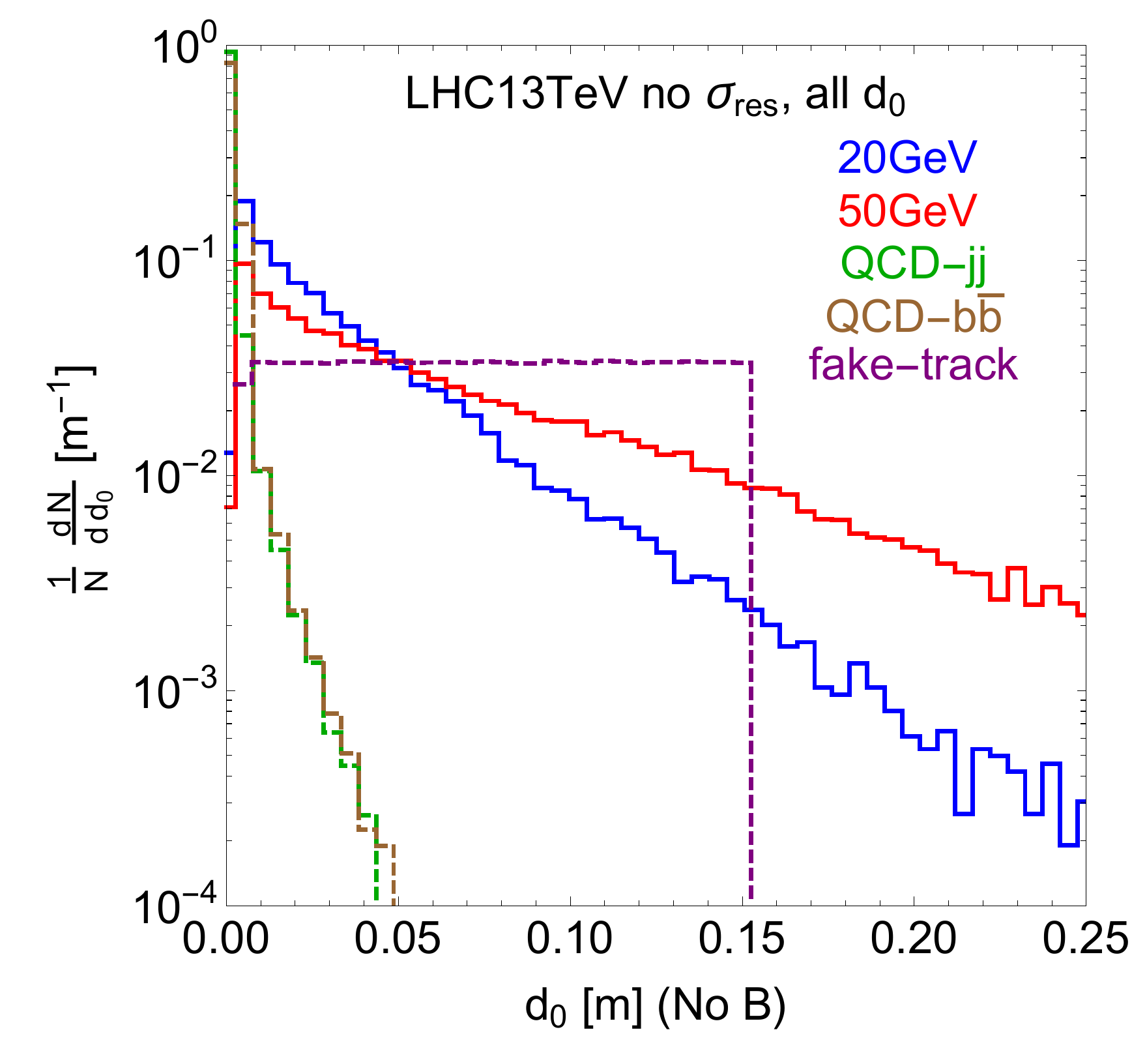} &       
    \includegraphics[width=0.48 \columnwidth]{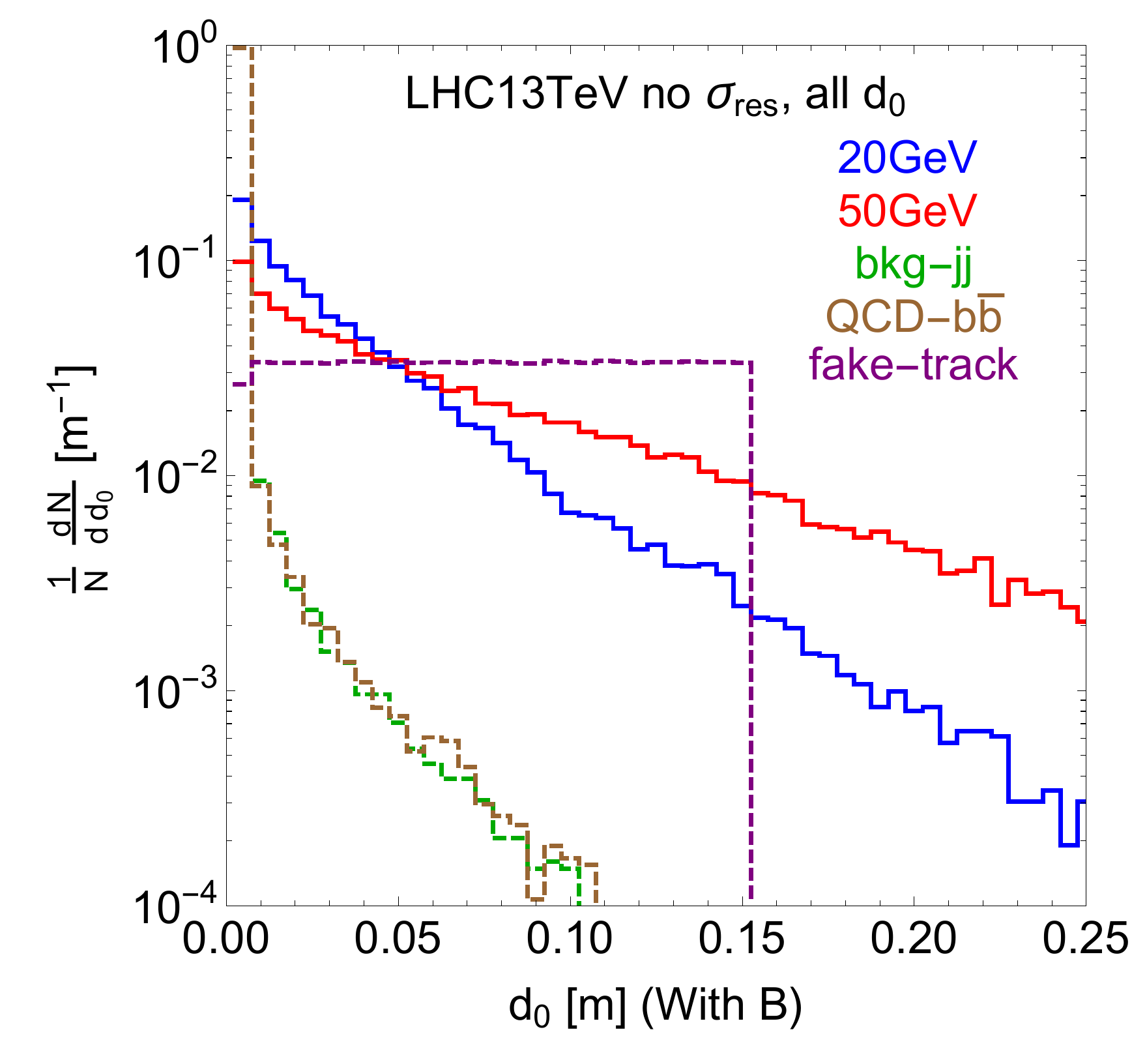} \\
        (c) & (d)       \\
    \includegraphics[width=0.48 \columnwidth]{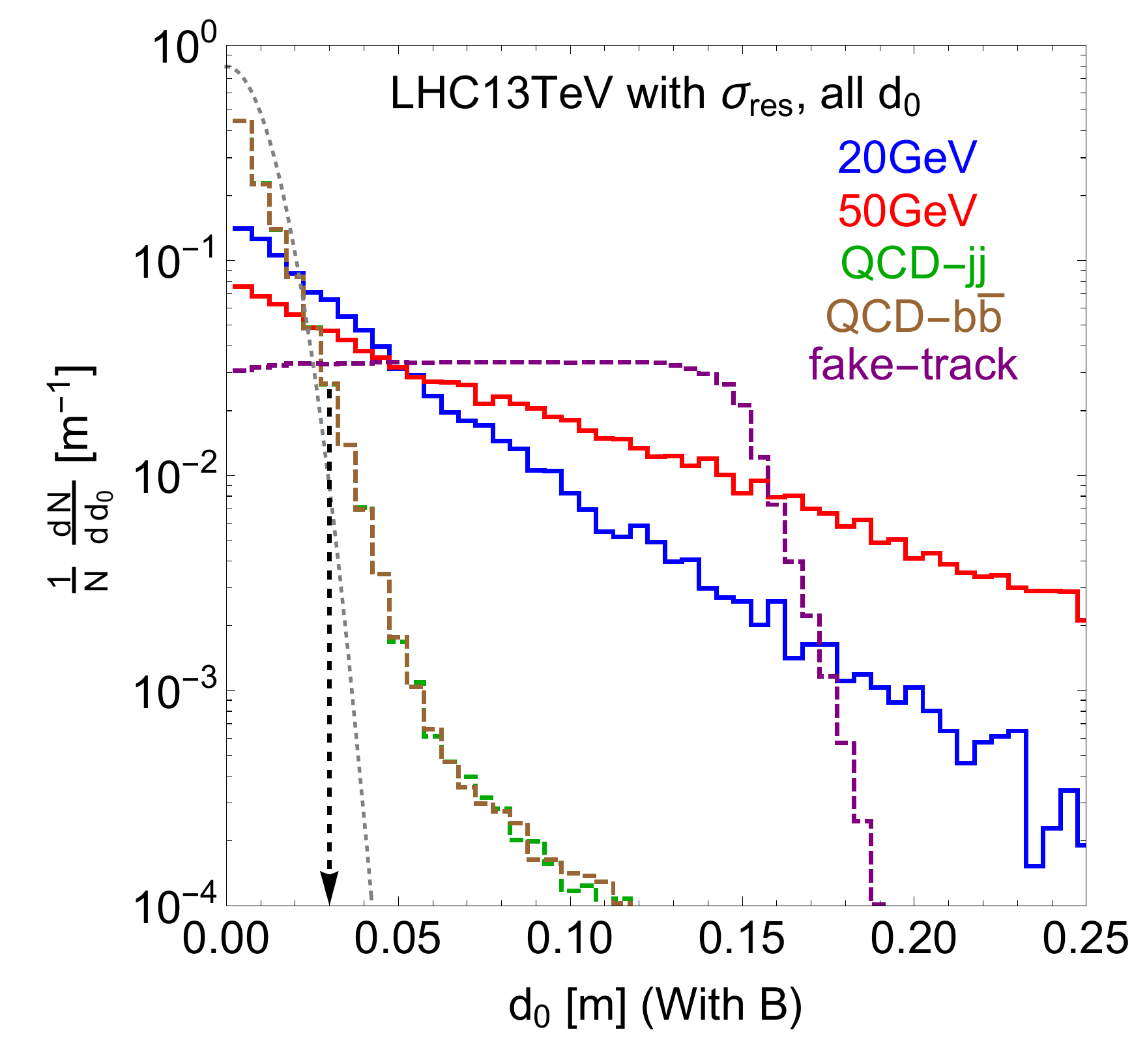} &
    \includegraphics[width=0.48 \columnwidth]{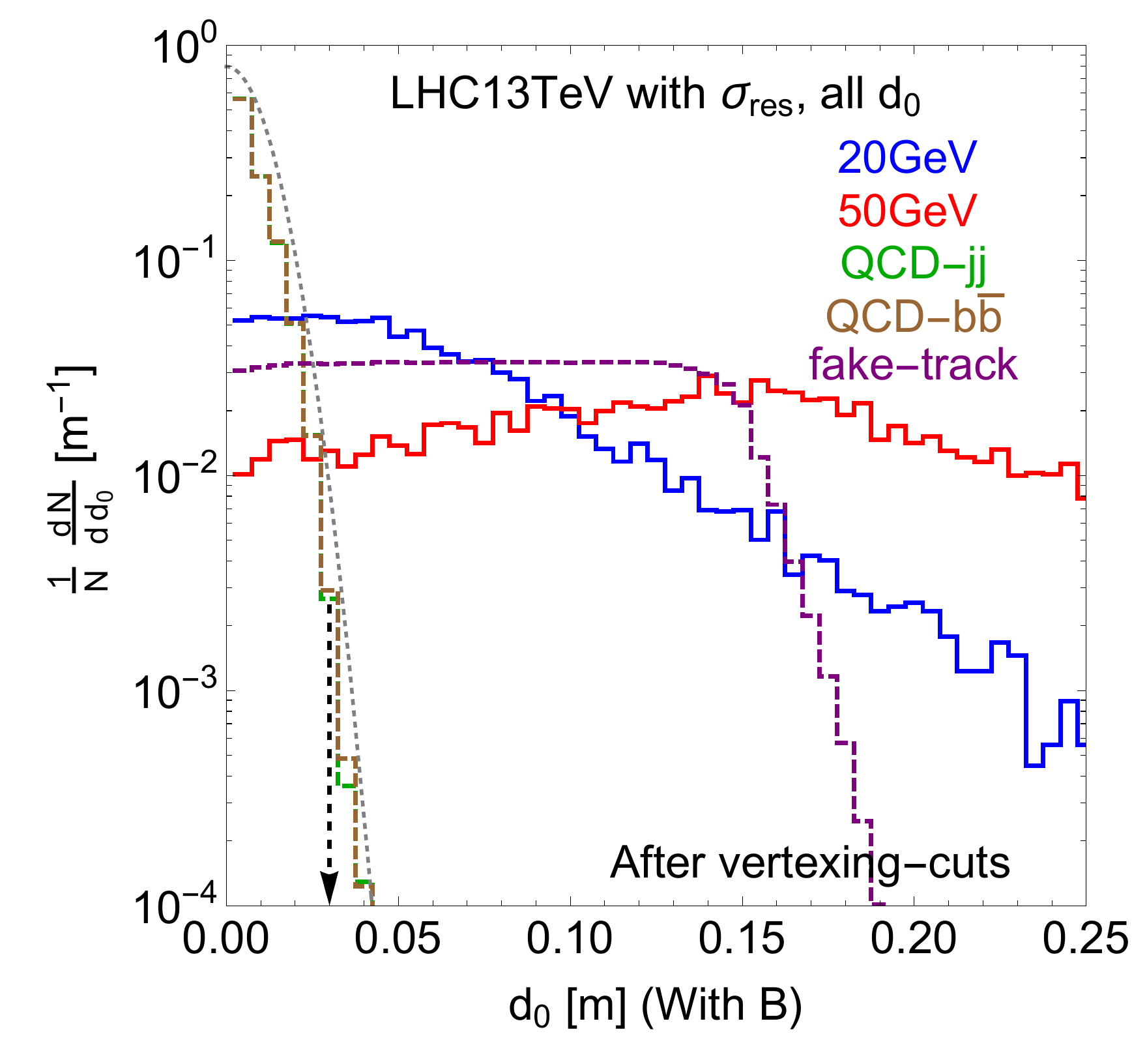} 
    \end{tabular}
    \caption{The distributions of the transverse impact parameter $d_0$ for the QCD background, the fake-track background and the signal.
    Panel (a) has no angular resolution effect and no magnetic field. Panel (b) has no angular resolution effect but with a magnetic field of $3.8$ T. Panel (c) has the angular resolution effect and the magnetic field. Panel (d) has both effects, and with \textit{vertexing-cuts} imposed. The dotted gray lines are the Gaussian function with a spread of 0.015~m, corresponding to the angular resolution times the $z$ coordinate of the HGCAL.}
    \label{fig:d0T}
\end{figure}

The $d_0$ distributions of the five tracks for the QCD background, fake-track background, and the signal are given in Fig.~\ref{fig:d0T}.
In Fig.~\ref{fig:d0T} (a),  the magnetic field is set as zero, and the angular resolution effect is not included either. The fake tracks have a flat $d_0$ distribution from its definition. The signal has a broad distribution due to the delayed decay of  $X$. 
Moreover, the lighter $X$ has a slightly narrower distribution since its decay products are more boosted. 
The QCD dijet background peaks at $d_0 = 0$ m, with a tail from the long-lived hadron decay.

In Fig.~\ref{fig:d0T} (b),  the effect of the magnetic field is included. 
Comparing with Fig.~\ref{fig:d0T} (a), the QCD background from long-lived hadron  are broadened, while the signal is less affected since the displacement before the $X$ decay is more important. The fake-track background is almost flat in $d_0$ by definition.

Both the magnetic field and the angular resolution effects are included in Fig.~\ref{fig:d0T} (c). Comparing with Fig.~\ref{fig:d0T} (b), the signal is almost unchanged. The QCD background is broadened with a spread of $0.015$ m. The spread can be estimated by $\sigma_{\theta} |z| \sim 0.015$~m, where $|z|$ is taken to be 3.2~m, the distance to HGCAL. The fake-track background is still  flat , with its edge smeared by the angular resolution. 

In Fig.~\ref{fig:d0T} (d), we have included both the magnetic field and angular resolution effect after applying \textit{vertexing-cuts}. Importantly, the distribution of the QCD background is trimmed to be a Gaussian shape. This is expected since the outliers with large $d_0$ come from  the decay of long-lived hadron, which fails the DV fitting (and thus fail to pass the \textit{vertexing-cuts}).
Comparing panel (c) and panel (d), we can see that the \textit{vertexing-cuts} improve the per-track QCD background rejection power from $5\times10^{-2}$ to $< ~10^{-3}$ level. For the signal, 
the vertex cuts $ \rm{r_{DV}}>0.16~m , ~ \bar{t} >1  ~ns$ , and $\left| \rm{\bar{z}} \right|>0.4 $ m, selects events in which the $X$ particle decays far from the origin. Any track with small $d_0$ is significantly affected.
The \textit{vertexing-cuts} do not affect the distribution of $d_0$ of the fake-track background.

\subsection{Correlations between the selection cuts}
\label{sec:independence}

Due to the limited statistics of our simulation in some cases, we estimate cut efficiencies by the product of efficiencies of different subsets of cuts.
To validate this approach, we study the correlations between those cuts.

To quantify the correlations among different cuts, we use the following function
\begin{align}
{\rm \rho}_{A,B} \equiv  \frac{\epsilon(A) \epsilon(B)}{ \epsilon(A\& B)},
\end{align}
where $A$ and $B$ are different cut variables. $\epsilon(A)$ is the efficiency for imposing the corresponding cut $A$, while $\epsilon(A\& B)$ is the cut efficiency when both $A$ and $B$ cuts are applied. ${\rm \rho}_{A,B} =1$ means $A$ and $B$ cuts are \textit{completely independent}. If ${\rm \rho}_{A,B} \approx {\mathcal{O}}(1)$, A and B are  \textit{approximate independent}. 
When ${\rm \rho}_{A,B} \ll 1$, $\epsilon(A) \epsilon(B)$ underestimates $\epsilon(A\& B)$. In this case, using $\epsilon(A) \epsilon(B)$ is \textit{inappropriate} for selection efficiency estimation. For ${\rm \rho}_{A,B} \gg 1$, $\epsilon(A) \epsilon(B)$ is a 
\textit{conservative} estimate for background. 
In summary, if ${\rm \rho} \gtrsim 1$, using the product of the individual cuts is a reasonable estimate.

\begin{table}[htb]
    \begin{center}
    \begin{adjustbox}{max width=\textwidth}
        \begin{tabular}{|c|c|c|c|c|c|c|}
            \hline
            $jj$ dijets &    $ \rm{r_{DV}}>0.16$ m $(*)$ & $\Delta  {\rm D}_{\min} <0.02 $ m & $\bar{t} >1$ ns $(*)$ &  $\rm{\sigma_t} < 0.3$ ns & $\left| \rm{\bar{z}} \right| >0.4 $ m $(*)$  & $\rm{\sigma_{z}} < 0.05$ m  \\  \hline
            $(d_0>0.01 {\rm m})^1$ & 0.70& 1.3     & 0.78 & 1.1 & 0.77   & 1.2 \\
            \hline
            $(d_0>0.03 {\rm m})^1$ & 0.25& 8.6   & 0.37 & 1.4  & 0.40 &1.8  \\
            \hline
            $(d_0>0.05 {\rm m})^1$ & 0.09 & 35.0  & 0.18 & 2.4  & 0.19 & 3.2 \\
            \hline
            $b\bar{b}$ dijets &    $ \rm{r_{DV}}>0.16$ m $(*)$ & $\Delta  {\rm D}_{\min} <0.02 $ m & $\bar{t} >1$ ns $(*)$ &  $\rm{\sigma_t} < 0.3$ ns & $\left| \rm{\bar{z}} \right|>0.4 $ m  $(*)$ & $\rm{\sigma_{z}} < 0.05$ m  \\  \hline
            $(d_0>0.01 {\rm m})^1$ & 0.71 & 1.3    & 0.78 & 1.1  & 0.77 & 1.2 \\
            \hline
            $(d_0>0.03 {\rm m})^1$ & 0.21 & 8.8   & 0.36 & 1.4    & 0.36 &1.8   \\
            \hline
            $(d_0>0.05 {\rm m})^1$ & 0.07 & 47.0   & 0.16  & 2.6   & 0.17 & 4.0  \\
            \hline
        \end{tabular}
    \end{adjustbox}
\end{center}
    \caption{The correlations $\rho (\text{vertexing-cuts}, ~d_0)$  for QCD $jj$ and $b\bar{b}$ backgrounds. 
    	 The columns with $(*)$ are not used to calculate the final selection efficiencies.
    }
    \label{tab:jjbb-precuts-d0T}
\end{table}

We begin with the QCD $jj$ and $b\bar{b}$ backgrounds. 
First of all, the correlations among \textit{vertexing-cuts} variables are not needed, since we have enough simulated events to compute the efficiency without relying using the product of efficiencies of the individual cuts. However, the \textit{vertexing-cuts} are not enough to suppress the background; we further require multiple tracks with large $d_0$. Here, we are limited by the statistics. Hence, we need to check the   the correlation between \textit{vertexing-cuts} and $d_0$ cuts, and the correlation among different $d_0$ cuts. 

The correlations  between \textit{vertexing-cuts} and the $d_0$ cut are given in Table~\ref{tab:jjbb-precuts-d0T}.
With higher cut threshold of $d_0$, the correlation between single $d_0$ cut and $\rm{r_{DV}}$, $\rm{\bar{t} }$ and $\left| \rm{\bar{z}} \right|$ becomes stronger. This is expected. For $jj$ and $b\bar{b}$ QCD backgrounds, the event with large transverse impact parameter is also likely to have large values for $\rm{r_{DV}}$, $\bar{t}$ and $\left| \rm{\bar{z}} \right|$. Therefore, we will not use these cuts when calculating the final selection efficiencies.
In this way, we avoid double counting and remain \textit{conservative} because all the remaining columns have $\rho > 1$ .

\begin{table}[h]
    \begin{center}
    \begin{adjustbox}{max width=\textwidth}
        \begin{tabular}{|c|c|c|c|c|c|}
            \hline
            $jj$ dijets &  $d_0>0.01$ m  & $d_0>0.015$ m & $d_0>0.02 $ m &   $d_0>0.025$ m &   $d_0>0.03$ m \\ \hline
            $ {\rm \rho }_{d}^1$  &    $0.970 \pm 0.016$    & $0.990 \pm 0.027$  & $1.000 \pm 0.056$   & $1.10 \pm 0.15$  & $1.40 \pm 0.45$ \\
            \hline
            $ {\rm \rho}_{d}^2$ &   $1.20 \pm 0.04$  &  $1.20 \pm 0.10$ &     $0.69\pm 0.17$       & - & - \\
            \hline
            $ {\rm \rho}_{d}^3$ &   $1.30 \pm 0.11$  &  $1.20 \pm0.35$ &  -        &-   & -\\
            \hline
            $ {\rm \rho}_{d}^4$ &   $1.60 \pm 0.30$ &  - &  -     & -            & -\\
            \hline
            $ {\rm \rho}_{d}^5$ &   $1.80 \pm 0.83$ &  - &  -     & -            & -\\
            \hline
            $b\bar{b}$ dijets &  $d_0>0.01$ m  & $d_0>0.015$ m & $d_0>0.02$ m &   $d_0>0.025$ m &   $d_0>0.03$ m \\
            \hline
            $ {\rm \rho}_{d}^1$ & $1.000 \pm 0.017$  & $1.000 \pm 0.029$   &  $1.000 \pm 0.054$   &  $1.30 \pm 0.17$  & $1.80 \pm 0.60$\\
            \hline
            $ {\rm \rho}_{d}^2$ &$1.100 \pm 0.041$&  $1.10 \pm 0.09$   & $1.00 \pm 0.29$     & -     & -\\
            \hline
            $ {\rm \rho}_{d}^3$ &  $1.100 \pm 0.087$  &  $0.84\pm 0.22$ & -   & -& -\\
            \hline
            $ {\rm \rho}_{d}^4$ &   $1.00\pm 0.15$  &  $-$    & -  & -       & -  \\
            \hline
            $ {\rm \rho}_{d}^5$ &   $0.62\pm 0.16$ &  -  & - & -        &- \\
            \hline
        \end{tabular}
    \end{adjustbox}
\end{center}
    \caption{The correlation (including the statistical uncertainty) from our simulation, for multiple $d_0$ tracks for QCD dijet backgrounds after applying \textit{vertexing-cuts}. For the entries with ``-", there are not enough statistics to make a reliable estimate. 
     For higher tracks multiplicity and $d_0$ threshold, the results suffer from larger fluctuations due to limited statistics.     }
    \label{tab:jjbb-d0T}
\end{table}

Next, we would estimate the cut efficiency on the QCD dijet background by the product of single track efficiency of $d_0 >  0.03 ~{\rm m}$. We would like to show that the consecutive $d_0$ cuts are approximately independent. This is expected since the large $d_0$ tracks are mainly from detector resolution effects, which are independent between tracks.
As shown in Fig.~\ref{fig:d0-first-track} in the Appendix, the $d_0$ distribution of the leading track is the same as the ensemble of the five tracks shown in Fig.~\ref{fig:d0T}. This indicates that we could apply the transverse impact parameter cut on multiple tracks {\it independently}.~\footnote{This independence of the tracks is true for both prompt QCD background from smearing effects and for the fake-track background. For the displaced tracks from long-lived hadrons, there is a certain level of correlations which is already removed by our \textit{vertexing-cuts}. Hence, we ignore these minor correlations here.} 
To quantify this further, we define the following function to study the correlations between different  $d_0$ cuts,  
\begin{align}
{\rho }_{d}^n \equiv  \frac{\epsilon^n(\text{1 track } d_0 > 0.03 {\rm m}  ) }{ \epsilon(\text{n tracks } d_0 > 0.03 {\rm m}  )},
\end{align}
where ${d_0} > 0.03$ m is chosen as an example. 
Note the tracks in numerator are randomly chosen, while in the denominator they are the $n$ hardest tracks. 
The correlation ${\rho}_{d_n}$ for QCD $jj$ and $b\bar{b}$ backgrounds \textit{after} imposing \textit{vertexing-cuts} are given in Table~\ref{tab:jjbb-d0T}.

In Table~\ref{tab:jjbb-d0T}, from ${\rm \rho}_{d}^1$ to ${\rm \rho}_{d}^{5}$, the correlations are mostly around 1, implying the $d_0$ cuts for different tracks are indeed \textit{independent}~\footnote{${\rm \rho}_{d}^1$ is not exactly 1, because the track in the numerator is randomly picked, while the track in the denominator is the leading track. Thus, the fact that its value is close to 1 is a kind of proof that different tracks are independent.}. 
After applying the \textit{vertexing-cuts} and requiring $d_0>0.03$ m, we are again limited by the statistics of our simulation. For this reason, $\rho_d^1$ deviates significantly from 1 here. Similarly, for the entries with ``-", there are not enough statistics to make a reliable estimate.  We also check the correlation for multiple $d_0$ tracks without applying \textit{vertexing-cuts}. The results are shown in Table~\ref{tab:jjbb-d0T-noprecuts} in the Appendix. As expected, they are \textit{approximately independent}.

\begin{table}[htb]
    \begin{center}
    \begin{adjustbox}{max width=\textwidth}
        \begin{tabular}{|c|c|c|c|c|c|c|}
            \hline
            fake tracks &  $ \rm{r_{DV}}>0.16$ m & $\Delta  {\rm D}_{\min} <0.02 $ m & $\bar{t} >1$ ns &  $\rm{\sigma_t} < 0.3$ ns & $\left| \rm{\bar{z} } \right|>0.4 $ m   & $\rm{\sigma_{z}} < 0.05$ m  \\  \hline
            $ \rm{r_{DV}}>0.16$ m           &  & $ 0.49 \pm 0.04$        & $0.85 \pm 0.01$ & $2.49 \pm 2.51$ & $0.156 \pm 0.001$  & $216.0 \pm 216.0$ \\
            \hline
            $\Delta  {\rm D}_{\min} <0.02 $ m  &  $ 0.49 \pm 0.04$ &      & $0.95 \pm 0.04$ & - & $0.62 \pm 0.04$ & $2.12 \pm 0.95 $ \\
            \hline
            $\bar{t} >1$ ns                              & $0.85 \pm 0.01$ & $0.95 \pm 0.04$     &      & $0.69 \pm 0.17$ & 
            $0.80 \pm 0.01$ & $1.05 \pm 0.03  $ \\
            \hline
            $\rm{\sigma_t} < 0.3$ ns       & $2.49 \pm 2.51$ & -    & $0.69 \pm 0.17$     &    & $0.87 \pm 0.45$ 
            & $0.25 \pm 0.25$ \\
            \hline
            $\left| \rm{\bar{z} } \right|>0.4 $ m     & $0.156 \pm 0.001$ & $0.62 \pm 0.04$    & $0.80 \pm 0.01$      & $0.87 \pm 0.45$     &       & $18.86 \pm 4.72 $ \\
            \hline
            $\rm{\sigma_{z}} < 0.05$ m                 & $216.0 \pm 216.0$ & $2.12 \pm 0.95$    & $1.05 \pm 0.03  $     & $0.25 \pm 0.25$      & $18.86 \pm 4.72 $      &    \\
            \hline
        \end{tabular}
    \end{adjustbox}
\end{center}
    \caption{The correlation table for \textit{vertexing-cuts} variables for the fake-track background. 
        The entries with ``-" contain too few events to make a reliable estimate and the statistical uncertainties 
        are given after $\pm$.
    }
    \label{tab:mis-presix}
\end{table}

Next, we discuss the correlations of the cuts for the fake-track background. Firstly, among the \textit{vertexing-cuts} variables, Table~\ref{tab:mis-presix} shows that most of them are close to 1, which means \textit{approximately independent}. Two of the correlations are much larger than 1, which means using the product of the individual cuts
efficiency is a reasonable estimate.
However, both of them have large statistical errors. Moreover, due to limited statistics, there is no reliable estimate 
for the correlations between $\rm{\sigma_{t}}$--$\Delta  {\rm D}_{\min}$. 
As a further check, we evaluated the correlations with a looser set of cuts thus containing more statistics, shown in Table~\ref{tab:mis-presix-weak} in the Appendix. For example, we relaxed the maximum $\rm{\sigma_t} $ cut to $ 0.5$ ns rather than $0.3$ ns. In this case, we see that all the variables are \textit{approximately independent}. 
As a result, we conclude that the total cut efficiency for the \textit{vertexing-cuts} variables estimated by using the product of the single cut efficiencies is reasonable.

\begin{table}[h]
    \begin{center}
    \begin{adjustbox}{max width=\textwidth}
        \begin{tabular}{|c|c|c|c|c|c |c|}
            \hline
            fake tracks &    $ \rm{r_{DV}}>0.16$ m & $\Delta  {\rm D}_{\min} <0.02 $ m & $\bar{t} >1$ ns &  $\rm{\sigma_t} < 0.3$ ns & $\left| \bar{z} \right|>0.4 $ m   & $\rm{\sigma_{z}} < 0.05$ m \\  \hline
            $(d_0>0.01 {\rm m})^1$  &    0.97 & 1.00  & 1.00 & 0.94 & 0.98 & 1.0  \\
            \hline
            $(d_0>0.03 {\rm m})^1$  &    0.91 & 1.10   & 1.00 & 1.20 & 0.94 & 1.0 \\
            \hline
            $(d_0>0.05 {\rm m})^1$  &    0.85 & 1.10 & 1.00 & 1.00 & 0.91 & 1.1 \\
            \hline
            $(d_0>0.03 {\rm m})^5$  & 0.65 & 1.00 & 0.99 &  0.76 & 0.77 & 1.2 \\
            \hline
        \end{tabular}
    \end{adjustbox}
\end{center}
    \caption{The correlation  between \textit{vertexing-cuts} and $d_0$ cuts for fake-track background.}
    \label{tab:mis-six-d0T}
\end{table}
We note that  there is enough statistics in the fake-track background to calculate the efficiency of the multiple $(d_0 > 0.03 {\rm m})$ cuts without approximation. Hence, there is no need to check the correlations among individual $d_0$ cuts here. We are left to check the independence between \textit{vertexing-cuts} and $d_0$ cuts, which is given in Table. \ref{tab:mis-six-d0T}. The first three rows show the correlations between the \textit{vertexing-cuts}  and various single $d_0$ cut from $0.01$ m to $0.05$ m. In the fourth row, we use the exact $d_0$ cuts for five tracks. The result shows the correlations between \textit{vertexing-cuts} and full $d_0$ cuts are \textit{approximate independent}.

\section{The Results}
\label{sec:results}

\subsection{Cut efficiencies}
\label{subsec:cut_eff}

\begin{table}[!h]
            \begin{tabular}{|c|c|c|c|c|c|}
                \hline 
                 cut conditions  & $jj$ dijet & $b\bar{b}$ dijet &  fake-track  & ggF $m_s=20$ GeV & ggF $m_s=50$ GeV \\
               \hline
                    $N_{\rm ini}$  & $5.1 \times 10^{14}$ &   $1.1 \times 10^{13}$ &$ 1\times 10^{12} $ & $1.3 \times 10^8 \times {\rm BR}$ & $1.3 \times 10^8 \times {\rm BR}$  \\
            \hline 
            5 tracks &  $8.7\times 10^{-1}$ &  $8.4\times 10^{-1}$ & 1.0 &  $8.3 \times 10^{-2}$ & $ 2.1\times 10^{-1} $  \\
            \hline \hline
            $ \rm{r_{DV}}>0.16~m$  &  $9.2\times 10^{-3}$ $(*)$ & $7.5 \times 10^{-3}$ $(*)$ &  $4.5\times 10^{-2}$ & $4.8\times 10^{-1}$ &  $3.1\times 10^{-1}$ \\
            \hline
            $\Delta  {\rm D}_{\min} <0.02 $ &   $6.1\times 10^{-1}$ &  $6.1\times 10^{-1}$ & $2.2\times 10^{-3}$ & $8.7\times 10^{-1}$  &$8.9\times 10^{-1}$ \\
            \hline
             $\bar{t} >1$~ns &  $3.3\times 10^{-2}$ $(*)$ & $2.8\times 10^{-2}$ $(*)$ & $2.8\times10^{-2}$  & $9.9\times10^{-1}$ & $9.9\times 10^{-1}$ \\
             \hline
             $\rm{\sigma_t} < 0.3$~ns & $7.1\times 10^{-1}$  & $7.2\times 10^{-1}$ & $4.5\times10^{-5}$   &$9.6\times 10^{-1}$ &  $9.8\times 10^{-1}$  \\
             \hline
             $\left| \bar{z} \right|>0.4 $ m &  $3.4\times 10^{-2}$ $(*)$  &$2.8\times 10^{-2}$ $(*)$ & $6.4\times10^{-2}$ & $9.9\times 10^{-1}$  &  $9.9\times 10^{-1}$\\ 
             \hline
              $\rm{\sigma_z} < 0.05$ & $4.9\times 10^{-1}$  &$4.9\times 10^{-1}$ &  $4.9\times10^{-3}$  & $8.5\times 10^{-1}$  &  $8.8\times 10^{-1}$  \\
              \hline \hline
              $\epsilon_{\rm vtc}$ &   $2.1\times 10^{-1}$ &$2.1 \times 10^{-1}$ & $4.0\times 10^{-13}$ & 
              $ 3.4 \times 10^{-1}$ & $2.4\times 10^{-1}$ \\
              \hline
              $(d_{0}>0.03 ~{\rm m})^5$  &$(5.7\times 10^{-4})^5$ &$(6.8\times 10^{-4})^5$ & $3.4\times10^{-1}$ & $2.6\times 10^{-1}$ &$8.1\times 10^{-1}$ \\
              \hline 
              $N_{\rm fin}$  &  $5.7\times 10^{-3}$ &$2.9\times 10^{-4}$  &$1.4 \times 10^{-1}$ & $ 9.7\times 10^5 \times {\rm BR}$ &  $5.3\times 10^6 \times {\rm BR}$ \\
              \hline
            \end{tabular}
            \caption{The cut-flow table for the QCD background, the fake-track background and the signal. $N_{\rm ini}$ and $N_{\rm fin}$ are the
    initial and final  event numbers before and after imposing the cuts. These numbers correspond to an integrated luminosity of $\mathcal{L} = 3~ {\rm ab}^{-1}$ at the HL-LHC. ``5 tracks" requires each track has  $p_T > 1$ GeV and at least 5 tracks arrive at HGCAL. ``$\epsilon_{\rm vtc}$" is the total efficiency for the \textit{vertexing-cuts}  except those with $(*)$. The efficiency of the $d_0$ cuts is calculated after applying the \textit{vertexing-cuts}. We used the two signal benchmarks with $m_X=$  20 and 50 GeV,  and lifetime $c\tau_X= 1$~m.  }
    \label{tab:DValg}
\end{table}

In this section, we present the efficiencies of cuts we adopt in this analysis in Table~\ref{tab:DValg}. $N_{\rm ini}$
is the initial event number from the cross-section only. 
$N_{\rm fin}$ is final event numbers after imposing the trigger and the cuts in the table at the HL-LHC with 13 TeV center-of-mass energy and 3~ab$^{-1}$ integrated luminosity. The row ``5 tracks" comes from the requirement that at least five tracks that arrive HGCAL and the trigger requirement. For the signal, it is the combination of the geometric probability for $X$ decay inside the $|z| < 1.5$ m region and the efficiency for tracks arriving HGCAL.
The QCD backgrounds have a better efficiency for tracks arriving HGCAL, because their tracks are more forward than the signal (see the upper panel of Fig.~\ref{fig:otherparameters}). Furthermore, the background jets are usually more energetic thus containing more tracks than the signal, which makes it much easier to satisfy the requirement.
The single-cut efficiencies for \textit{vertexing-cuts} DV fitting variables are listed. 
The variables $\overline{\Delta t}$ and $\rm{\sigma_{\Delta t}}$ are highly degenerate with $\bar{t}$ and $\rm{\sigma_{ t}}$, and are not used here~\footnote{For a general discussion on effectiveness of time-delay variable for a broad class of LLP signatures, see Ref.~\cite{Liu:2018wte}.}. After multiplying $N_{\rm ini}$ by the cut efficiencies in the ``5 tracks" row, the ``$\epsilon_{\rm vtc}$" row and the ``$(d_{0}>0.03 {\rm m})^5$" row, we obtain the final event number $N_{\rm fin}$.

For the QCD background, we apply a partial set of \textit{vertexing-cuts} on $ \Delta  {\rm D}_{\min},~  \rm{\sigma_t} $ , and $\rm{\sigma_z} $. The cuts with $(*)$ are correlated with transverse impact parameter $d_0$ cut. Hence, they are not included in ``$\epsilon_{\rm vtc}$" to avoid double counting~\footnote{One can apply them  in an experimental search and it will help to further suppress the QCD background. }. 
Furthermore, we apply the single cut efficiency $\epsilon(\text{1 track} ~d_0>0.03 \text{m}) $ five times as an estimate of the efficiency requiring all the five tracks with $d_0>0.03 $ m.
We found the background the number of the events for $jj$ and $b\bar{b}$ are $5.7 \times 10^{-3}$ and $2.9 \times 10^{-4}$ respectively. 
We have demonstrated that $d_0$ cuts on different tracks are \textit{approximate independent}, as discussed in detail in the previous section and the appendix. Nevertheless, one might still worry that cutting on five tracks is too aggressive. 
We also consider, as an alternative, cutting on only four tracks together with a stronger cut $d_0>0.05 $ m. In this case, the single-cut efficiency for $d_0 > 0.05$ m is about $2.5 \times 10^{-5}$ for QCD backgrounds. After applying $(d_0 > 0.05 ~\rm{m})^4$, QCD background can be suppressed down to $\sim 10^{-5}$, which works equally well.

For the fake-track background, we multiply the individual efficiency for each variable in the \textit{vertexing-cuts} and obtain $\epsilon_{\rm vtc} = 4.0 \times 10^{-13}$.  Requiring all five tracks with $d_0>0.03 $~m can suppress the background further by a factor of 0.34, leaving only $0.14$ events. 
We note that, even though we did not include it in this analysis, the fake-track has to match the track information with the HGCAL calorimeter energy deposit~\cite{Hook:2019qoh}, which can further suppress the fake-track background.

In summary, both the QCD background and the fake-track background can be suppressed to be smaller than one event during the lifetime of the HL-LHC. The suppression for the QCD background mainly comes from requiring large track displacement, while displaced vertex reconstruction is mainly responsible for suppressing the fake-track background.

For the signal, the full set of \textit{vertexing-cuts} are applied with a total efficiency of $\epsilon_{\rm vtc} = 0.34 $ and $0.24$ for $m_X = 20$ GeV and 50 GeV, respectively. Applying $d_0$ cuts on all the tracks reduces the signal further. 
Multiplying the sub-sequential overall cut efficiency from the columns with 5 tracks, $\epsilon_{\rm vtc}$ and $(d_0 > 0.03 \text{m})^5$, one obtains the total cut efficiency for the signal. This is the exact signal efficiency for applying all the cuts. 
The remaining signal events as a function of branching ratio $\text{BR}(h \to X X)$ is given in the last row. Heavier $X$ has higher efficiency for several reasons. First, 
heavier $X$ moves slower, leading to a larger probability of decaying before reaching HGCAL for a fixed proper lifetime. Second, lighter $X$ has only a slightly better efficiency under the $\epsilon_{\rm vtc}$ cut.  Last,  lighter $X$ has a lower $d_0$ cut efficiency, because the tracks tend to be collimated with the direction of $X$. Therefore, the search is more sensitive to heavier $X$. 
For the VBF channel, the distributions of \textit{vertexing-cuts} variables in Fig. \ref{fig:VBF-precuts} and transverse impact parameter $d_0$ in Fig. \ref{fig:VBF-d0T} are similar to those of the ggF signal. 
Comparing with the ggF signal, the sensitivity in the VBF channel is weaker by about two orders of magnitude due to the smaller cross-section and the stringent VBF trigger threshold.

\subsection{The reach}
\label{subsec:reach}

The results in the previous sections allow the determination of the potential in the search for new physics. In this section, we present the results for both the ggF and VBF channels. 
In Fig.~\ref{fig:ggF}, we show the projected sensitivity in the Higgs exotic decay into LLPs branching fractions, $BR(h\to XX)$, as a function of the proper lifetime of the LLP for both channels.
The VBF search, shown in the left panel, represents a very conservative strategy with the existing VBF trigger. The ggF search, on the right panel, requires a dedicated displaced trigger. The solid line on the bottom of the color shaded region indicates the reach using a 5-displaced-track trigger, while for the solid line on the top of the shaded region, an additional $H_T > 100$ GeV cut is employed. It represents a more conservative version of the displaced trigger, and consequently, it decreases the sensitivity by a factor of 10. The best reach for VBF channel is
about BR$(h\to XX ) \sim \mathcal{O}(10^{-4})$, with the LLP lifetime of  $c\tau_X \sim 0.1$--$1$ meters, while for the ggF
channel it is about BR$(h\to XX ) \sim \mathcal{O}(10^{-5}\text{--}10^{-6})$ for a similar lifetime. Alternatively, for an LLP with $c\tau_X \sim 10^3$ meters, the HGCAL based search should be able to probe  BR$(h\to XX)$ down to a few $\times 10^{-4}$($10^{-2}$) in the ggF (VBF) channels, respectively.

For comparison, we show the limits from existing searches for our benchmark signal model in Fig.~\ref{fig:ggF}.
For very small $c \tau_X$, the best limits come from the ATLAS search for the prompt $h \to X X \to 4b$, at 13 TeV with $36.1 ~\rm{ fb^{-1}}$ \cite{Aaboud:2018iil}. 
A short lifetime of $X$ is allowed by the b-tagging algorithm, with maximal sensitivity for $c\tau_X \sim 0.5$ mm. 
For $c\tau_X$ between $\{10^{-2}, 10^3 \}$ m, there are several ATLAS searches using 13 TeV data. One is based on the  muon spectrometer (MS)  with $36.1 ~\rm{ fb^{-1}}$ \cite{Aaboud:2018aqj}.  The other uses the low-$E_T$ calorimeter energy ratio trigger, with $10.8 ~\rm{ fb^{-1}}$ \cite{Aaboud:2019opc}. 
In the gap for LLP lifetime around cm, the displaced jet searches can be sensitive. 
A recent CMS search based on displaced vertex in the tracker system with $139 ~\rm{ fb^{-1}}$ obtained limits
at the level of $10^{-1}$--$10^{-2}$ \cite{CMS-PAS-EXO-19-021} for LLP decay $X \to \bar{b}b$.
Though its limit on $X \to \bar{d}d$ is about 10 times better due to one reconstructed secondary vertex requirement.

Since we are using HL-LHC with integrated luminosity $3~\text{ab}^{-1}$, it is not a fair comparison for the existing limits. 
One can scale up the results of  those search to  $3~\text{ab}^{-1}$. The sensitivity gain is proportional to the square root of the luminosity increase since those searches have non-zero backgrounds. As a result, the gain from luminosity ranges from 9.1 to 4.8. 
The other improvements we have compared to the existing searches are from both trigger and background suppression. 
The trigger efficiencies are $0.033$, $0.040$ and $0.21$ for signals $m_X = 50$ GeV for VBF channel, ggF channel with $H_T$ cut and ggF channel without $H_T$ cut respectively. They have included the requirement of track arrival at HGCAL.
Therefore, it is easy to see that the novel trigger from \cite{Gershtein:2017tsv} provides an improvement about a factor of 5.
The last improvement comes from the ability of driving the background to a negligible level.
From Fig.~\ref{fig:ggF}, for long lifetime case (e.g. $c\tau = 1 ~{\rm km}$), the sensitivity of ggF channel with the novel trigger is better than the existing limit (e.g. ``ATLAS-MS") by a factor of about 1000. The improvement from the high luminosity is about 9, and the novel trigger contributes a factor of 5. The improvement from the background suppression using vertexing and track information at HGCAL contributes a factor of about 20, which is one of the dominant factors of our enhanced sensitivity.

\section{Conclusion}
\label{sec:conclusion}

High granularity calorimeters offer new opportunities for the search of the  long-lived particle. In this work, we study the potential reach for the long-lived particle signal based upon a new search mainly relying on the HGCAL upgrade of the CMS detector. 
We present results based on both the more conservative traditional VBF trigger and a pair of novel displaced track triggers. Based on a simplified modeling of the signal and background of this new approach, we carefully devised kinematical cuts and estimated the size of the leading backgrounds.   HGCAL provides the shower direction and timing information with unprecedented precision, enabling us to view them as ``tracks". 
We find that the QCD background is mostly prompt, which can be suppressed effectively by requiring a large
transverse impact parameter for multiple tracks, after applying the vertex-cuts removing the SM metastable mesons. Another major source of background is the fake-track background, which comes from mis-connected hits in the detector. The resulting tracks have
a random distribution, typically with a large transverse impact parameter and hence requires additional selection. We take advantage of the fact that these tracks to rarely fit in a common vertex and design
a set of corresponding vertexing-cuts to suppress such backgrounds. Using these selections, combined with our different trigger considerations, we obtained our projections of the HL-LHC sensitivities for Higgs decaying to LLPs at HGCAL that improves the current reach by 2--4 orders of magnitude.

Finally, we note here our study is rather conservative in many aspects. leaving potentially large room for further improvement.
\begin{itemize}
\item For the QCD background and the signal, the most relevant parameter of the HGCAL detector is its angular resolution. In this study, we use the standalone angular resolution from HGCAL. In practice, the track trajectory can be detected by both the tracking system and the tracker inside the HGCAL. Combining the two can further improve the angular resolution. This will result in a better DV fitting and enhance the suppression of the QCD background. 
\item 
We require the LLP to decay way before reaching the HGCAL detector, leaving at least five hits in the outer layer of the tracking system. With a detailed understanding of the showering behavior of the background, novel searches for LLP decaying \textit{within} the HGCAL can also be sensitive. This will enable an HGCAL standalone trigger, and enlarge the decay volume for the LLP (hence the reach in $c \tau_X$)  by a factor of a few.
\item 
In our selection cuts, we have left a large room for improvement. For instance, we did not fully utilize the timing information of the displaced vertices. This is due to our lack of understanding of fake-track behavior in the timing dimension. A full-fledged 4D vertex fit could result in a much more powerful suppression of the background. 
\item For LLPs with lower lifetime, our cuts are not optimal. 
Three of our cuts are mainly responsible for reducing the signal efficiency at lower lifetimes: these are
$d_0>0.05$~m, which is 10-50 times larger than the normal cuts on displaced tracks;
$\bar t > 1$~ns, which effectively requires the signal to decay after traveling 30 cm;
$|\bar z|>$0.4~m, which again requires the signal to decay after traveling more than 40 cm. Many of these cuts can be adjusted and make the search more effective.
\end{itemize}

\section*{Acknowledgement}
We thank Jared Evans, Yuri Gershtein, Simon Knapen, Felix Kling for helpful discussion.
JL acknowledges support by an Oehme Fellowship. ZL is supported in part by the NSF under Grant No. PHY1620074 and by the
Maryland Center for Fundamental Physics. XPW is supported by the U.S. Department of Energy under Contract No. DE-AC02-06CH11357. LTW is supported by the DOE grant DE-SC0013642.
ZL and LTW acknowledge the hospitality of the Kavli Institute for Theoretical Physics, UC Santa Barbara, during the “Origin of the Vacuum Energy and Electroweak Scales” workshop supported by the NSF grant PHY-174958.
ZL and LTW would also like to thank Aspen Center for Physics (supported by NSF
grant PHY-1607611) for support from their programs
and providing the environment for collaboration.

\section{Appendix}
\label{sec:app}

We put the supportive figures and tables in the Appendix to avoid redundancy in the main text while keeping helpful information to the readers.

\begin{figure}[ptb]
    \centering
    \begin{tabular}{cc}
                (a) & (b)  \\
        \includegraphics[width=0.33 \columnwidth]{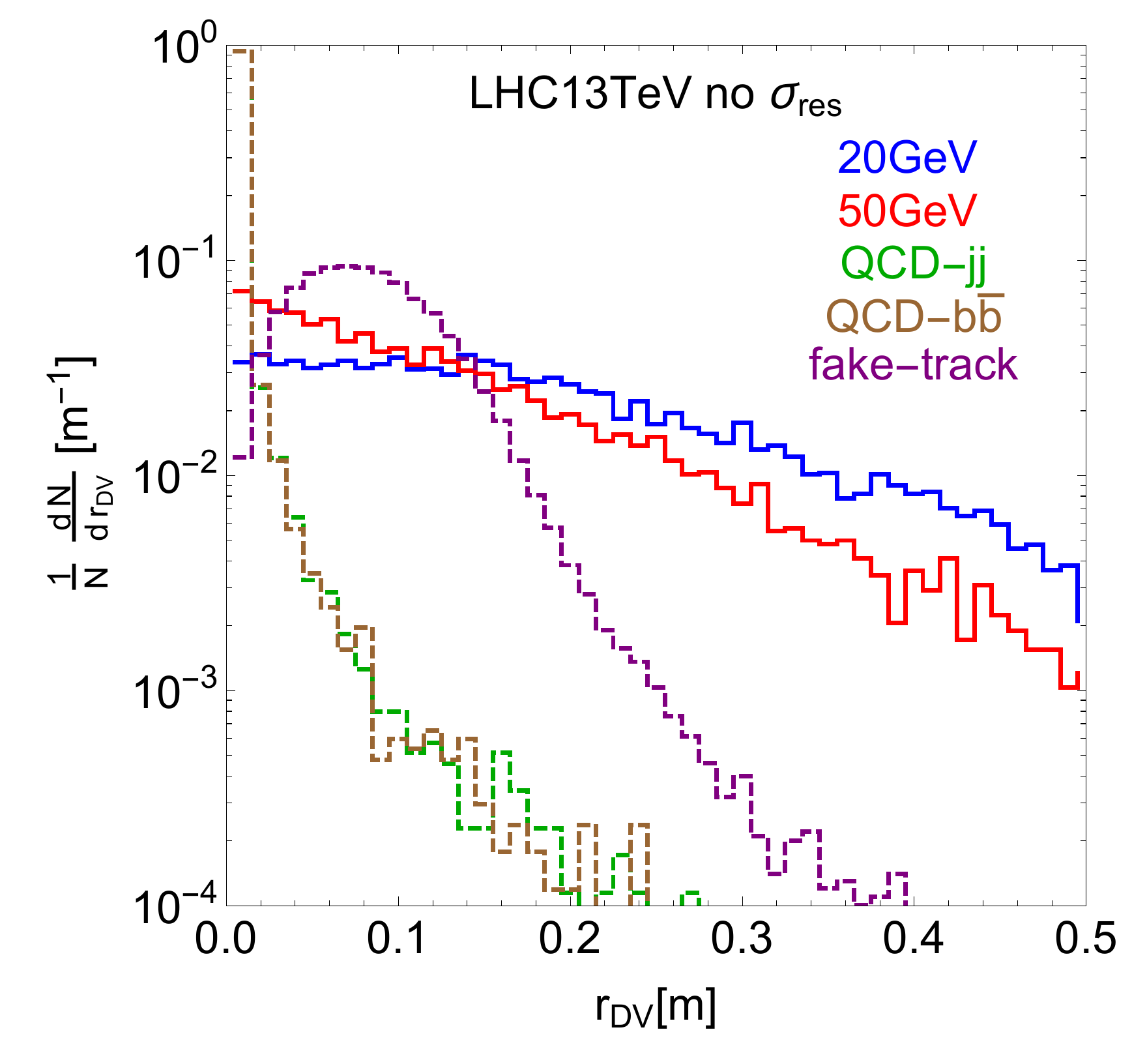} &
        \includegraphics[width=0.33 \columnwidth]{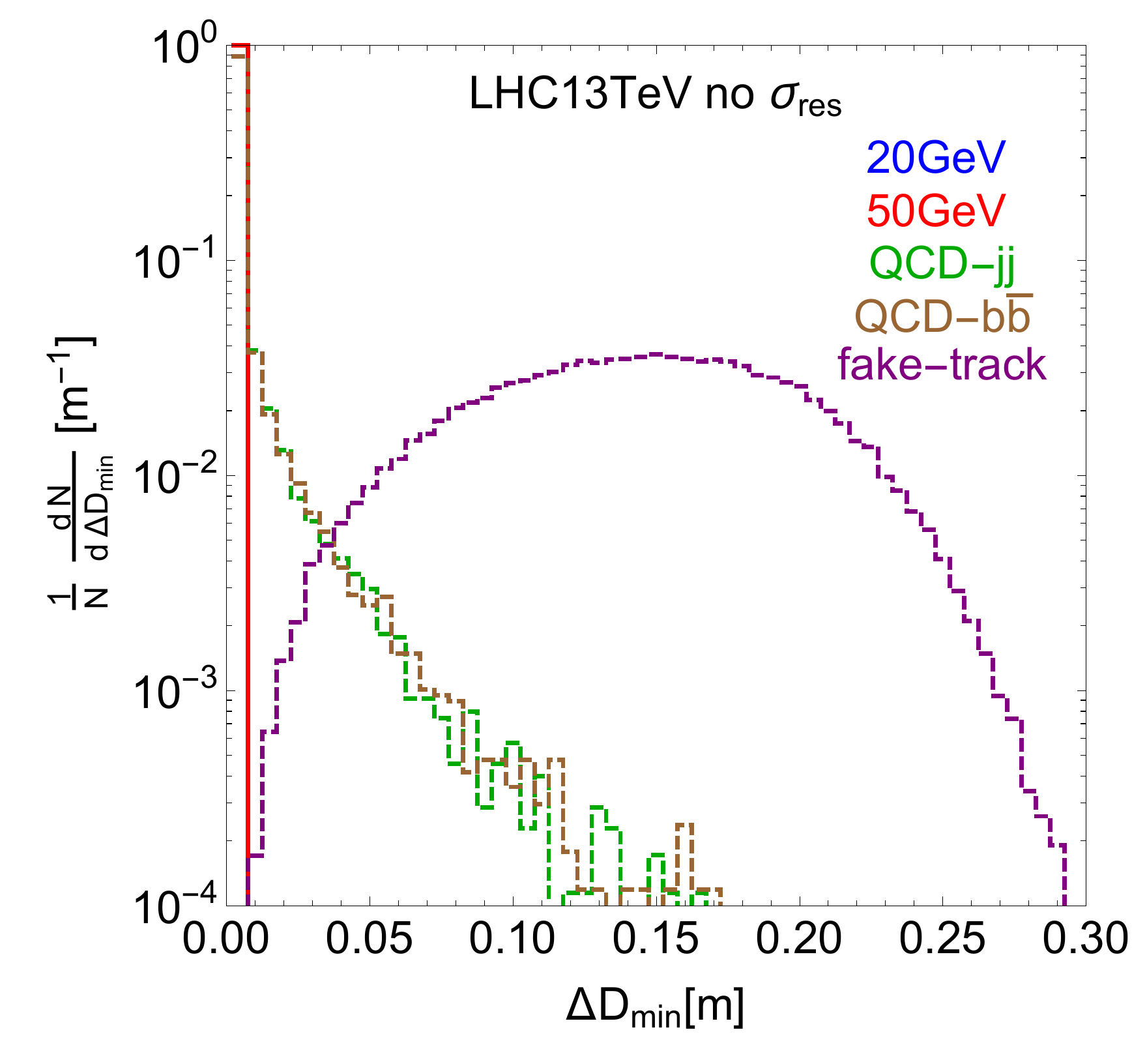}    
    \end{tabular}
    \begin{tabular}{ccc}
                (c) & (d) & (e)  \\
        \includegraphics[width=0.33 \columnwidth]{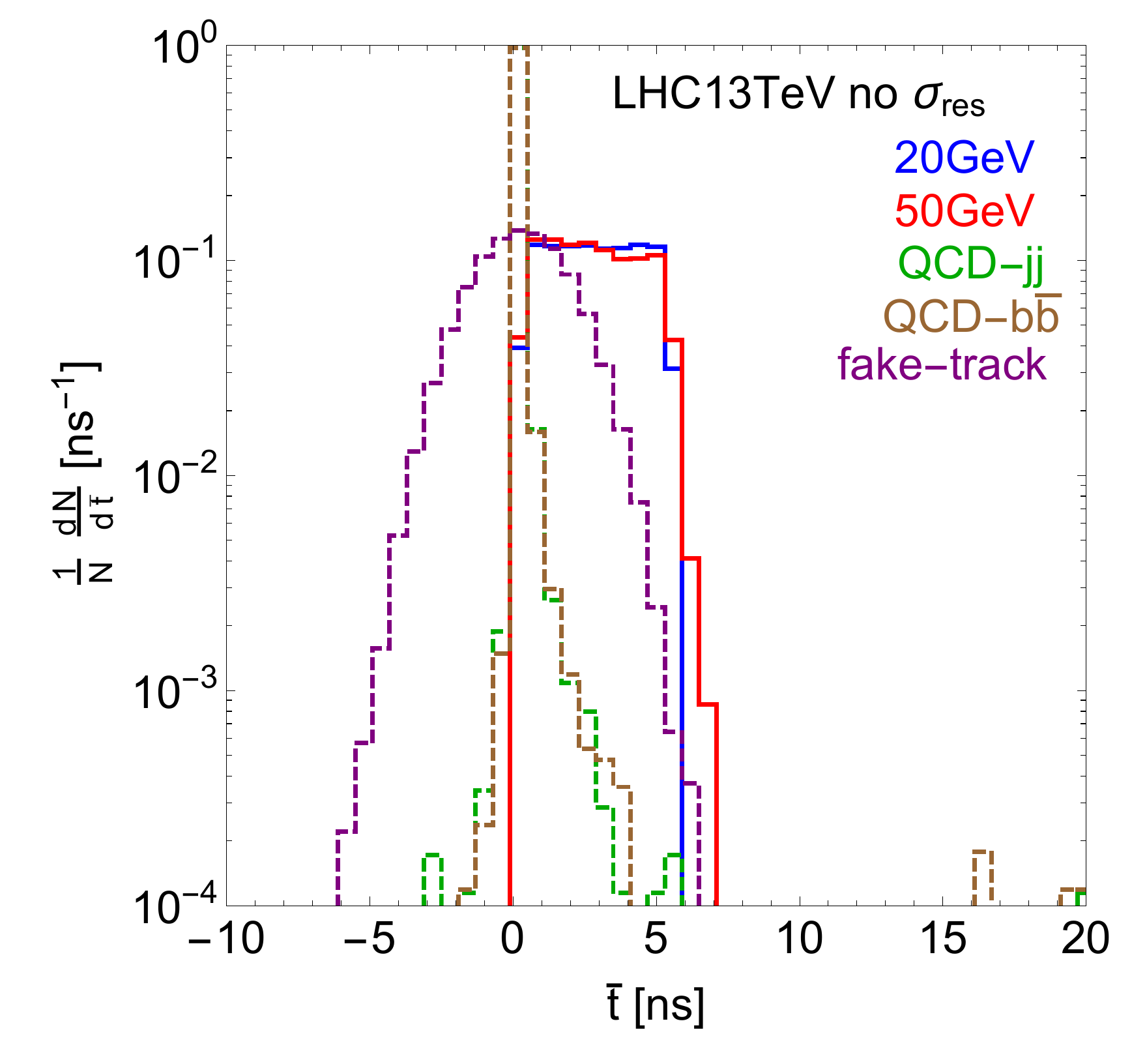} &
        \includegraphics[width=0.33 \columnwidth]{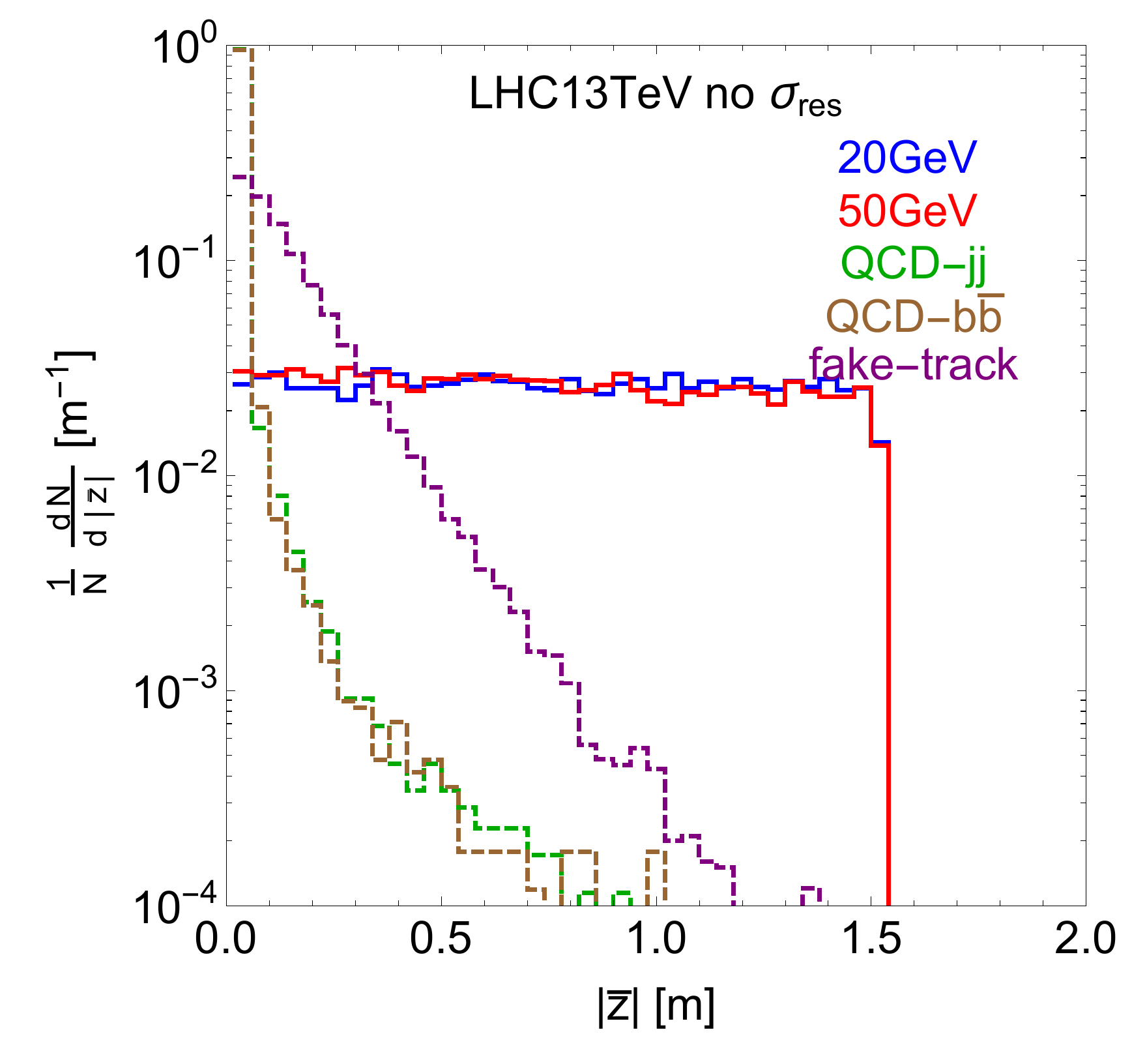}  &
        \includegraphics[width=0.33 \columnwidth]{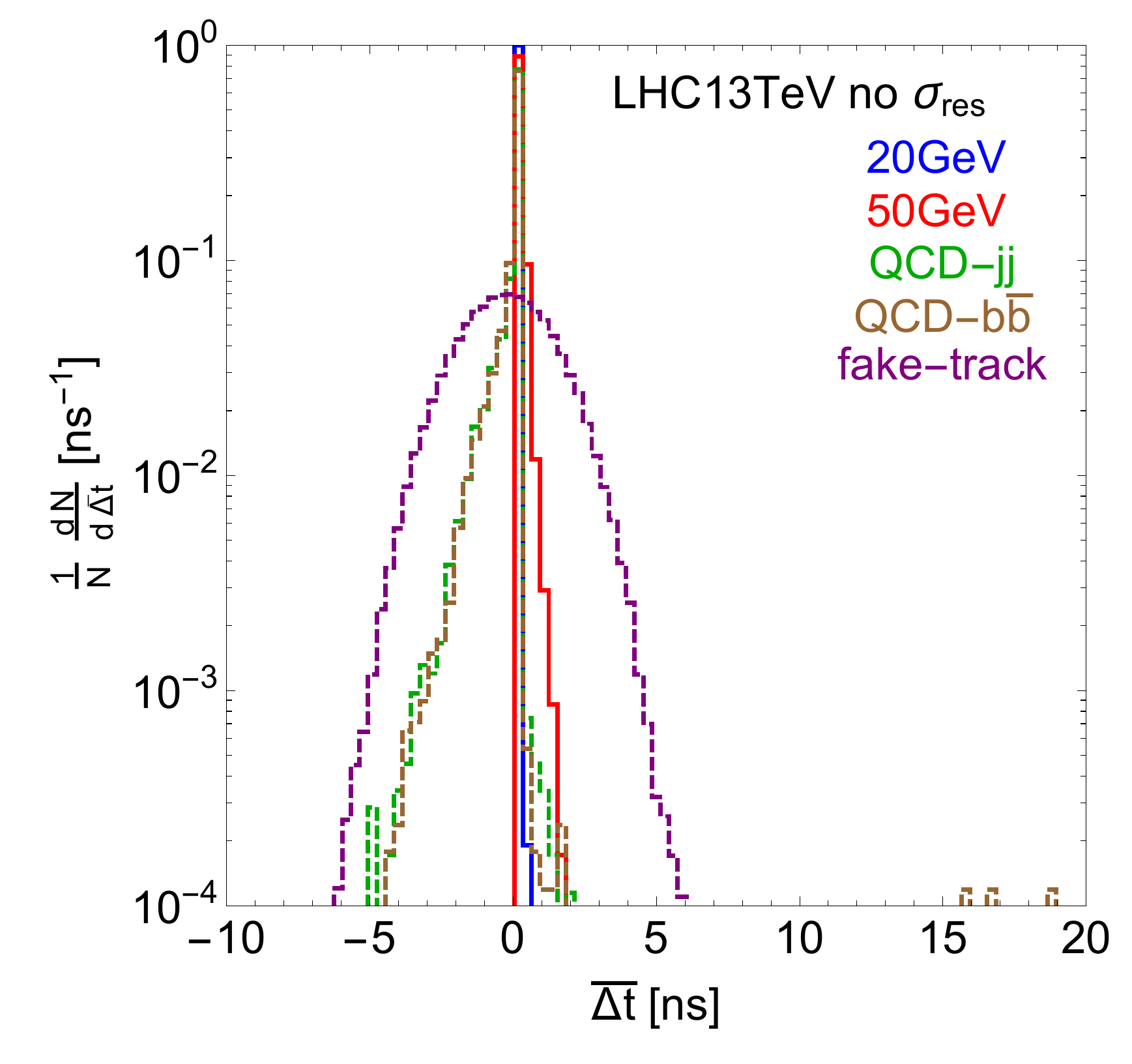}  \\        
                (f) & (g) & (h) \\
        \includegraphics[width=0.33 \columnwidth]{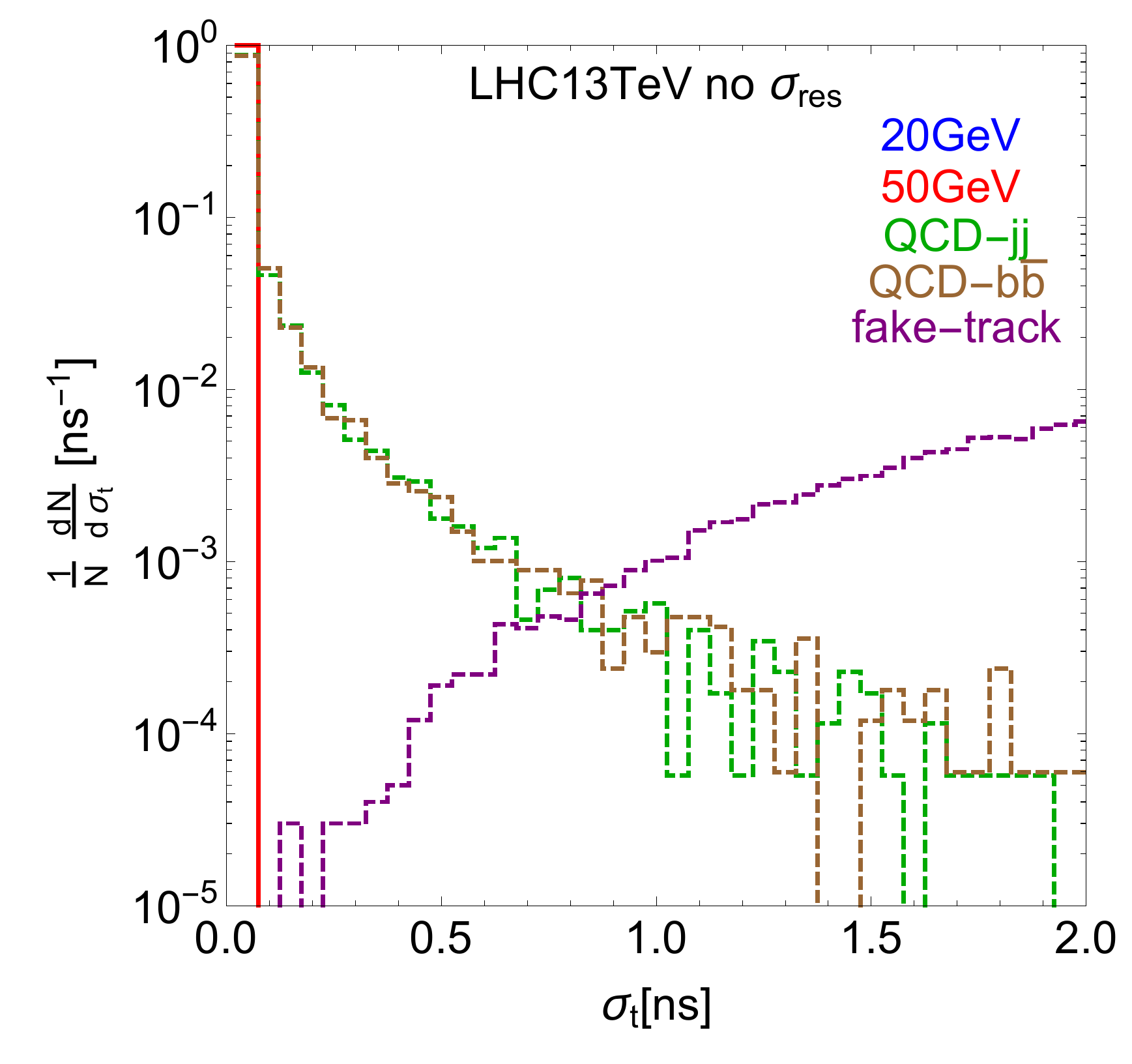} &
        \includegraphics[width=0.33 \columnwidth]{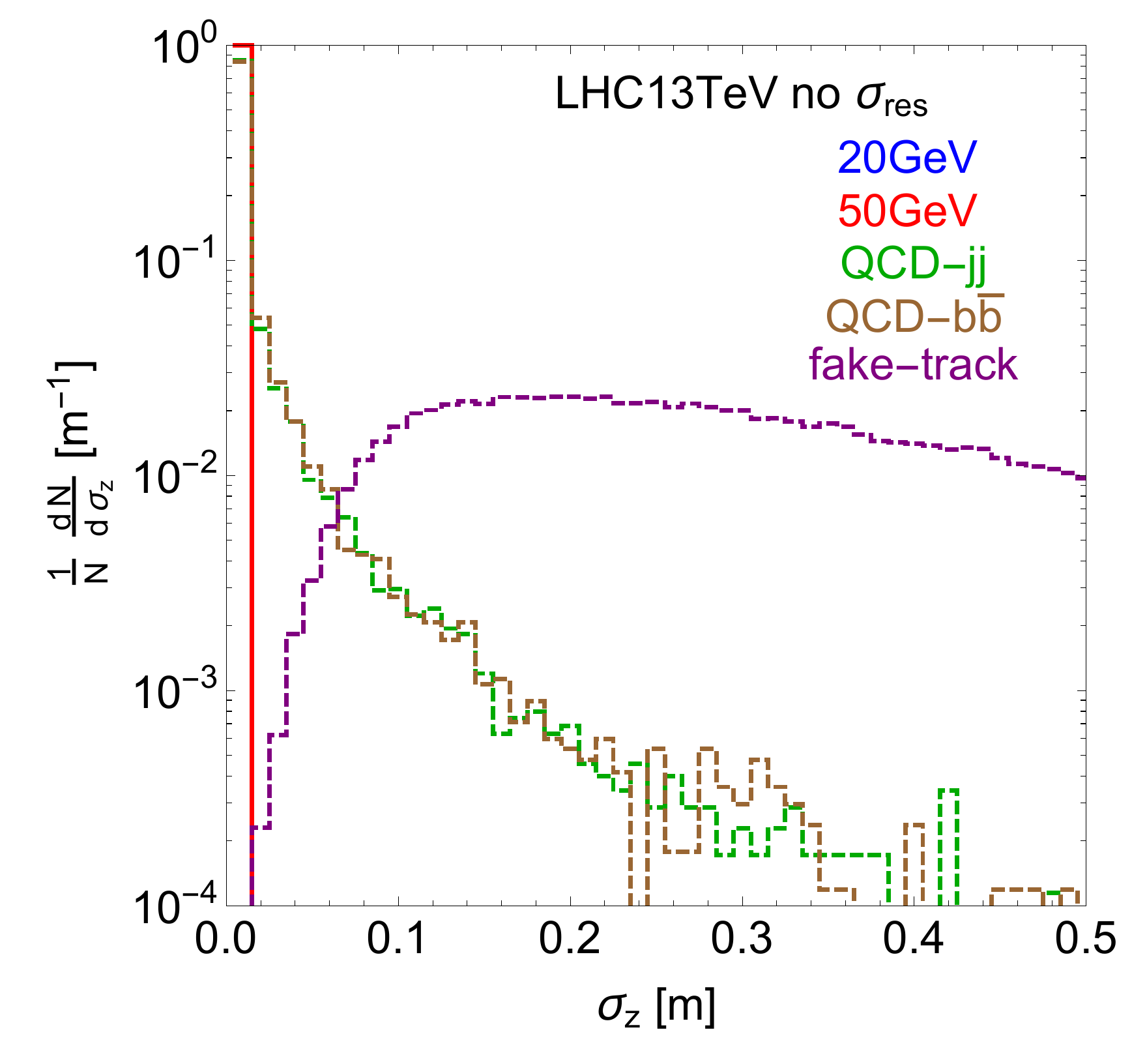} &
        \includegraphics[width=0.33 \columnwidth]{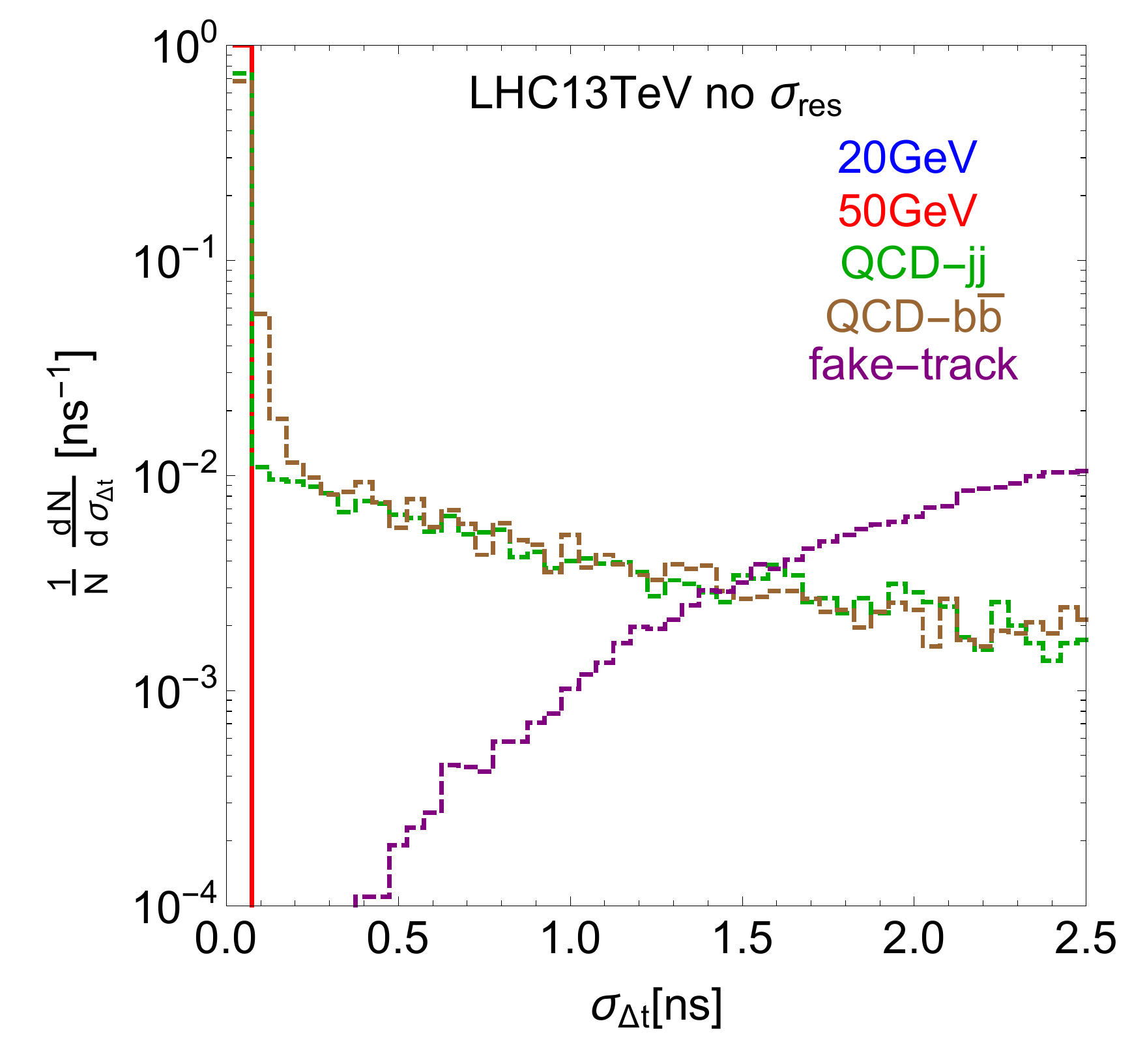}  
    \end{tabular}
    \caption{The kinetic variable distributions for the QCD background, fake-track background and the signal \textit{without} the angular resolution effect included. The variables and definitions are the same as Fig. \ref{fig:9varswithRES}.
    }
    \label{fig:9varswithoutRES}
\end{figure}

In Fig.~\ref{fig:9varswithoutRES}, kinetic variable distributions for the QCD background, fake-track background and the signal are shown \textit{without} the angular resolution effect included. This is a sanity check for Fig.~\ref{fig:9varswithRES} which \textit{has} included the angular resolution effect. For the distribution of $\Delta  {\rm D}_{\min}$, $\rm{\sigma_t} $, $\rm{\sigma_{z}} $ and $\rm{\sigma_{\Delta t}} $, the signals are exactly at 0 while the QCD background are peaked at 0. It shows that the DV fitting algorithm has worked well and found the expected true vertex.

\begin{figure}[ptb]
        \centering
    \begin{tabular}{cc}
        \includegraphics[width=0.35 \columnwidth]{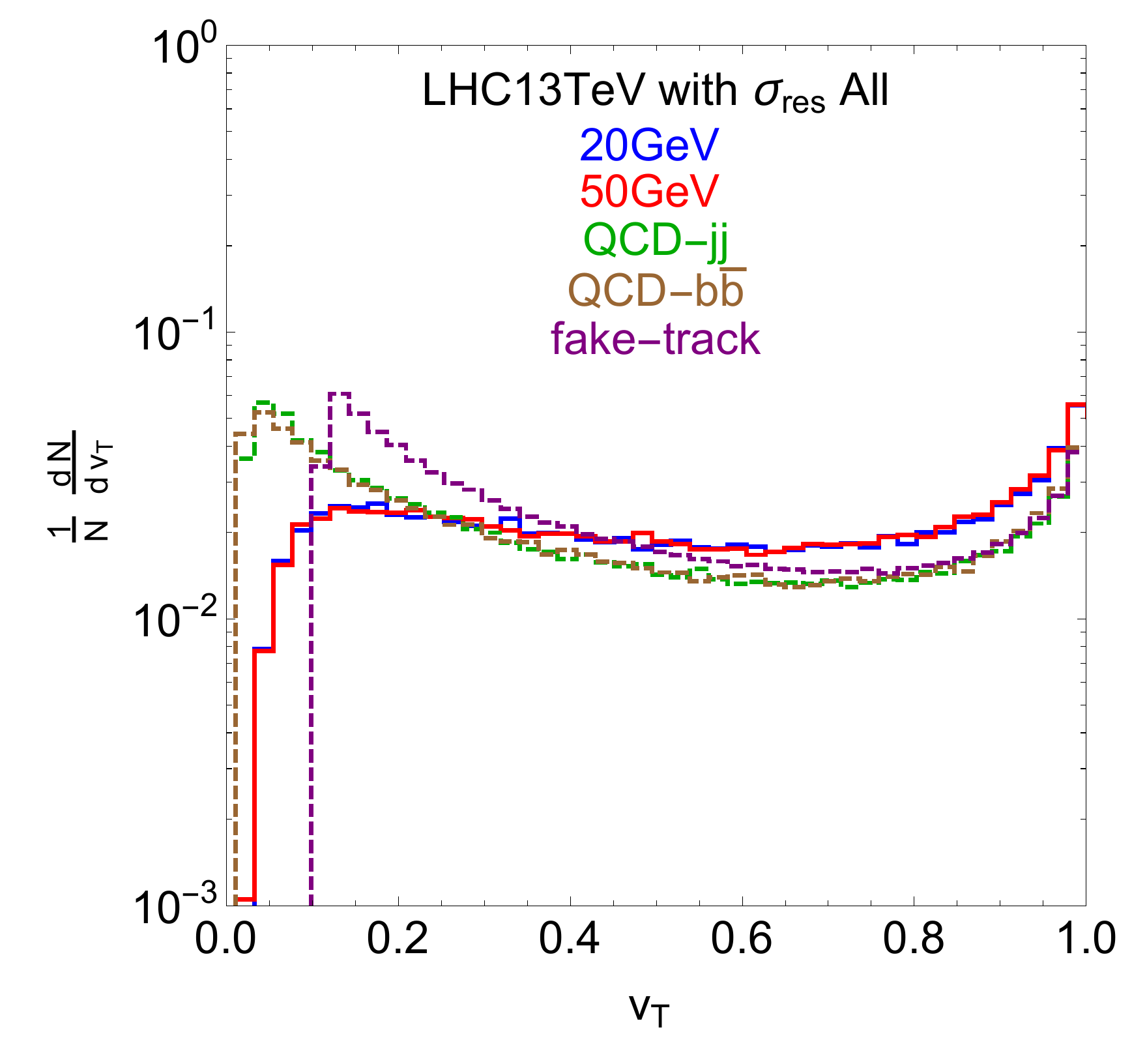} &
        \includegraphics[width=0.35 \columnwidth]{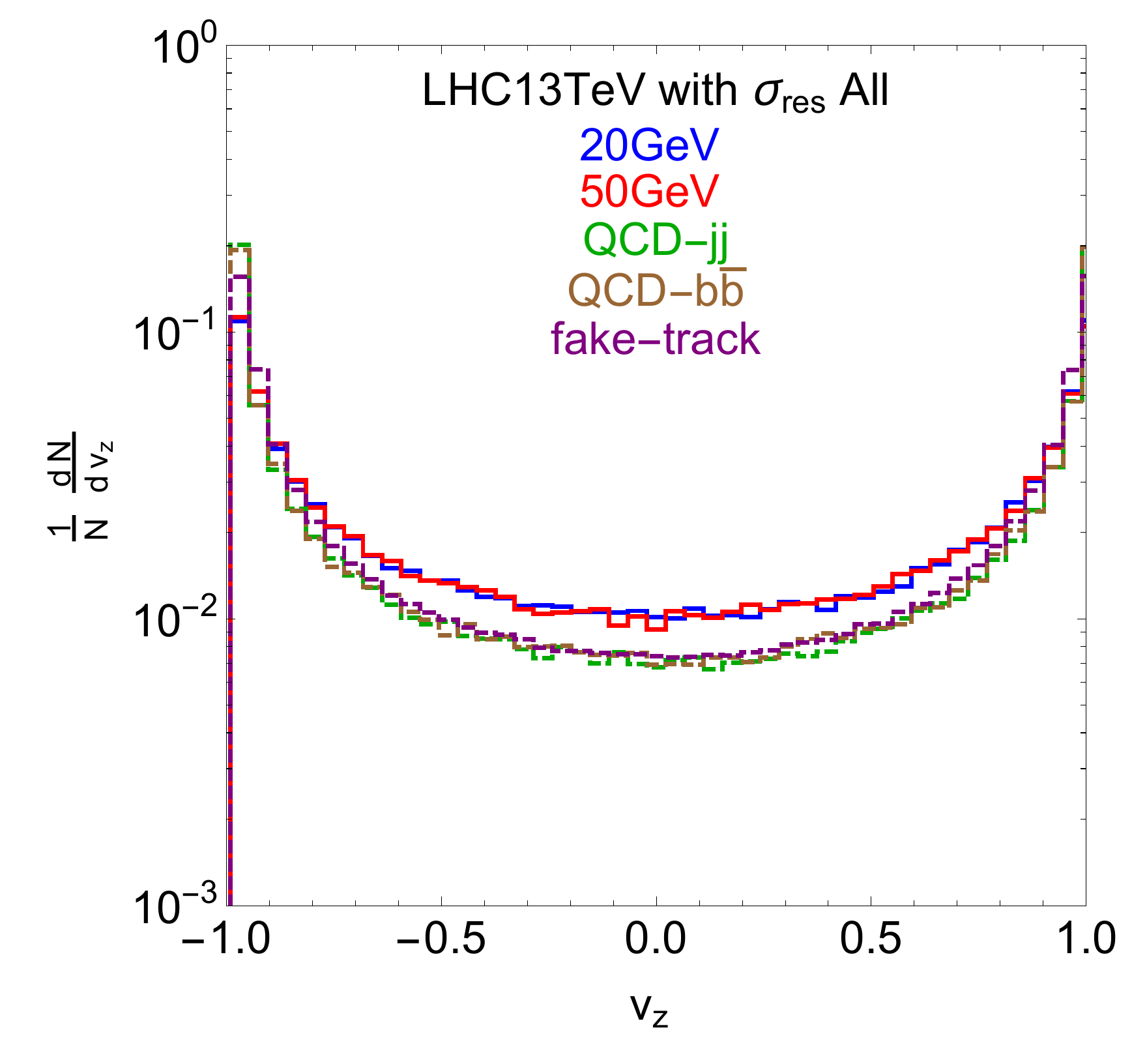}    \\
        \includegraphics[width=0.35 \columnwidth]{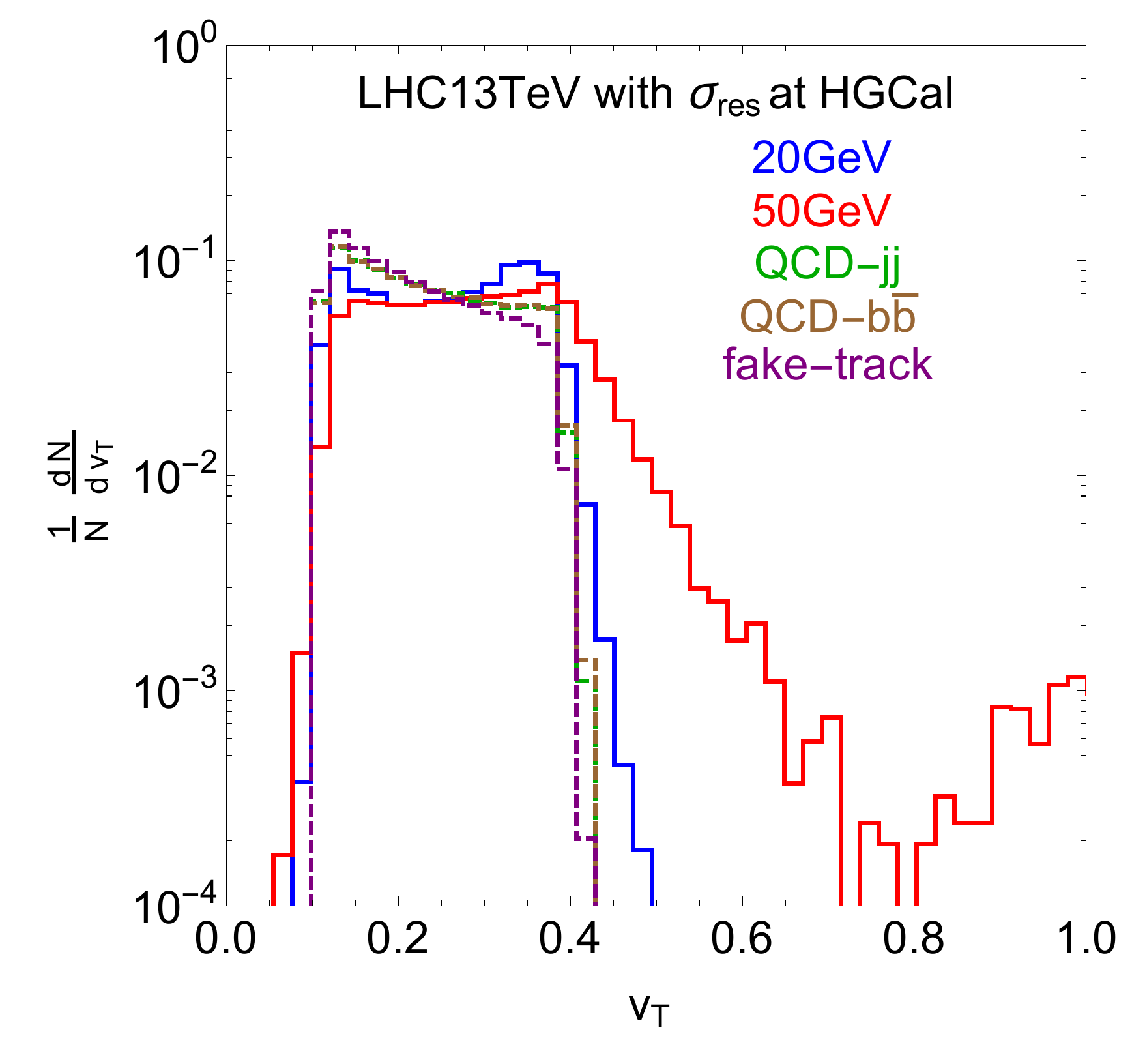} &
        \includegraphics[width=0.35 \columnwidth]{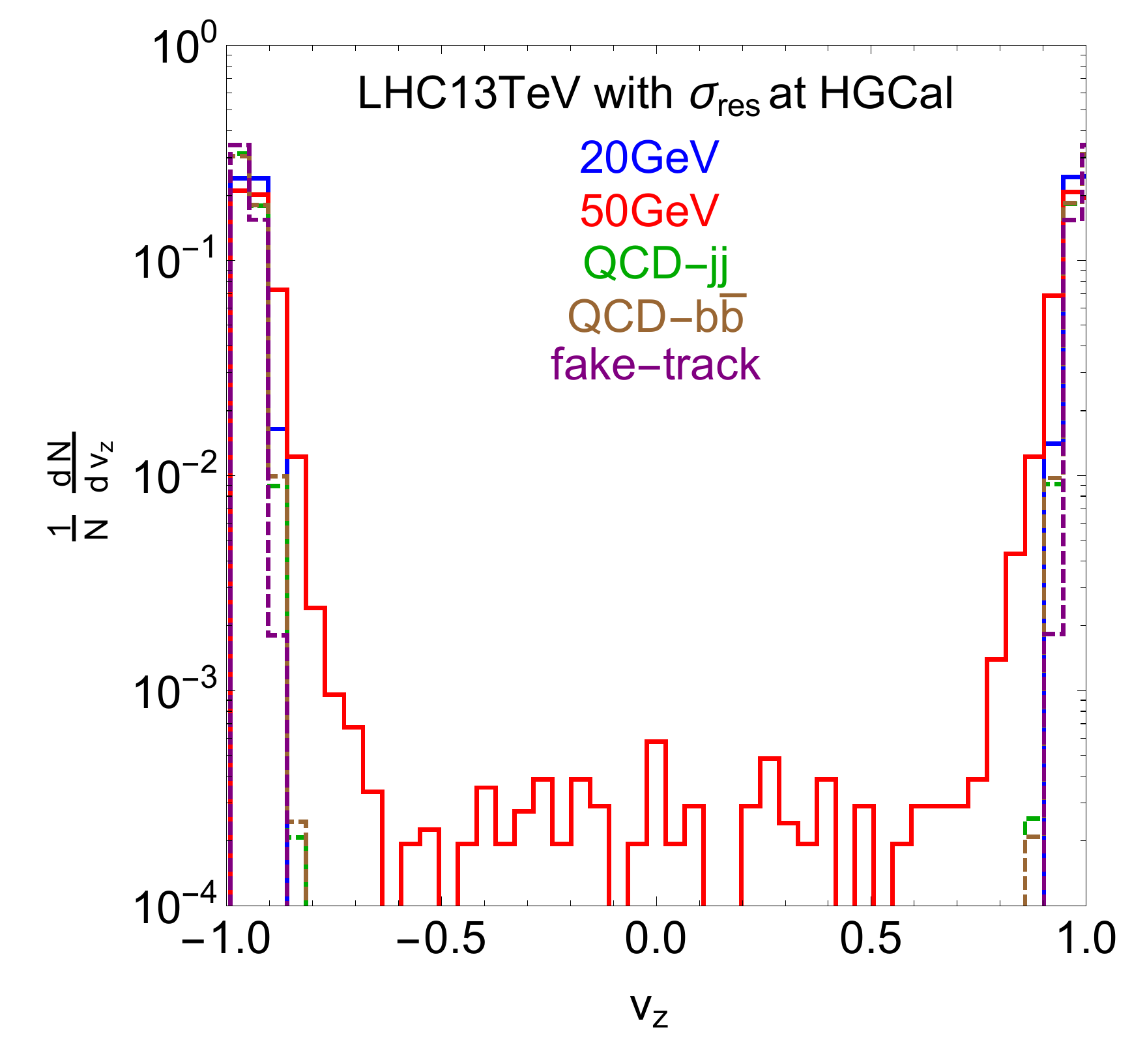}  
    \end{tabular}
    \caption{The distributions of transverse velocity $v_T$ and longitudinal velocity $v_z$ for the tracks \textit{without} (upper panel) and \textit{with} (lower panel) the requirement to arrive at HGCAL. 
    }
    \label{fig:otherparameters}
\end{figure}

In Fig.~\ref{fig:otherparameters}, we plot the distribution of $v_T$ and $v_z$ for the tracks \textit{with} and \textit{without} the requirement to arrive at HGCAL. Moreover, we require the track should not hit the barrel electrocalorimeter. 
In the upper panel, it is clear to see that \textit{without} requiring arriving at HGCAL, the $|v_z|$ distribution for all the signal and background have a peak around 1, while a flat valley in the middle. This reflects the distribution of the track zenith angle $\theta$. 
Once requiring arriving at HGCAL, we can see that the $v_T$ for signal and backgrounds are dominated by small values, e.g., $0.1 \sim 0.4$. The reason is that HGCAL is a forward detector, which picks the forward tracks. Therefore, the $v_T$ is forced to be small.

\begin{figure}[ptb]
    \centering
    \includegraphics[width=0.45 \columnwidth]{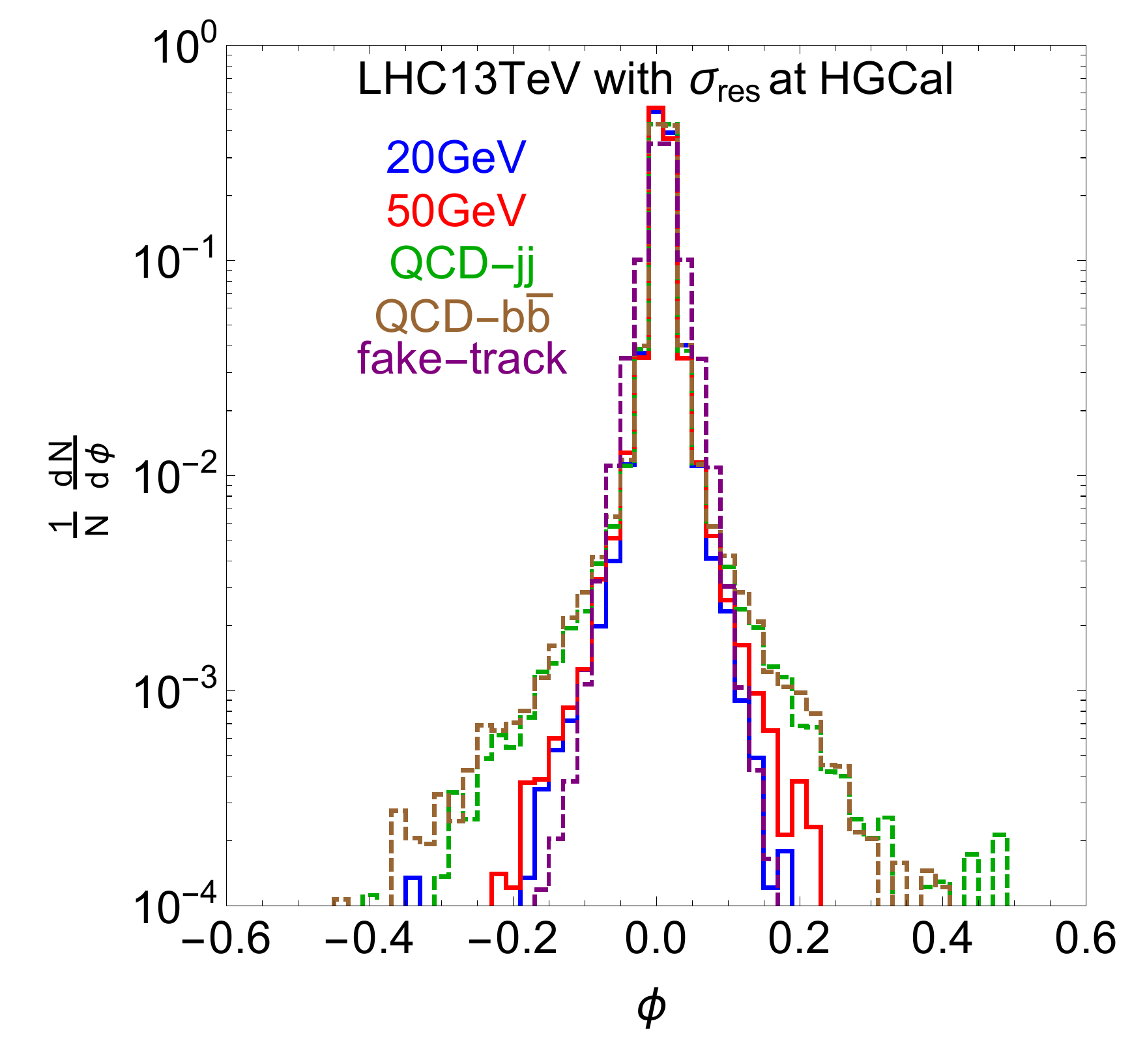}  
    \caption{The distribution of $\Delta \phi$ for the tracks in the DV fitting procedure, where $\Delta \phi$ is the azimuthal angle change when moving from the reference point to the fitted DV.}.
    \label{fig:deltaPhi}
\end{figure}

In Fig.~\ref{fig:deltaPhi}, we show the distribution of $\Delta \phi$ for the tracks in the DV fitting procedure. The QCD background and the signal have a similar distribution, peaked with $\Delta \phi = 0$ because they both have a common vertex. $\Delta \phi $ comes from the angular resolution effect of HGCAL, which has a spread of about 0.02, which is a few times the angular resolution $\sigma_{\theta}$. For fake-track background, the distribution of $\Delta \phi$ has a reason smaller than order 1. From the definition of $\Delta \phi$, its starting point (the reference point) is the closest point to the origin. Hence, the fitted DV should be enclosed within these reference points, as going far from the origin will lead to a bad fit. As a result, the movement in $\phi$ angle is not large from the starting point to DV.

In Table~\ref{tab:jjbb-d0T-noprecuts}, we show the independence correlation table for multiple $d_0$ tracks for QCD dijet backgrounds \textit{without} applying \textit{vertexing-cuts}. This is an auxiliary check for Table~\ref{tab:jjbb-d0T}. It has higher statistics and also shows the $d_0$ of different tracks are nicely \textit{independent} under this condition.

\begin{table}[ptb]
    \begin{center}
        \begin{adjustbox}{max width=\textwidth}
            \begin{tabular}{|c|c|c|c|c|c|}
                \hline
                $jj$ dijets &  $d_0>0.01$ m  & $d_0>0.015$ m & $d_0>0.02 $ m &   $d_0>0.025$ m &   $d_0>0.03$ m \\ \hline
                $ {\rm \rho}_{d}^1$  &    $1.0\pm 0.06$    & $1.0\pm 0.008$   & $1.0\pm 0.01$   & $1.0\pm 0.015$  & $1.0\pm 0.02$ \\
                \hline
                $ {\rm \rho}_{d}^2$ &  $1.0\pm 0.01$  &  $0.98\pm 0.016$ &    $0.96\pm 0.025$       & $0.88\pm 0.038$  & $0.74\pm 0.053$ \\
                \hline
                $ {\rm \rho}_{d}^3$ &  $ 0.99\pm 0.018$ &  $0.98\pm 0.032$ &  $0.90\pm 0.062 $        &$0.90\pm 0.15$  &$1.3\pm 0.65$\\
                \hline
                $ {\rm \rho}_{d}^4$ &   $0.97\pm 0.027$ &  $1.0\pm 0.07$ &  $0.75\pm 0.14$      & -            & -\\
                \hline
                $ {\rm \rho}_{d}^5$ &   $0.95\pm 0.04$  & $ 0.95\pm 0.14$ &  -      & -             & -  \\
                \hline
                $b\bar{b}$ dijets &  $d_0>0.01$ m  & $d_0>0.015$ m & $d_0>0.02$ m &   $d_0>0.025$ m &   $d_0>0.03$ m \\
                \hline
                $ {\rm \rho}_{d}^1$  &    $1.0\pm 0.06$    & $1.0\pm 0.008$   & $1.0\pm 0.01$   & $1.0\pm 0.015$  & $1.0\pm 0.02$ \\
                \hline
                $ {\rm \rho}_{d}^2$ &  $1.0\pm 0.01$  &  $1.0\pm 0.017$ &    $0.97\pm 0.026$       & $0.89\pm 0.40$  & $0.76\pm 0.056$ \\
                \hline
                $ {\rm \rho}_{d}^3$ &  $ 0.98\pm 0.018$ &  $0.95\pm 0.032$ &  $0.93\pm 0.066 $        &$0.69\pm 0.1$  &-\\
                \hline
                $ {\rm \rho}_{d}^4$ &   $0.97\pm 0.027$ &  $0.93\pm 0.06$ &  $1.1\pm 0.24$      & -            & -\\
                \hline
                $ {\rm \rho}_{d}^5$ &   $0.94\pm 0.04$  & $ 0.81\pm 0.11$ &  -      & -             & -  \\
                \hline
            \end{tabular}
        \end{adjustbox}
    \end{center}
    \caption{The correlation table for multiple $d_0$ tracks for QCD dijet backgrounds \textit{without} applying the \textit{vertexing-cuts}. The symbol ``-" means no events left and the number in () indicates the small number of statistics after the cuts. When increasing to multiple tracks and larger $d_0$ cuts, there are less events thus the result suffers from larger statistical fluctuations. It is an auxiliary check for Table~\ref{tab:jjbb-d0T} that is after applying the \textit{vertexing-cuts}. 
    }
    \label{tab:jjbb-d0T-noprecuts}
\end{table}

The Table~\ref{tab:mis-presix-weak} shows the independence correlation table for \textit{vertexing-cuts} variables for fake-track backgrounds, but with a weaker set of cuts comparing with Table \ref{tab:mis-presix}. We can see that most of the correlations are around 1 (\textit{approximate independent}), with some results are $4.8$ and $20$ which are \textit{conservative}. With the Table~\ref{tab:mis-presix} and Table~\ref{tab:mis-presix-weak}, it indicates that the estimate of fake-track background by multiplying each of these efficiency should be considered as \textit{conservative}.

\begin{table}[ptb]
    \begin{center}
        \begin{adjustbox}{max width=\textwidth}
            \begin{tabular}{|c|c|c|c|c|c|c|}
                \hline
                fake-track &  $ \rm{r_{DV}}>0.05$ m & $\Delta  {\rm D}_{\min} <0.05 $ m & $\bar{t} >2$ ns &  $\rm{\sigma_t} < 0.5$ ns & $\left| \bar{z} \right|>0.4 $ m   & $\rm{\sigma_{z}} < 0.1$ m  \\  \hline
                $ \rm{r_{DV}}>0.05$ m             &  & $0.814 \pm 0.006$ & $0.95  \pm 0.004$ & $1.03 \pm 0.08$ & 
                $0.751 \pm 0.004$  & $5.58 \pm 0.07 $ \\
                \hline
                $\Delta  {\rm D}_{\min} <0.05 $ m  & $0.814 \pm 0.006$ &       & $0.95 \pm 0.02$ & $1.19 \pm 0.36$ 
                & $0.65\pm 0.01 $ & $2.37 \pm 0.09 $ \\
                \hline
                $\bar{t} >2$ ns                                & $0.95  \pm 0.004$ & $0.95 \pm 0.02$     &       & $0.40 \pm 0.04 $ 
                & $0.533 \pm 0.005 $ & $1.19 \pm 0.02 $ \\
                \hline
                $\rm{\sigma_t} < 0.5$ ns                & $1.03 \pm 0.08$ & $1.19 \pm 0.36$     & $0.40 \pm 0.04 $     &      & $2.49 \pm 0.80 $ & $0.77 \pm 0.16 $ \\
                \hline
                $\left| \bar{z} \right| >0.4 $ m       & $0.751 \pm 0.004$ & $0.65\pm 0.01 $    & $0.533 \pm 0.005 $      & $2.49 \pm 0.80 $     &       & $16.23 \pm 0.94$ \\
                \hline
                $\rm{\sigma_{z}} < 0.1$ m             & $5.58 \pm 0.07 $  & $2.37 \pm 0.09 $     & $1.19 \pm 0.02 $      & $0.77 \pm 0.16 $     & $16.23 \pm 0.94$      &    \\
                \hline
            \end{tabular}
        \end{adjustbox}
    \end{center}
    \caption{The correlation table for \textit{vertexing-cuts} variables for fake-track backgrounds.
        These cuts are weaker than the cuts in Table \ref{tab:mis-presix}.
    }
    \label{tab:mis-presix-weak}
\end{table}

\begin{figure}[ptb]
    \centering
    \begin{tabular}{cc}
    \includegraphics[width=0.4 \columnwidth]{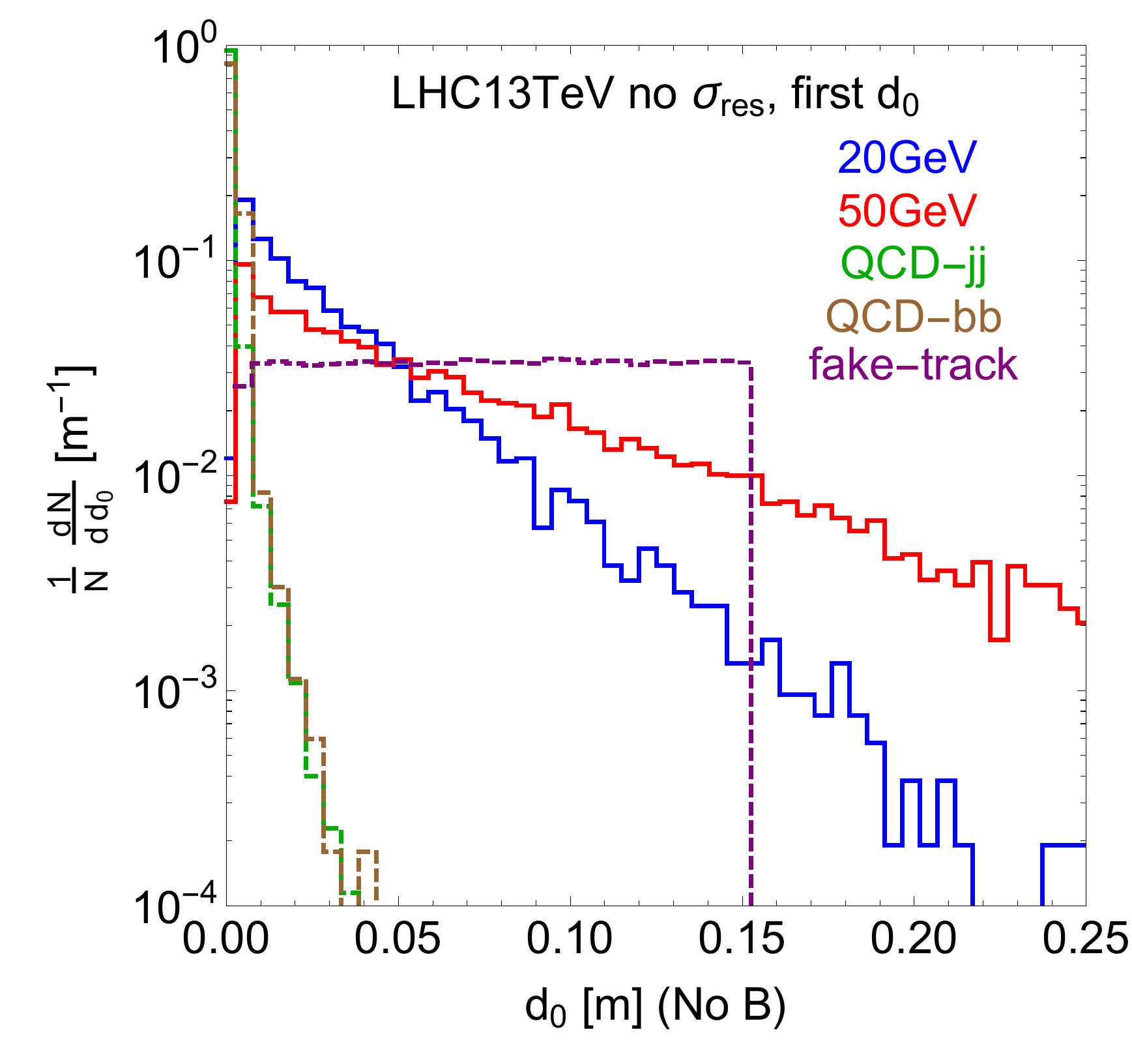} &
    \includegraphics[width=0.4 \columnwidth]{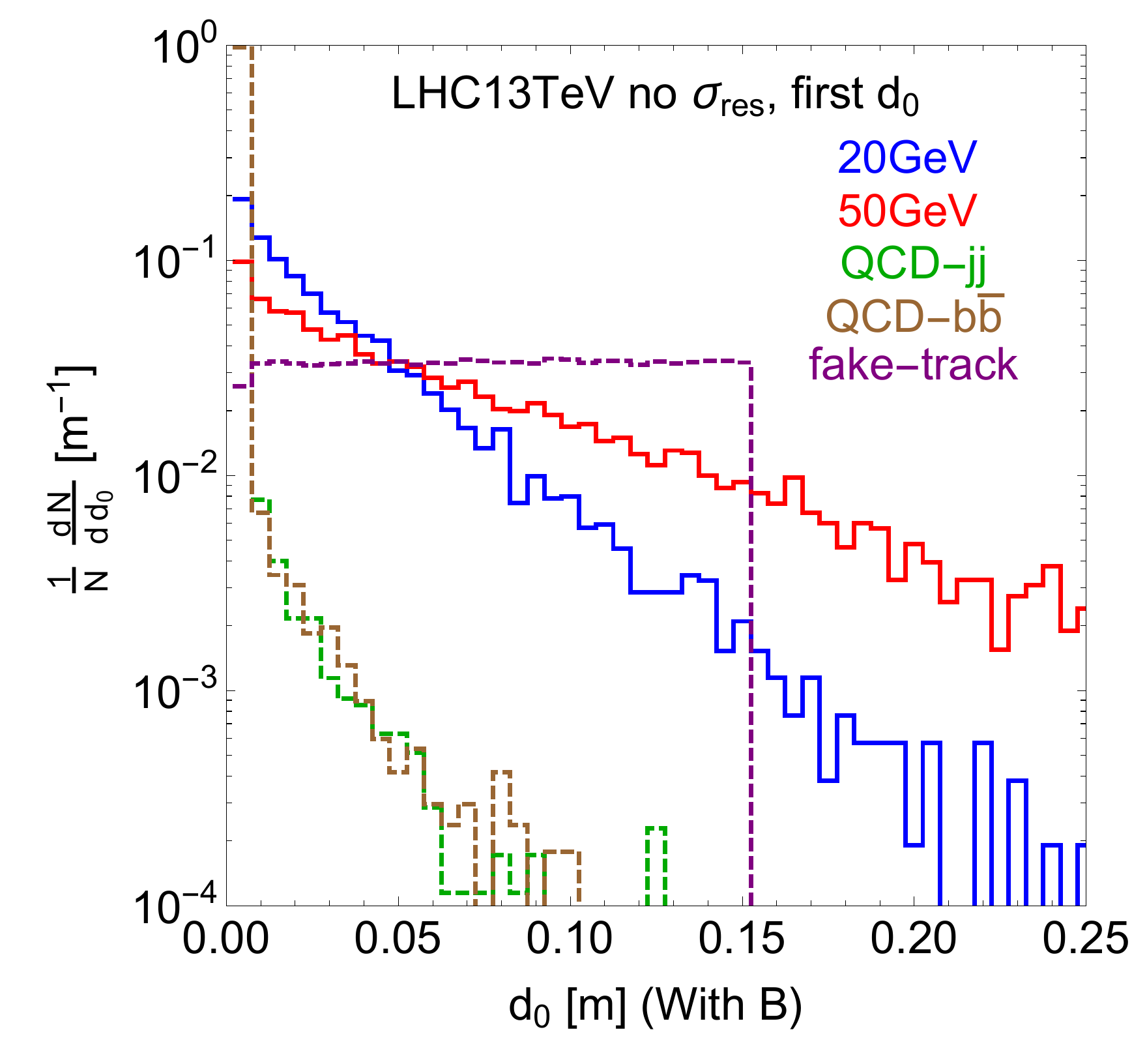}  \\
    \includegraphics[width=0.4 \columnwidth]{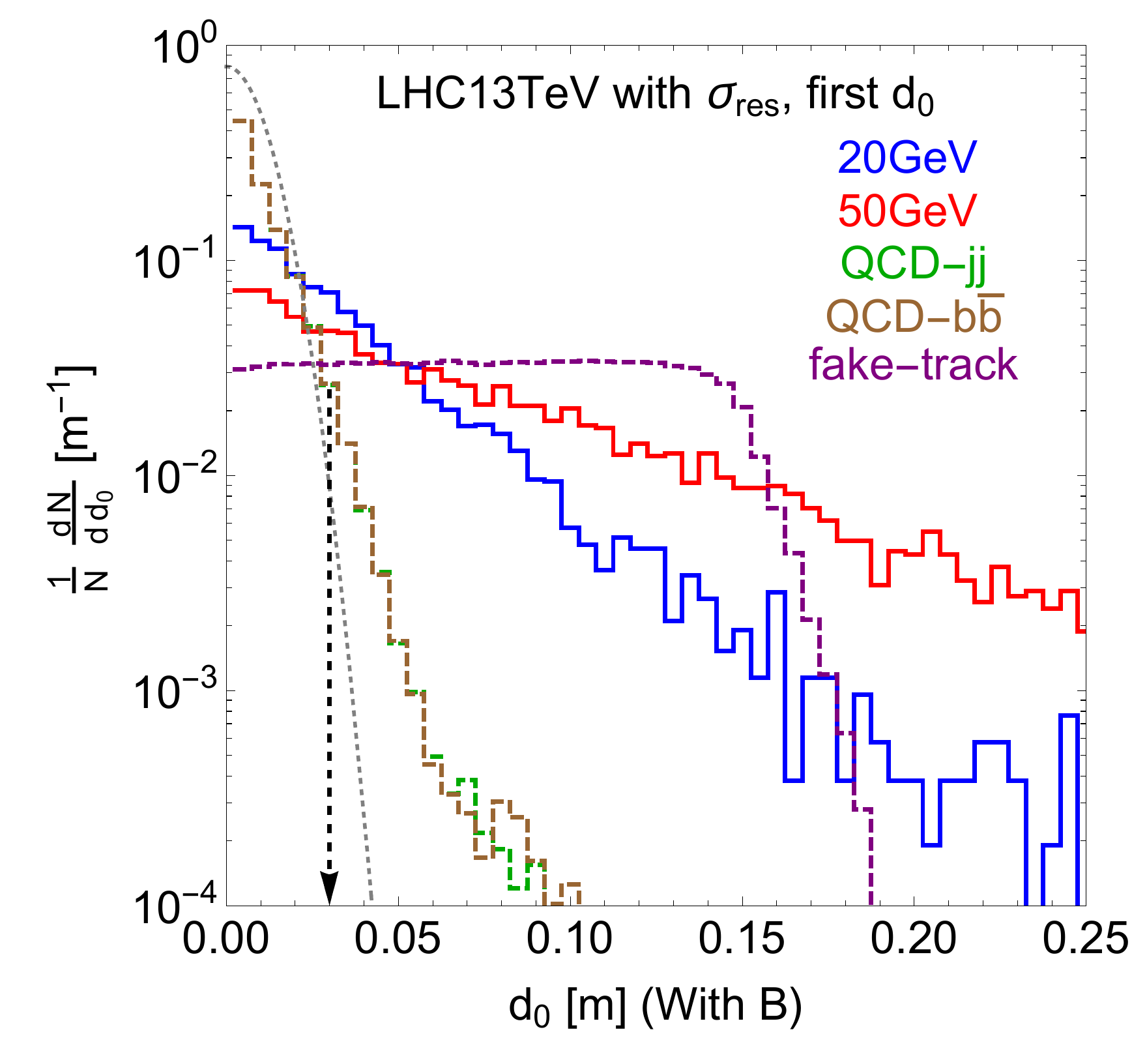} &
    \includegraphics[width=0.4 \columnwidth]{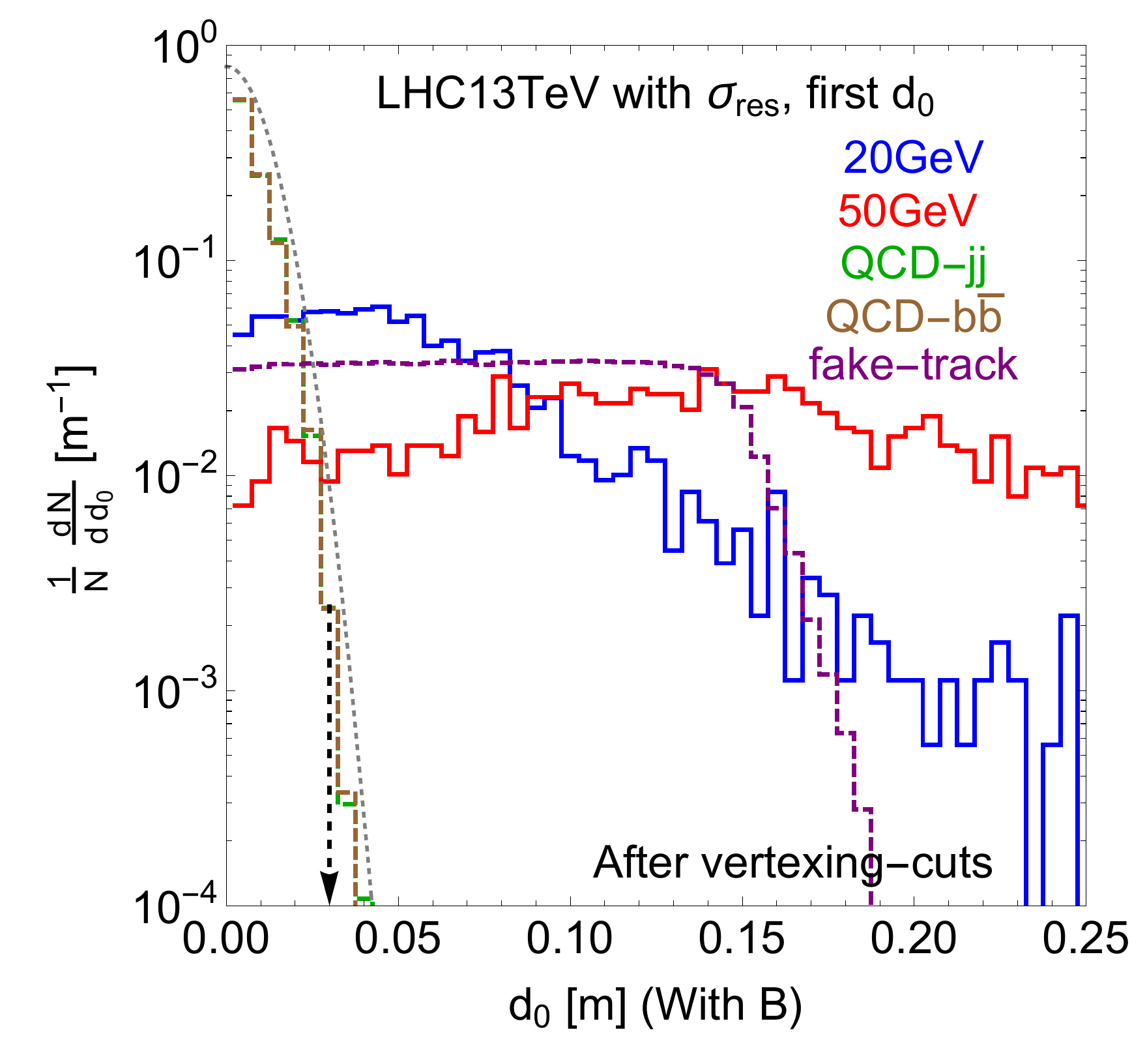} 
    \end{tabular}
    \caption{The transverse impact parameter $d_0$ distributions for QCD background, fake-track background and the signal. This figure is similar to Fig.~\ref{fig:d0T}, but only the leading track distribution is displayed. }
    \label{fig:d0-first-track}
\end{figure}

In Fig.~\ref{fig:d0-first-track}, we show the transverse impact parameter $d_0$ distribution of the leading track for QCD background, fake-track background, and the signal. This figure is similar to Fig.~\ref{fig:d0T}, but with only the leading track included. 

\begin{figure}[ptb]
    \centering
    \begin{tabular}{cc}
        \includegraphics[width=0.33 \columnwidth]{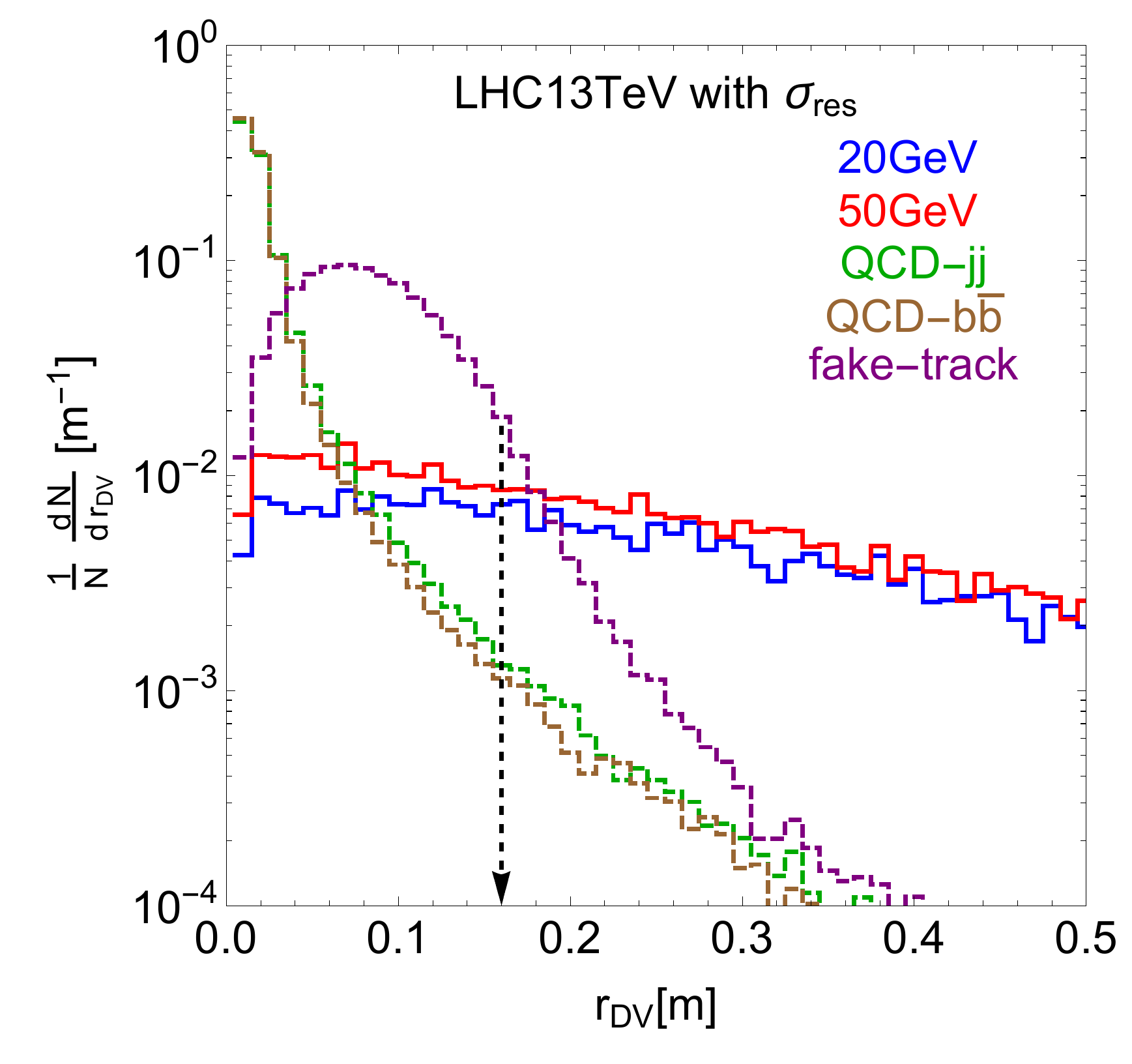} &
        \includegraphics[width=0.33 \columnwidth]{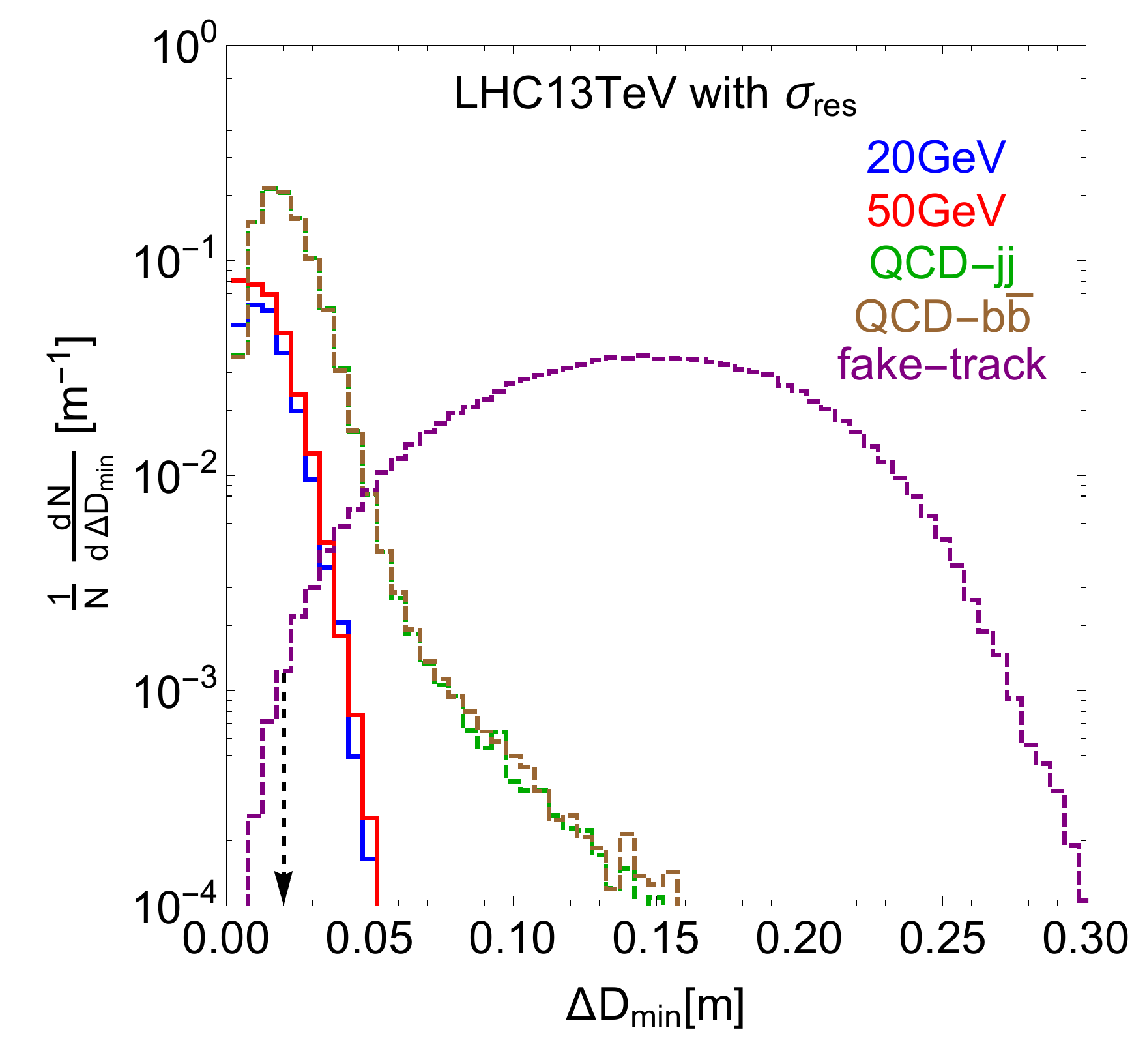}    
    \end{tabular}
    \begin{tabular}{ccc}
        \includegraphics[width=0.33 \columnwidth]{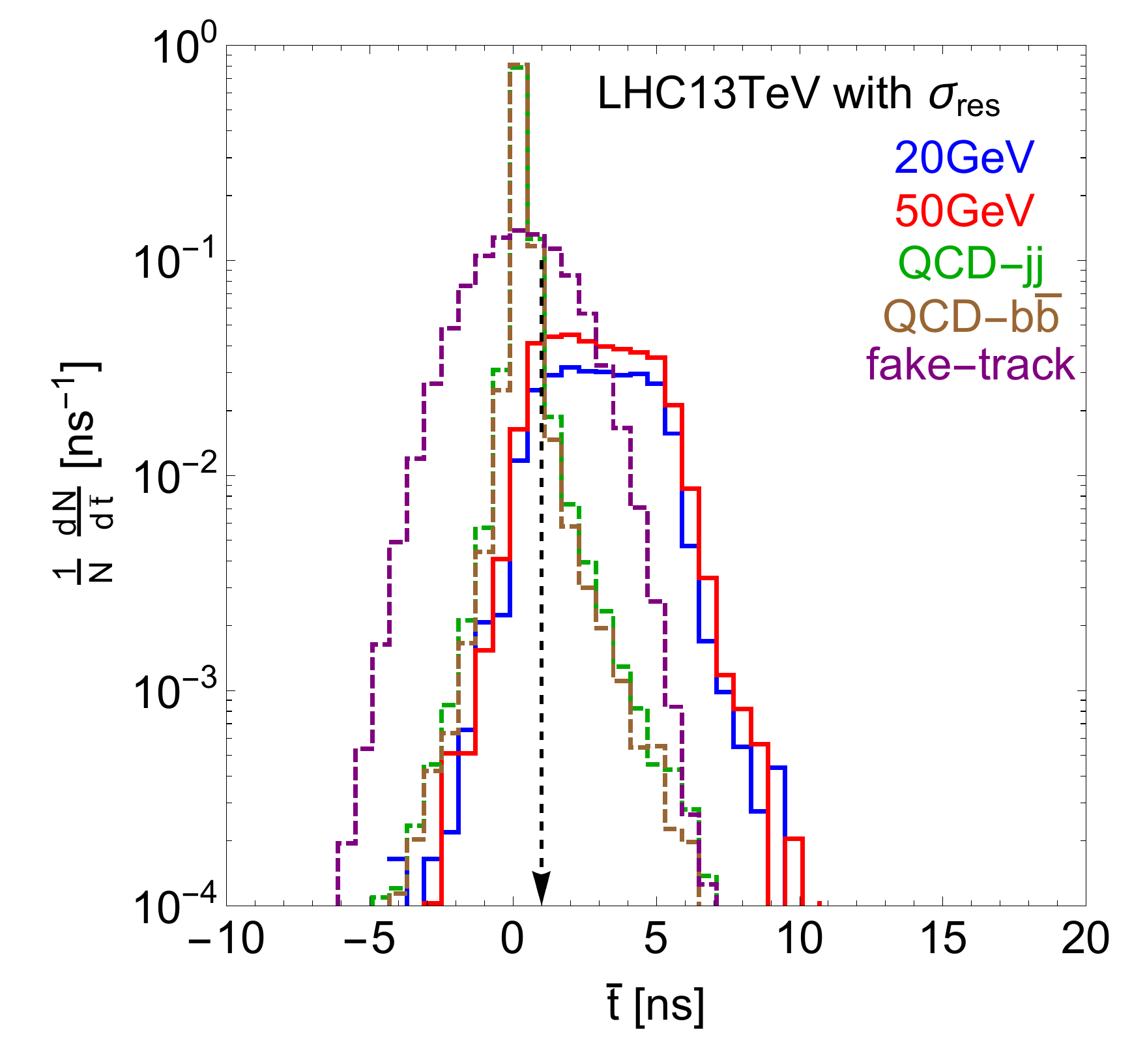} &
        \includegraphics[width=0.33 \columnwidth]{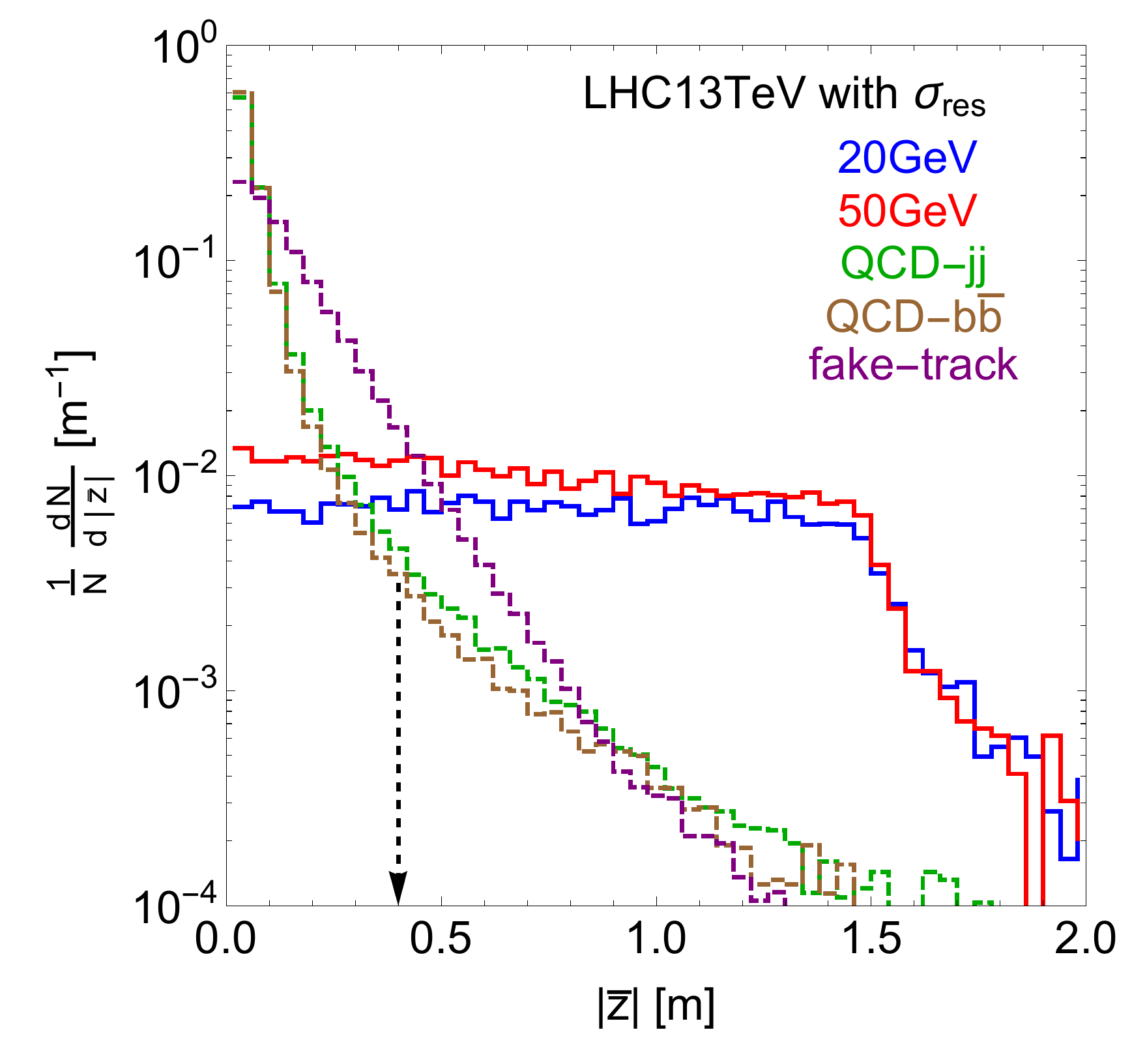}  &
        \includegraphics[width=0.33 \columnwidth]{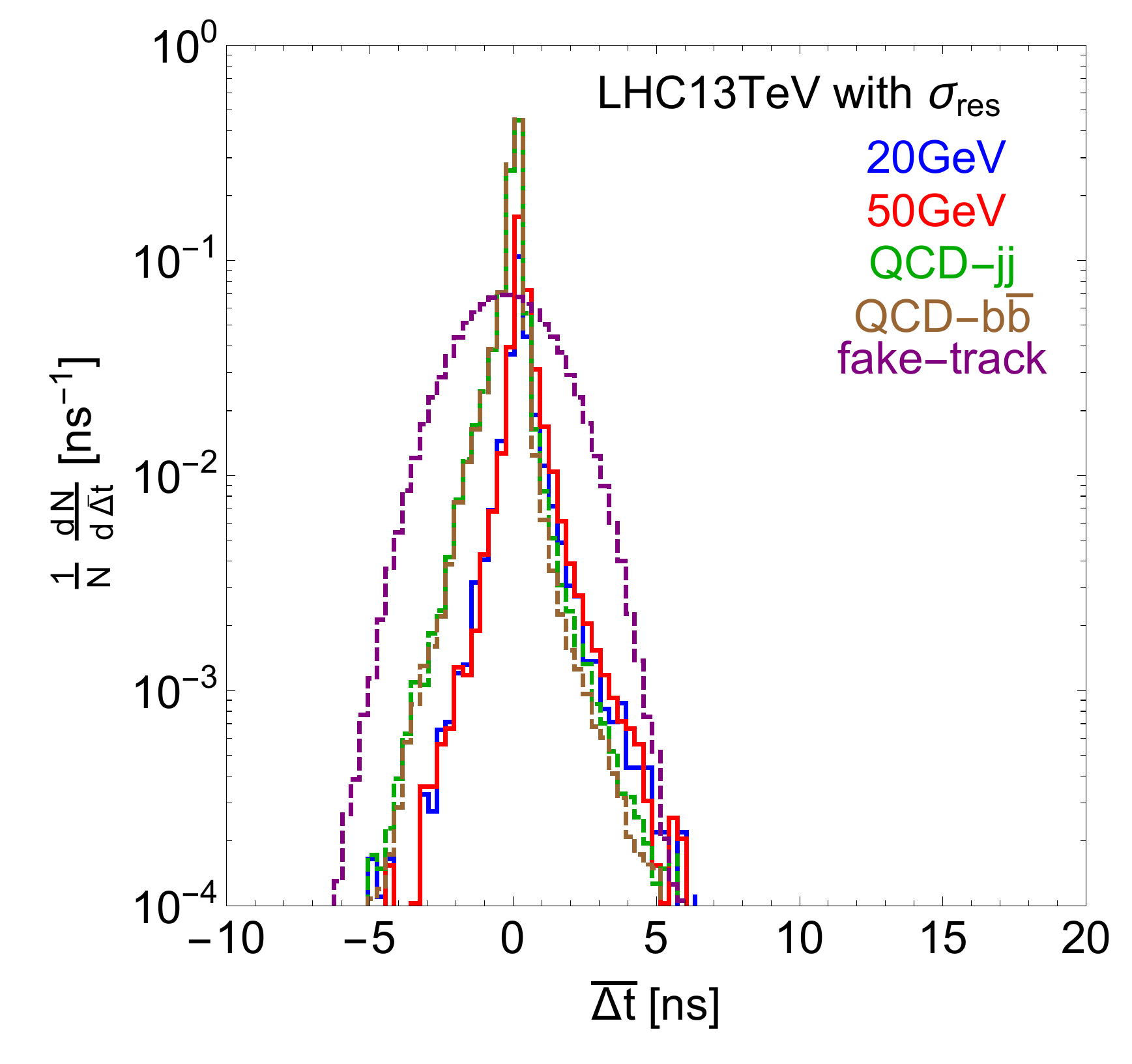}  \\
        \includegraphics[width=0.33 \columnwidth]{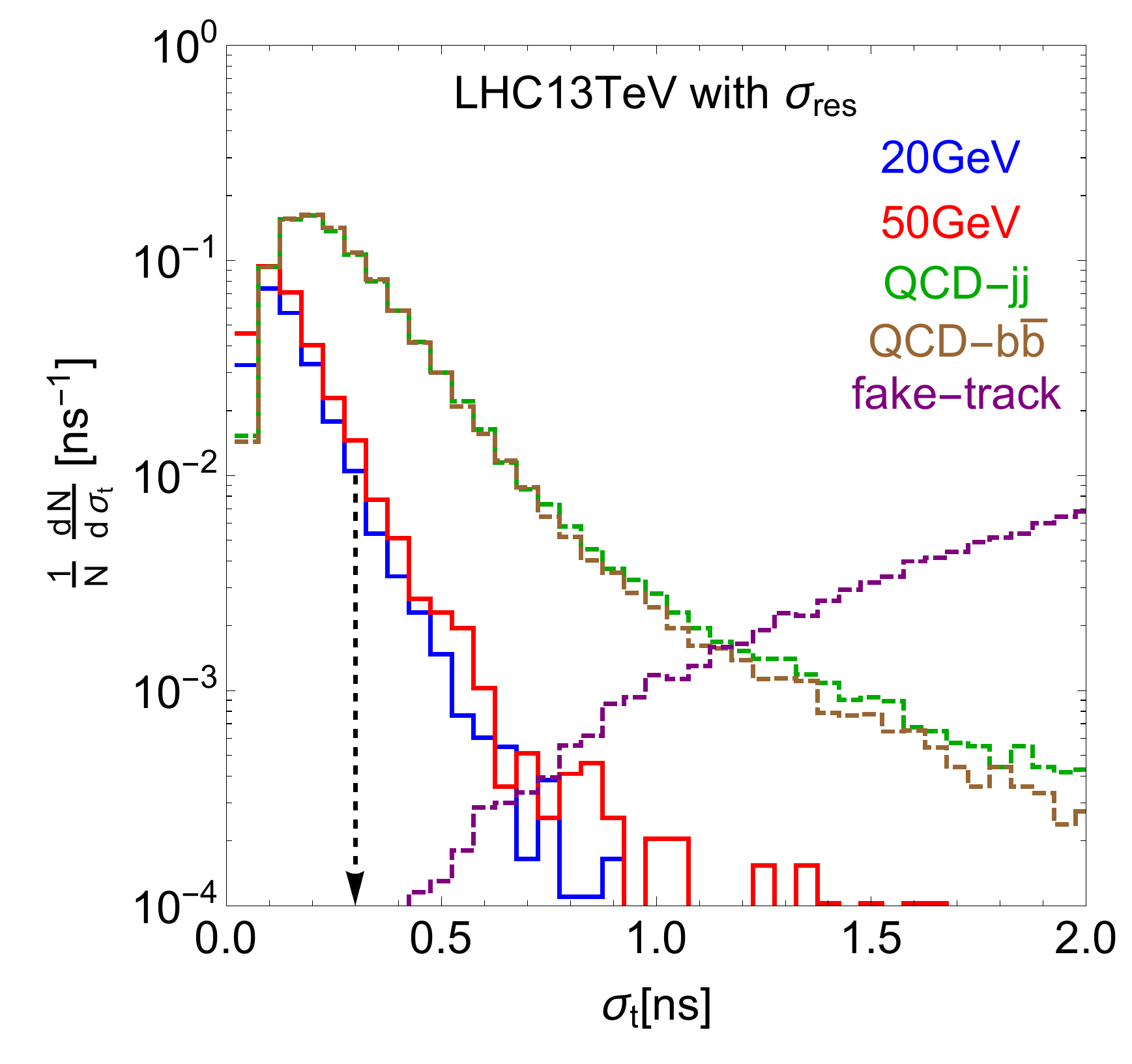} &
        \includegraphics[width=0.33 \columnwidth]{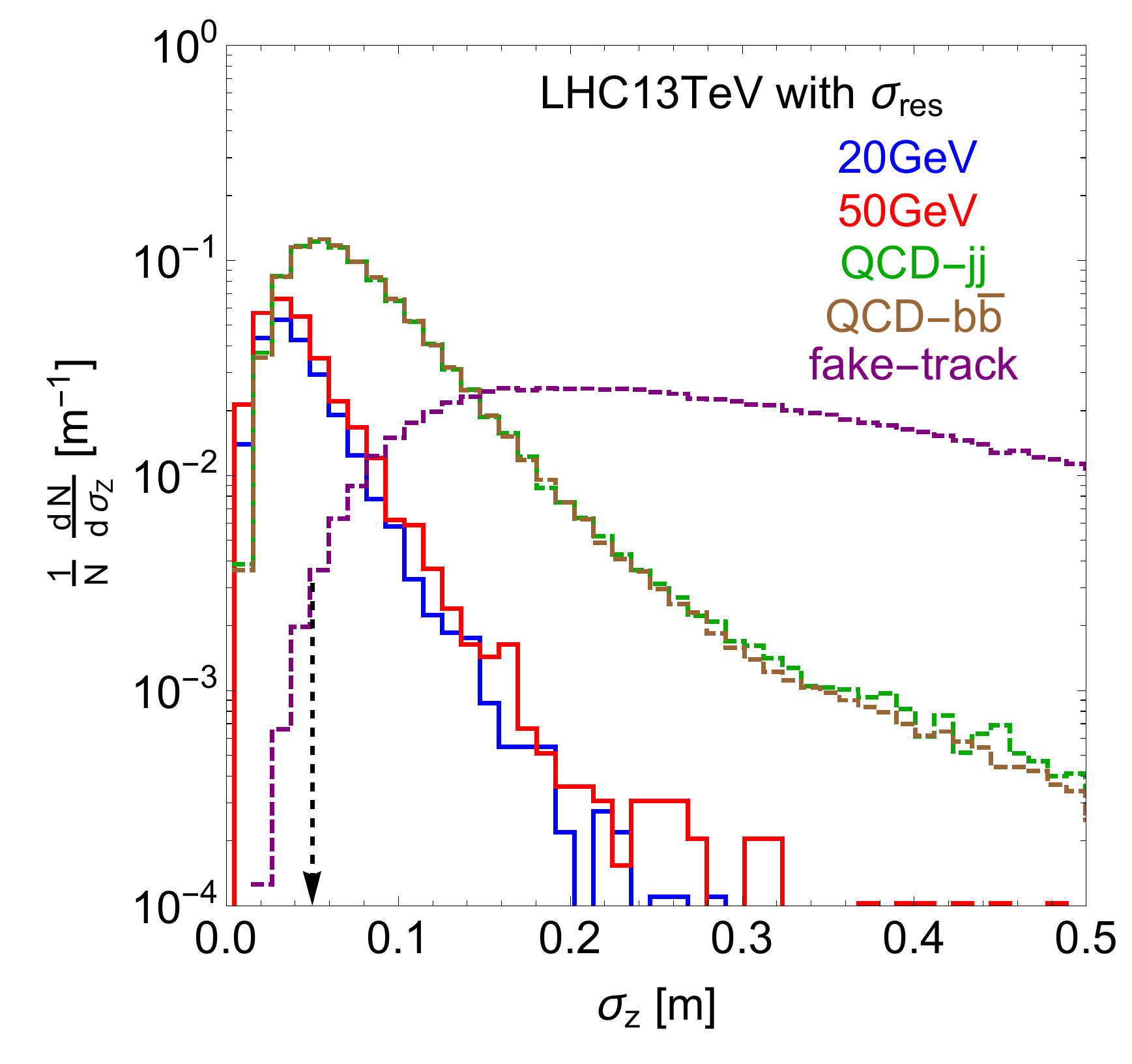} &
        \includegraphics[width=0.33 \columnwidth]{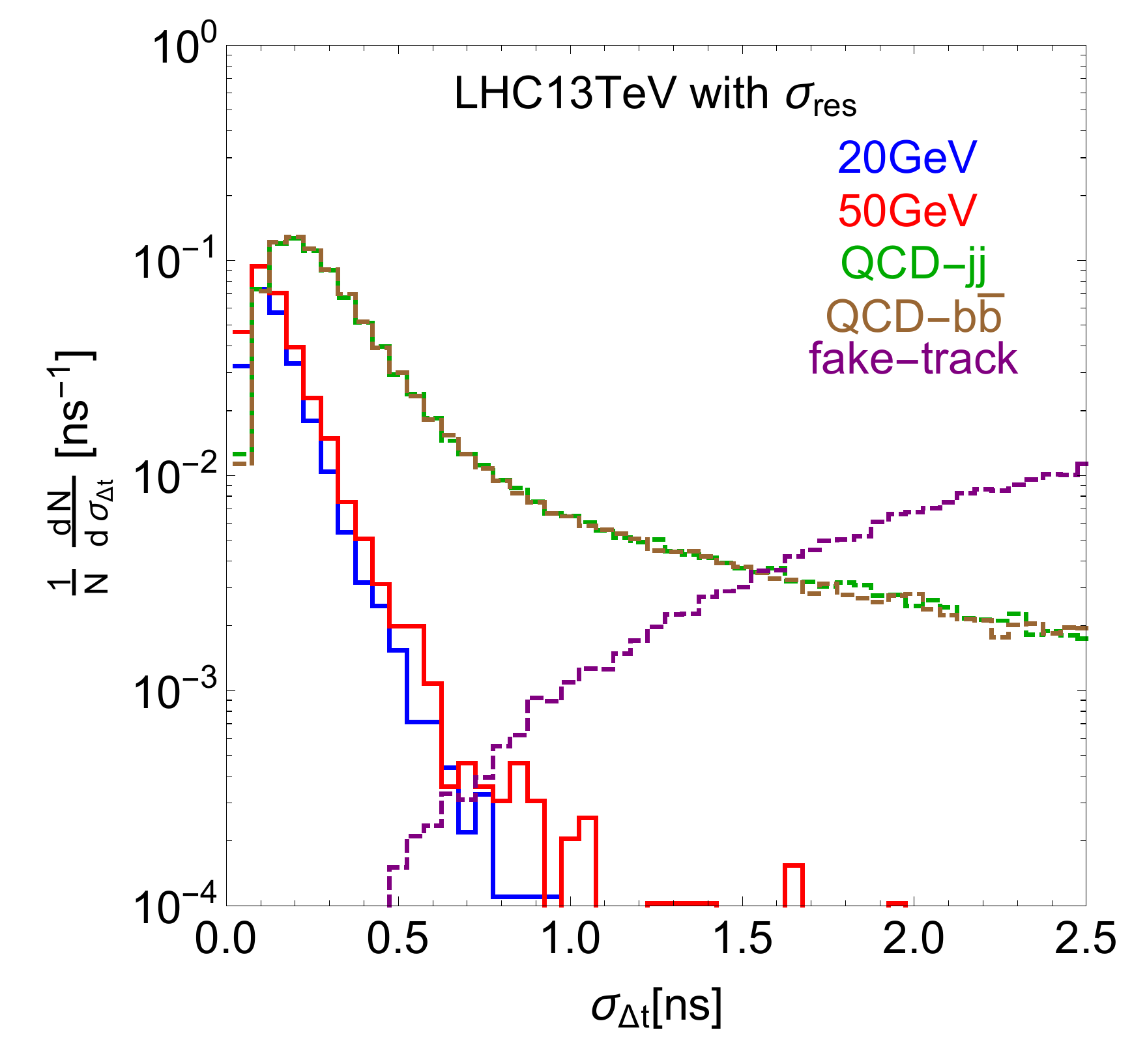}  
    \end{tabular}
    \caption{(VBF channel) The kinetic variable distributions for the QCD background, fake-track background and the signal with angular resolution effect included. The parameter setup is the same as the ggF channel in Fig. \ref{fig:9varswithRES}.}
    \label{fig:VBF-precuts}
\end{figure}

\begin{figure}[ptb]
    \centering
    \begin{tabular}{cccc}
        \includegraphics[width=0.24 \columnwidth]{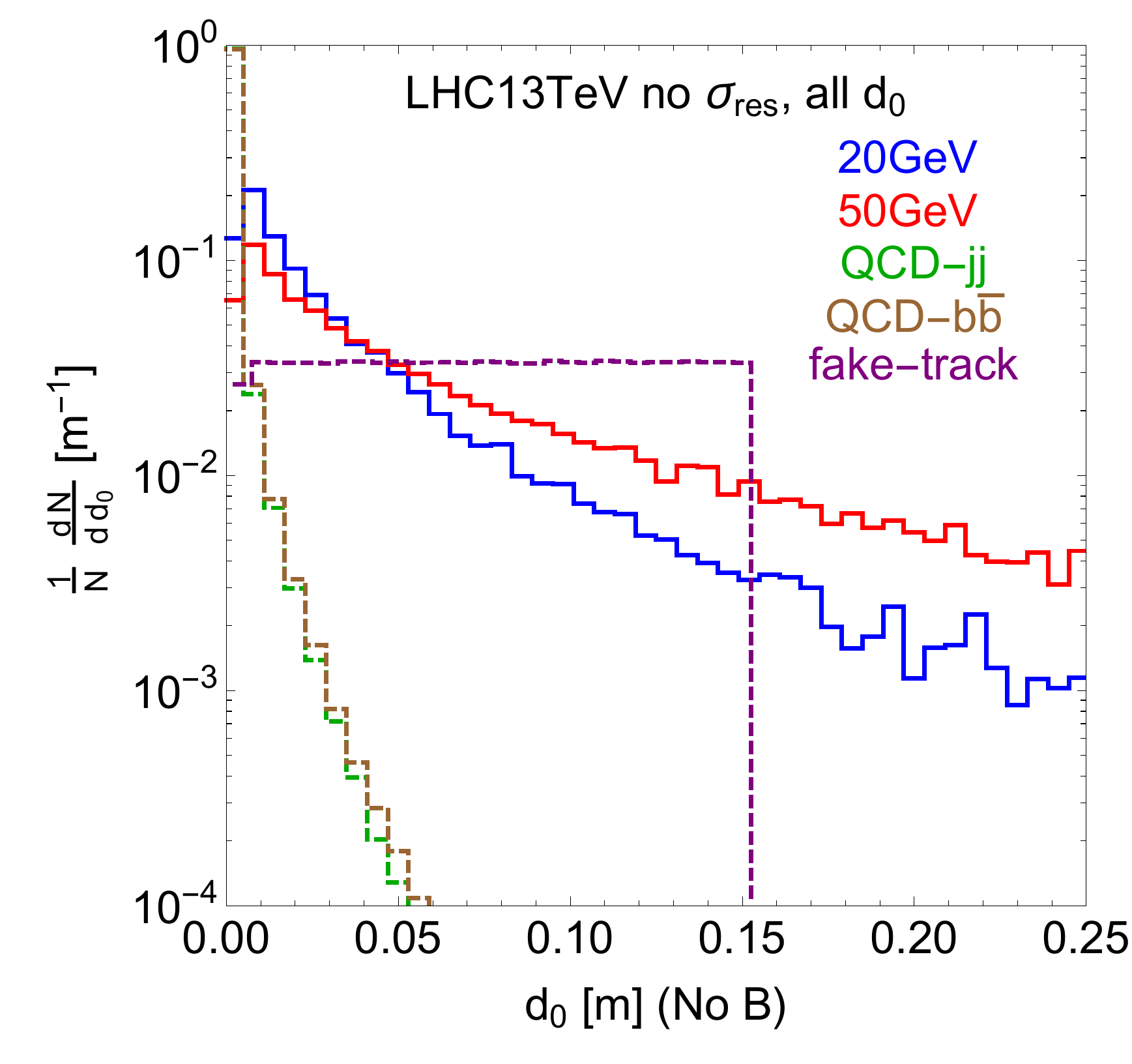}  & 
        \includegraphics[width=0.24 \columnwidth]{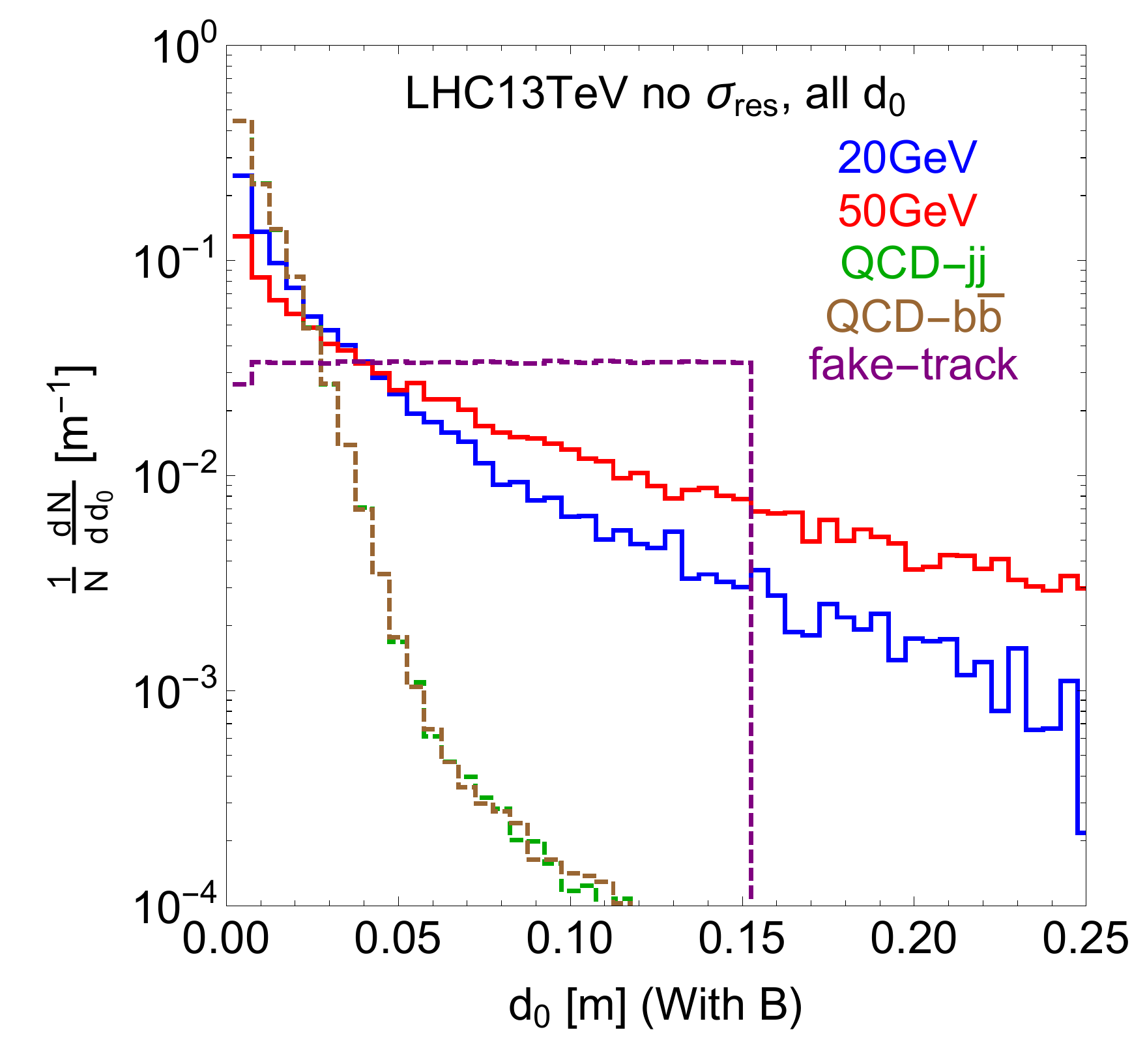}  
        &
        \includegraphics[width=0.24 \columnwidth]{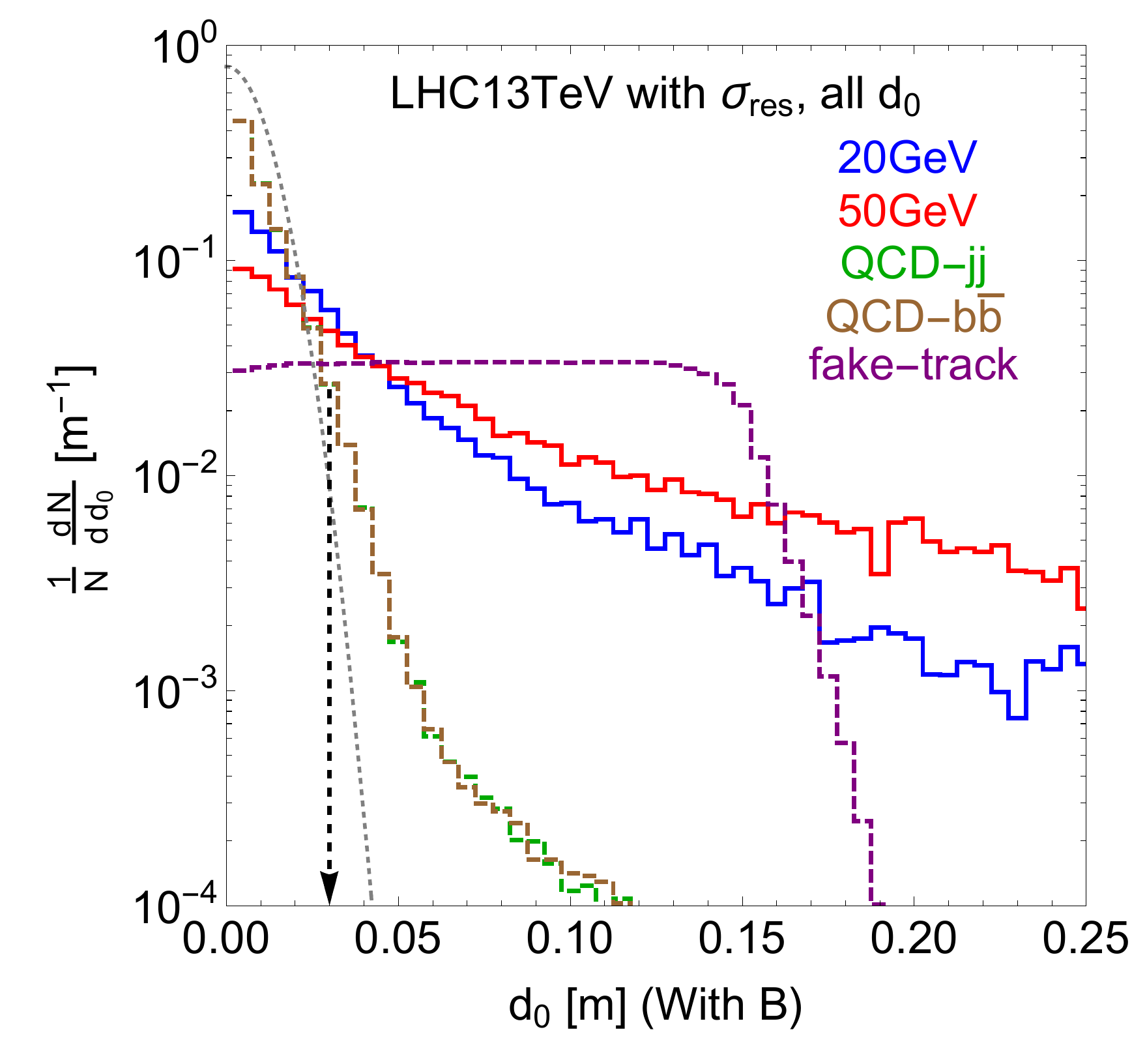}  &
        \includegraphics[width=0.24 \columnwidth]{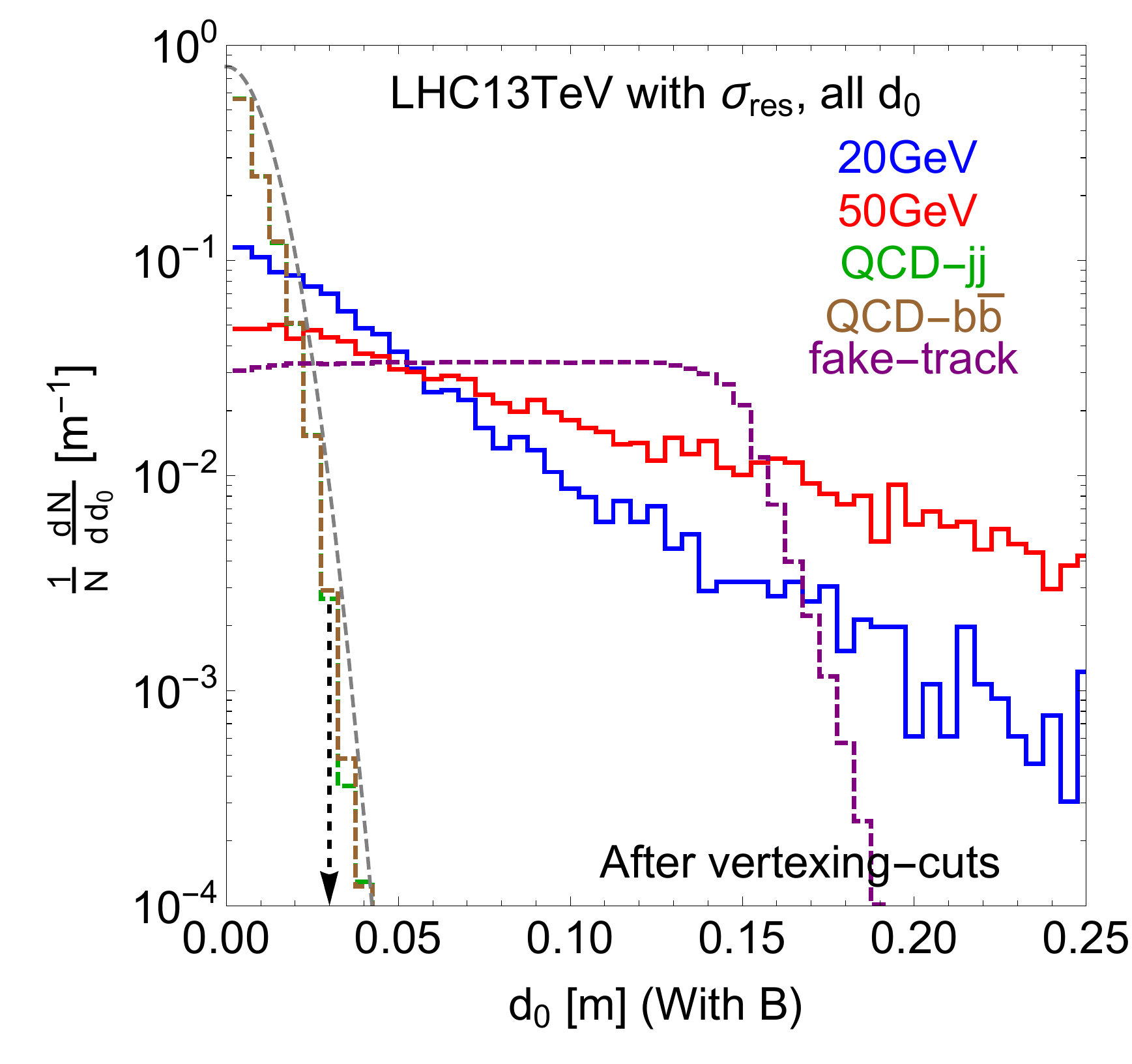} \\
        \includegraphics[width=0.24 \columnwidth]{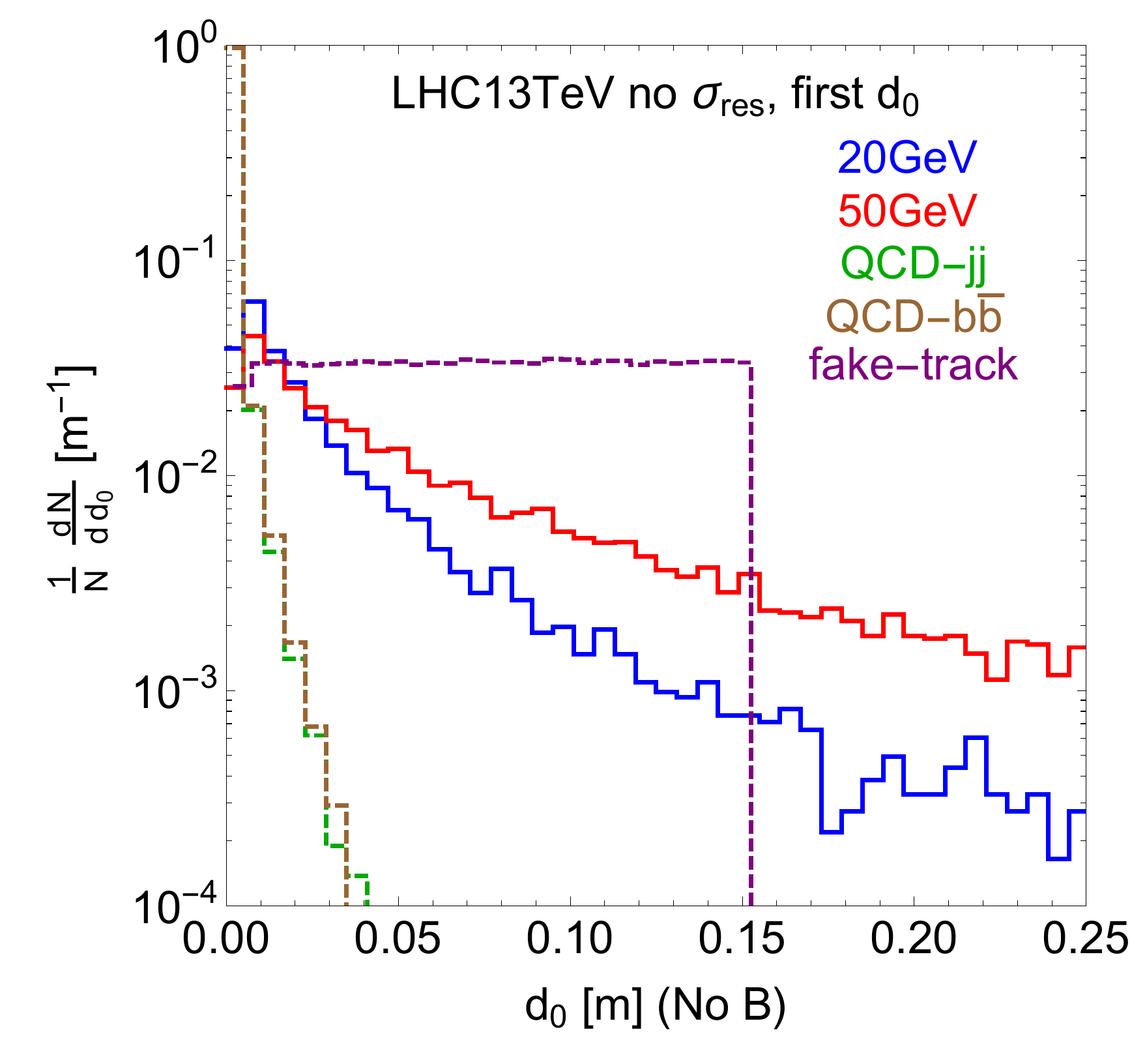}  & 
        \includegraphics[width=0.24 \columnwidth]{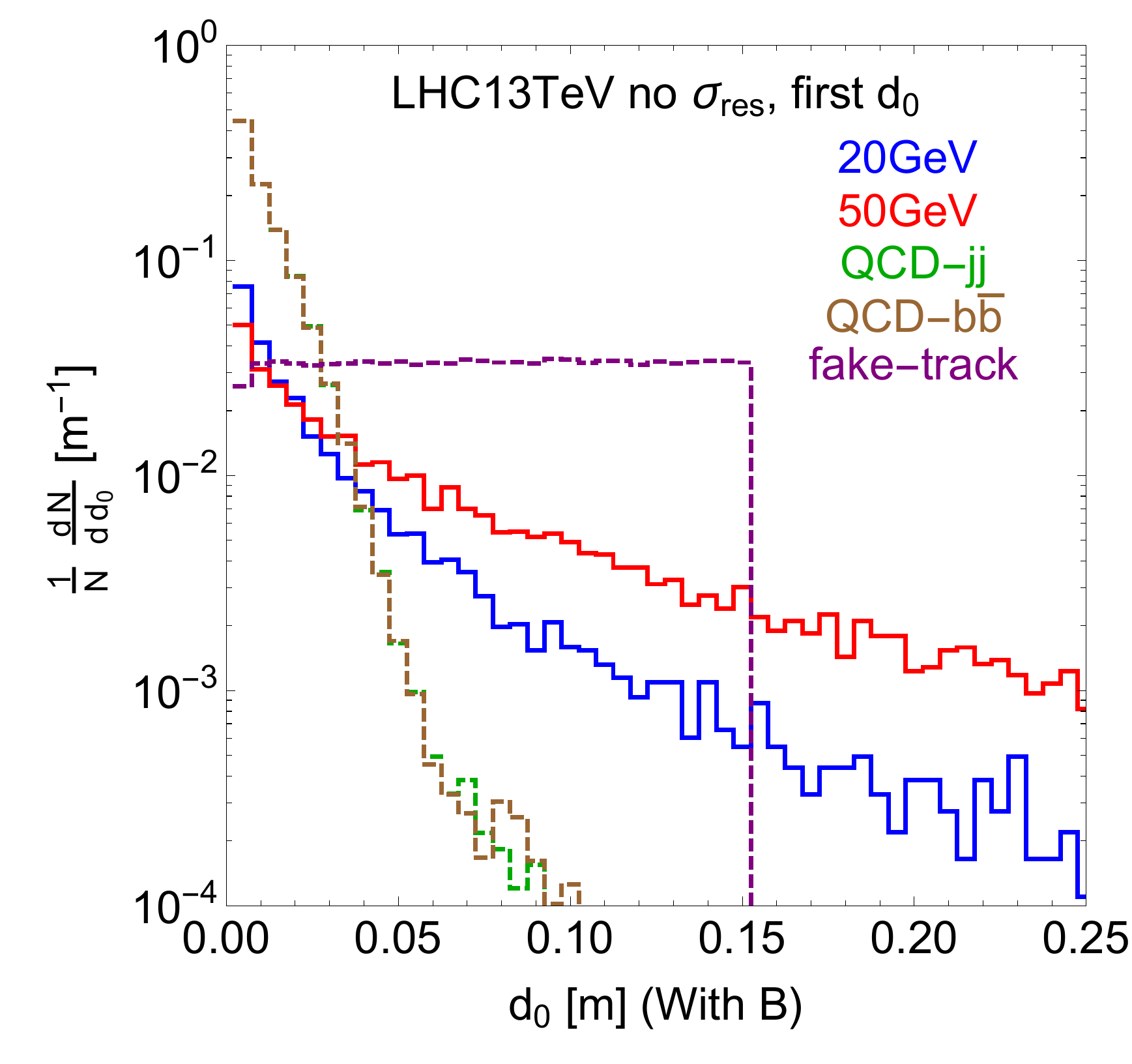} &
        \includegraphics[width=0.24 \columnwidth]{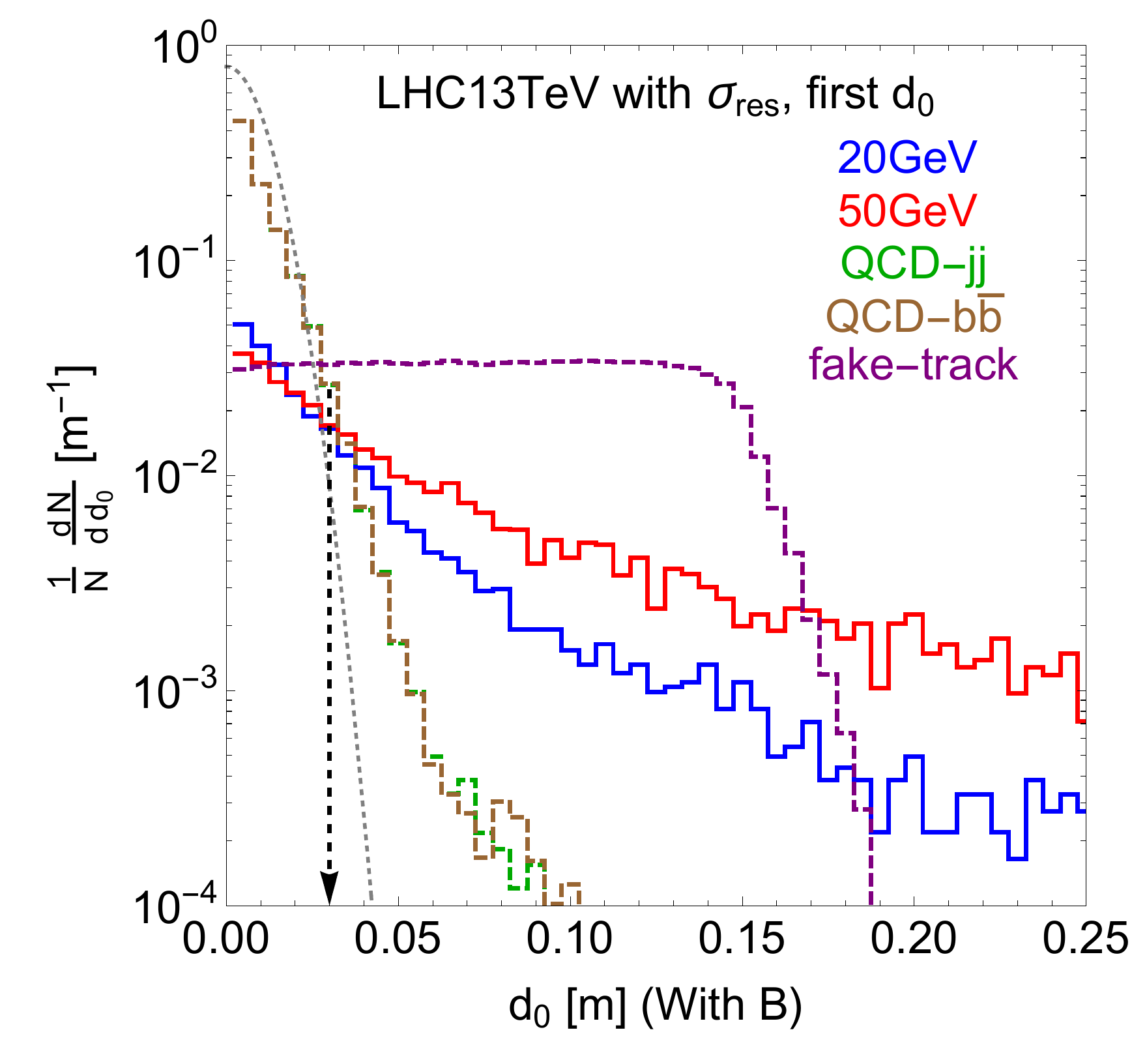}  &
        \includegraphics[width=0.24\columnwidth]{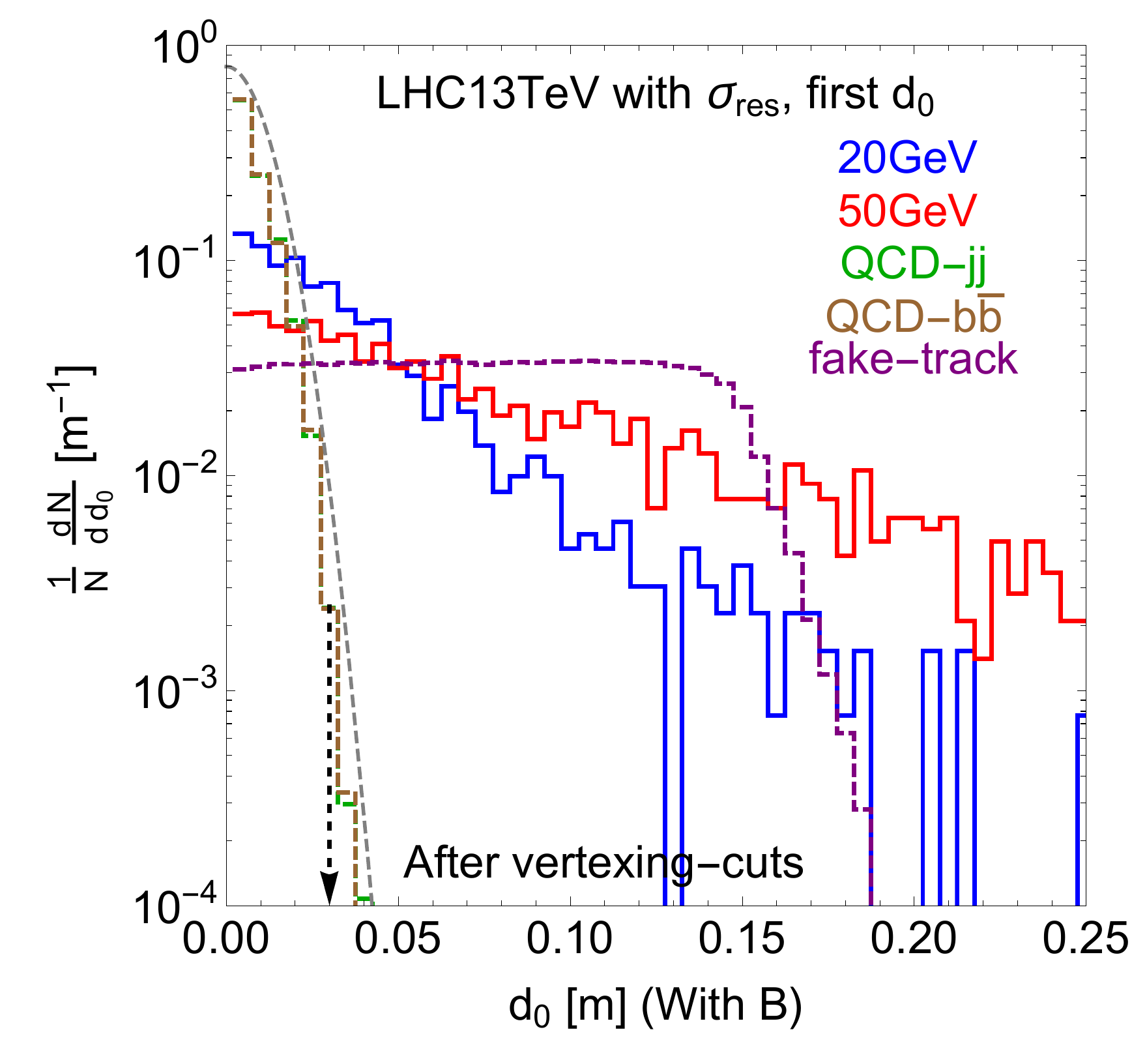} \\
    \end{tabular}
    \caption{(VBF channel) The distributions for transverse impact parameter $d_0$ for QCD background, fake-track background and the signal. The parameter setup is the same as the ggF channel in Fig.~\ref{fig:d0T} and Fig.~\ref{fig:d0-first-track}.}
    \label{fig:VBF-d0T}
\end{figure}

In Fig.~\ref{fig:VBF-precuts} and Fig.~\ref{fig:VBF-d0T}, the kinetic variables and $d_0$ distributions for VBF channel are given.
One can see that the distributions of the VBF channel are similar to the ggF channel.

\clearpage

\bibliographystyle{utphys}
\bibliography{ref}

\end{document}